\documentclass[fleqn,usenatbib]{mnras}

\usepackage{amsfonts}

\usepackage[T1]{fontenc}
\usepackage{ae,aecompl}
\usepackage[position=top]{subfig}
\usepackage{float}
\usepackage{comment}
\usepackage{bm}

\usepackage{grffile}
\usepackage{etoolbox}
\makeatletter 
  \patchcmd{\NAT@citex}
    {\@citea\NAT@hyper@{%
      \NAT@nmfmt{\NAT@nm}%
      \hyper@natlinkbreak{\NAT@aysep\NAT@spacechar}{\@citeb\@extra@b@citeb}%
      \NAT@date}}
    {\@citea\NAT@nmfmt{\NAT@nm}%
    \NAT@aysep\NAT@spacechar\NAT@hyper@{\NAT@date}}{}{}

  \patchcmd{\NAT@citex}
    {\@citea\NAT@hyper@{%
      \NAT@nmfmt{\NAT@nm}%
      \hyper@natlinkbreak{\NAT@spacechar\NAT@@open\if*#1*\else#1\NAT@spacechar\fi}%
        {\@citeb\@extra@b@citeb}%
      \NAT@date}}
    {\@citea\NAT@nmfmt{\NAT@nm}%
    \NAT@spacechar\NAT@@open\if*#1*\else#1\NAT@spacechar\fi\NAT@hyper@{\NAT@date}}
    {}{}
\makeatother

\usepackage{etoolbox}
 \makeatother
\usepackage{graphicx}	
\usepackage{amsmath}	
\usepackage{amssymb}	
\usepackage{pifont}
\title[Population III stellar systems]{Dynamical evolution of Population III stellar systems and the resulting binary statistics}

\author[B. Liu, G. Meynet and V. Bromm]{Boyuan Liu\textsuperscript{\href{https://orcid.org/0000-0002-4966-7450}{\includegraphics[width=2.5mm]{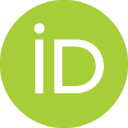}}\,}\thanks{E-mail: boyuan@utexas.edu}$^{1}$, 
Georges Meynet$^{2}$
and Volker Bromm$^{1}$
\\
$^{1}$Department of Astronomy, University of Texas, Austin, TX 78712, USA\\
$^{2}$Geneva Observatory, University of Geneva, Chemin des Maillettes 51, 1290 Sauverny, Switzerland
}

\date{Accepted XXX. Received YYY; in original form ZZZ}

\pubyear{2020}

\begin{document}
\label{firstpage}
\pagerange{\pageref{firstpage}--\pageref{lastpage}}
\maketitle

\begin{abstract}
We use N-body simulations to study the dynamical evolution of Population~III (Pop~III) stellar systems and the resulting binary statistics. We design a physically-motivated framework for the initial conditions of Pop~III star clusters, based on small-scale hydrodynamic simulations and the scale-free nature of disk evolution during Pop~III star formation. Our novel approach enables us to explore the dependence of binary statistics on initial conditions and arrive at more robust predictions for the signals of Pop~III X-ray binaries (XRBs) and binary black hole (BBH) mergers, compared to simple extrapolations of Pop~III protostar systems. We find that binary properties are highly sensitive to the initial cluster size and distribution of binary separation, while the effect of initial mass function is relatively minor. Our simulations predict less close binaries, and thus, significantly lower efficiencies (by a factor of $\sim 10-10^{4}$) for the formation and accretion of Pop~III XRBs, than found in previous studies, implying that the contribution of Pop~III XRBs to the cosmic X-ray background is negligible and their feedback effects are unimportant. We estimate the efficiency of Pop~III BBH mergers as $\sim 10^{-5}-10^{-4}\ \rm M_{\odot}^{-1}$, for which 3-body hardening by surrounding stars in dense star clusters or close binary interactions is required to facilitate in-spirals of BBHs. All simulation data, including catalogs of Pop~III binaries and multiple systems, are \href{https://drive.google.com/drive/folders/1JHBjhSBDPT3jdipdQtmr2IJ5TV6Xe6Id?usp=sharing}{publicly available}.
\end{abstract}



\section{Introduction}
\label{s1}
Stars are not alone. More than half of nearby stars are in binary/multiple systems, especially for massive stars (e.g. \citealt{sana2012binary,sana2013}). Binary interaction has significant and complex impacts on stellar evolution and nucleosynthesis, crucial for many aspects in astrophysics, as stars are the building blocks of galaxies. Therefore, in the past decades, astronomers developed the machinery of binary population synthesis (BPS; see \citealt{han2020binary} for a review), which is applied to numerous topics, such as carbon-enhanced-metal-poor (CEMP) stars (e.g. \citealt{izzard2009population,abate2013wind,hansen2016}), type Ia supernovae (SNe; e.g. \citealt{meng2009single,toonen2012supernova}), X-ray binaries (XRBs; e.g. \citealt{liu2006population,fragos2013xray,shao2015population}), synthetic spectra of stellar populations in galaxies (e.g. \citealt{han2007binary,stanway2016stellar}), reionization (e.g. \citealt{mirabel2011,gotberg2020contribution,secunda2020delayed}), and mergers of compact objects (e.g. \citealt{belczynski2016first,kruckow2018progenitors,mapelli2019properties,bavera2020origin}). 

In particular, it is also important to understand binaries of the first generation of stars in the Universe \citep{bromm2013}, including their remnants, the so-called Population~III (Pop~III), with zero or very low metallicities and top-heavy initial mass functions (IMFs), whose unique chemical and radiative feedback shapes the environments for the first galaxies \citep{bromm2011first} and the subsequent cosmic structure formation process \citep{dayal2018}, leaving imprints even in the local Universe (e.g. \citealt{frebel2015near,ji2015preserving}). For instance, Pop~III XRBs, if formed at a high enough efficiency, can have interesting feedback on star formation in the first galaxies (e.g. \citealt{jeon2014radiative,hummel2015first}), and contribute significantly to heating and ionization of the intergalactic medium (IGM) at $z\gtrsim 10$ (e.g. \citealt{xu2014heating,xu2016x,ryu2016formation}). Combined with the feedback from Lyman-Werner radiation, this leads to distinct signatures in the 21-cm signal from neutral hydrogen (see \citealt{fialkov2013complete,fialkov2014rich,fialkov2017constraining,schauer2019constraining,Mirocha2019,21cm2020}). 

Furthermore, \citet{kinugawa2014possible,kinugawa2020chirp} have shown via classical binary stellar evolution (for isolated binaries) that binary black hole (BBH) mergers originating from Pop~III remnants may constitute a significant fraction of the population observed by LIGO-Virgo \citep{abbott2019gwtc}, with good agreements in merger rate and chirp mass distribution. Pop~III binaries are also promising progenitors \citep{farrell2020gw190521,kinugawa2020formation,bl2020gw190521} for the recently reported BBH merger event GW190521 with unusual BH masses in the pulsational pair-instability mass gap \citep{abbott2020gw190521,abbott2020properties}. However, the uncertainties in Pop~III binary stellar evolution models are significant, resulting in up to two orders of magnitude discrepancy in the BBH merger rate (see e.g. \citealt{hartwig2016,belczynski2017likelihood,kinugawa2020chirp,tanikawa2020merger}). It is still debated how Pop~III stars contribute to the demographics of BBH mergers. Therefore, it is an ongoing effort to explore the parameter space, as well as alternative formation and evolution channels of Pop~III BBH mergers (e.g. \citealt{dvorkin2016metallicity,inayoshi2016gravitational,boyuan2020,bl2020gw190521}).

For all the aforementioned studies regarding Pop~III binaries (and BPS in general), a key input is the (initial) binary statistics, i.e. the faction of stars in binaries, distributions of binary separation, orbital eccentricity, mass ratio and total mass, in newly formed Pop~III star clusters. This serves as the starting point for all Pop~III BPS models. Unfortunately, these characteristics are not well constrained due to the lack of direct observations and limitations in computational power. Previous studies usually adopted idealized estimations, based on the properties of present-day stars (which in principle are very different from Pop~III stars), or Pop~III protostars in small-scale hydrodynamic simulations (e.g. \citealt{stacy2010first,stacy2012first,stacy2016building,greif2012formation,stacy2013constraining,susa2014mass,machida2015accretion,hirano2017formation,sugimura2020birth,chiaki2020}). A common caveat in these approaches is that they apply the binary parameters of low-mass systems (of only a few $\rm M_{\odot}$) to much more massive Pop~III systems (of a few $100\ \rm M_{\odot}$) by artificially scaling-up stellar masses, with other conditions unchanged. In reality, however, the system will expand during accretion due to angular-momentum conservation (see \citealt{sugimura2020birth} for an example of Pop~III wide binaries expanding with accretion). Therefore, such studies tend to overestimate the fraction of stars in close binaries with unphysically small sizes of clusters. 

In light of this, it is timely to revisit Pop~III binary statistics with self-consistent dynamical models for Pop~III star clusters. In this work, we use N-body simulations to study the dynamical evolution of Pop~III systems and the resulting binary statistics, with physically-motivated initial conditions of Pop~III star clusters. Our novel initial condition models (Sec.~\ref{s2}) are based on the fragment properties derived from small-scale hydrodynamic simulations (e.g. \citealt{beuther2019high,susa2019merge,clark2020emergent,oliva2020modeling}), and utilizing the (nearly) scale-free nature of disk evolution during Pop~III star formation. With a comprehensive series of simulations, we derive the general trends in Pop~III cluster evolution (Sec.~\ref{s3}), and systematically explore the dependence of binary statistics on initial conditions (Sec.~\ref{s4}). We further discuss the implications of our improved binary statistics for Pop~III XRBs and BBH mergers (Sec.~\ref{s5}). We offer conclusions and summarize our main findings in Section~\ref{s6}.

\section{Methodology}
\label{s2}

\begin{figure*}
\includegraphics[width=2\columnwidth]{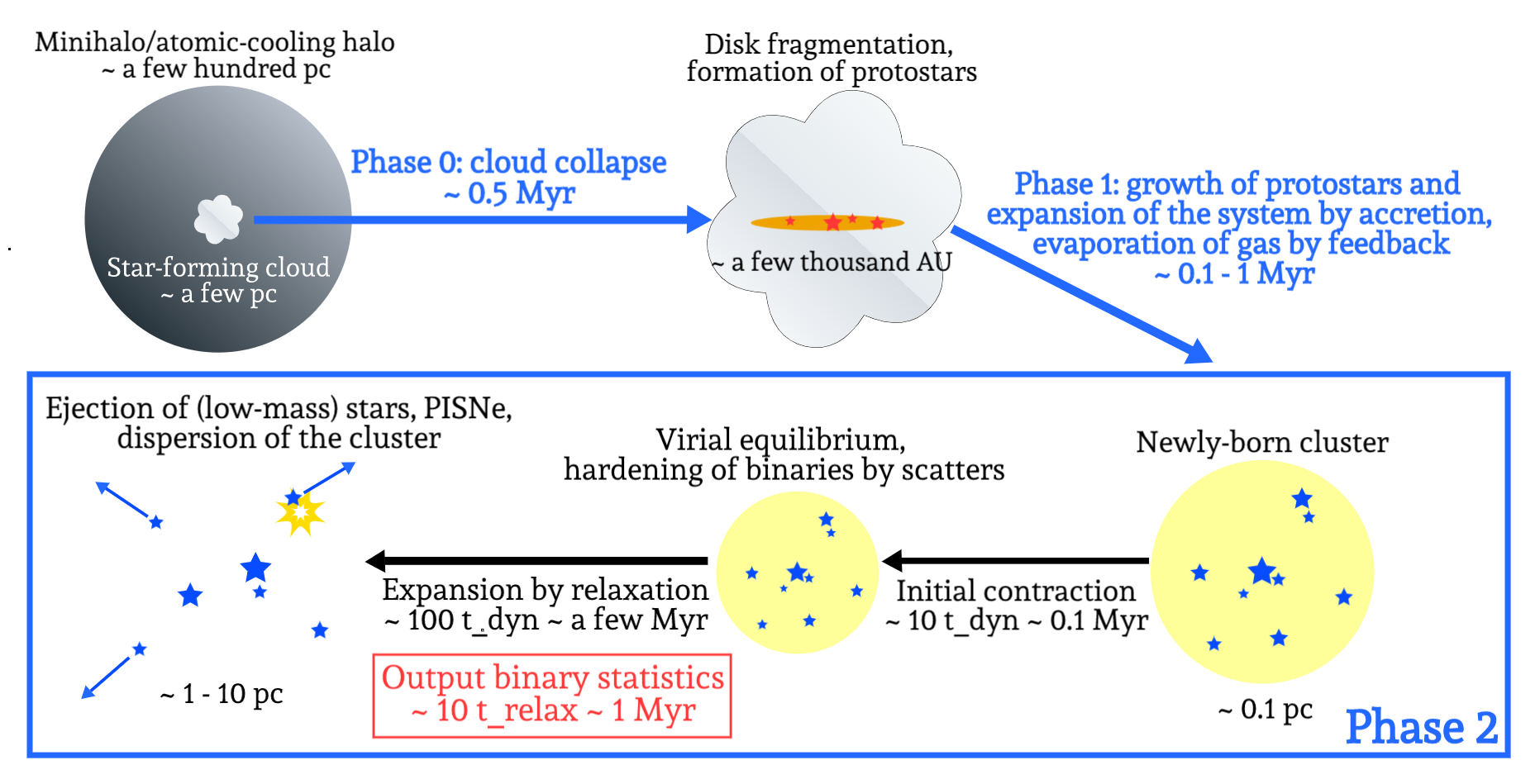}
\caption{Illustrating the three phases of Pop~III star formation at high redshifts (produced with sketchpad: \url{https://sketch.io/sketchpad/}). 
This work focuses on the final two phases, Phase 1 (Sec.~\ref{s2.1} and \ref{s2.2}) and Phase 2 (Sec.~\ref{s3}), whose detailed properties are shown in Fig.~\ref{f1}-\ref{f2} and \ref{qrt}-\ref{colrate}, respectively. Note that the specific values of physical quantities (e.g. the dynamical and relaxation timescales, $t_{\rm dyn}$ and $t_{\rm relax}$) during Phase 2 depend on the Phase 1 model. Here we show the case of the fiducial Phase 1 model \texttt{tf1e2ta1e5a1m1} (see Sec.~\ref{s3} for definition).}
\label{sketch}
\end{figure*}

In this section we introduce our approach of simulating Pop~III stellar systems by combining a novel semi-analytical model of star-forming disk evolution with pure N-body simulations. In general, our method characterizes the two phases of Pop~III binary formation following the initial cloud collapse and disk formation (Phase 0): disk fragmentation and accretion onto protostars (Phase 1) and gravitational interactions among newly-born stars (Phase 2). In Figure~\ref{sketch}, we illustrate the `standard' picture of high-$z$ Pop~III star formation in the context of these three phases.
Phase 0 and 1 have been intensively studied with hydrodynamic simulations, targeting cosmological star-forming clouds in minihaloes (e.g. \citealt{stacy2010first,stacy2012first,stacy2016building,greif2012formation,stacy2013constraining,susa2014mass,machida2015accretion,hirano2017formation,sugimura2020birth,chiaki2020}). However, these simulations adopt different numerical methods and resolution, most of which do not reach the end of Phase 1 (when feedback from protostars completely shuts down fragmentation and accretion) due to limited computational power, rendering the connection between Phase 1 and 2 unclear. 

As it is still computationally prohibitive to simulate the entire Phase 1 and meanwhile resolve the (optically thick) protostellar cores (reaching densities $n\gtrsim n_{\rm ad}\sim 10^{19}\ \rm cm^{-3}$, which is necessary to capture the detailed fragmentation process), we develop a semi-analytical model for Pop III star-forming disks to fill the gap, which includes two aspects: the global properties of the system (i.e. size, total mass and number of stars), and its internal configuration (i.e. distributions in mass and phase space). We relate the global properties to two characteristic timescales based on the scale-free nature of the basic equations of hydrodynamics and gravity during disk evolution (Sec.~\ref{s2.1}), and construct the configurations according to fragment properties derived in previous small-scale simulations (Sec.~\ref{s2.2}). The basic idea is that new fragments are formed around existing fragments which meanwhile grow via accretion and mergers, i.e. hierarchical fragmentation, which is supported by numerous simulations and observations (e.g. \citealt{beuther2019high,susa2019merge,clark2020emergent,oliva2020modeling}). Once the end products of Phase 1 are known, they are used to set up N-body simulations for Phase 2 (Sec.~\ref{s2.3}).

\subsection{Scale-free nature of disk evolution}
\label{s2.1}
As shown in \citet{susa2019merge}, although different small-scale simulations make different assumptions and resolution, it is possible to compare their results considering the scale-free nature of the system during (quasi-isothermal) disk evolution. It turns out that they are all consistent with one universal solution that describes the evolution of the disk/cluster size $R_{c}$, total number $N_{\star}$ and mass $M$ of \textit{surviving}\footnote{Most fragments are lost in mergers, tidal disruption or dynamical ejection during disk evolution, such that the number of surviving fragments is much (a factor of $\sim 10$) smaller than the number of fragments that have ever formed (e.g. \citealt{stacy2013constraining,hirano2017formation,susa2019merge,chiaki2020}).} (proto)stars/fragments, before protostellar feedback becomes effective.

This can be understood by rewriting the fundamental equations of hydrodynamics and gravity 
\begin{align}
    &\frac{\partial \rho}{\partial t}+\nabla\cdot(\rho \bm{v})=0\ ,\notag\\
    &\frac{\partial\bm{v}}{\partial t}+(\bm{v}\cdot\nabla)\bm{v}=-\frac{1}{\rho}\nabla P-\nabla\Phi\ ,\notag\\
    &P=\kappa\rho^{\gamma_{\rm eff}}\ ,\quad \nabla^{2}\Phi = 4\pi G\rho\ ,\notag
\end{align}
with a set of dimensionless variables $\tau$, $\bm{\xi}$, $\eta$, $\bm{\zeta}$, $\sigma$ and $\phi$, following the normalization
\begin{align}
    &t=t_{0}\tau\ ,\quad \bm{r}=r_{0}\bm{\xi}\ ,\quad \rho=\rho_{0}\eta\ ,\notag\\
    &\bm{v}=v_{0}\bm{\zeta}\ ,\quad P=P_{0}\sigma\ ,\quad \Phi=\Phi_{0}\phi\ .\notag
\end{align}
The form of the fundamental equations is preserved 
when the normalization constants satisfy five equations:
\begin{align}
\begin{split}
    &t_{0}=\frac{1}{\sqrt{4\pi G\rho_{0}}}\ ,\quad v_{0}=\sqrt{\kappa\rho_{0}^{\gamma_{\rm eff}-1}}\ ,\\
    & r_{0}=v_{0}t_{0}\ ,\quad \Phi_{0}=v_{0}^{2}\ ,\quad P_{0}=\kappa\rho_{0}^{\gamma_{\rm eff}}\ .
\end{split}\label{e1}
\end{align}
Here $\gamma_{\rm eff}=1.09$ is the effective polytropic index for star-forming primordial gas \citep{omukai1998}. Within this formalism, we can derive the following scaling relations for the size $R_{c}$ and mass $M$ of the system with time $t$:
\begin{align}
    r_{0}\propto t_{0}^{2-\gamma_{\rm eff}}\rightarrow &R_{c}\propto t^{2-\gamma_{\rm eff}}\ ,\notag\\
    \rho_{0}r_{0}^{3}\propto t_{0}^{4-3\gamma_{\rm eff}}\rightarrow &M\propto t^{4-3\gamma_{\rm eff}}\ .\notag
\end{align}
Furthermore, \citet{susa2019merge} found that the (average) total number of \textit{surviving} protostars ($N_{\star}$) also satisfies a simple power-law scaling with time, $N_{\star}\propto t^{0.3}$, when \textit{time in different simulations is normalized to the corresponding density thresholds, $n_{\rm th}$, for identifying protostars or forming sink particles}, such that the normalized time is $t_{\rm norm}\propto n_{\rm th}^{1/2}t_{\rm phs}$, given the original physical time $t_{\rm phys}$ from the simulation. 

\begin{figure*}
\includegraphics[width=1.7\columnwidth]{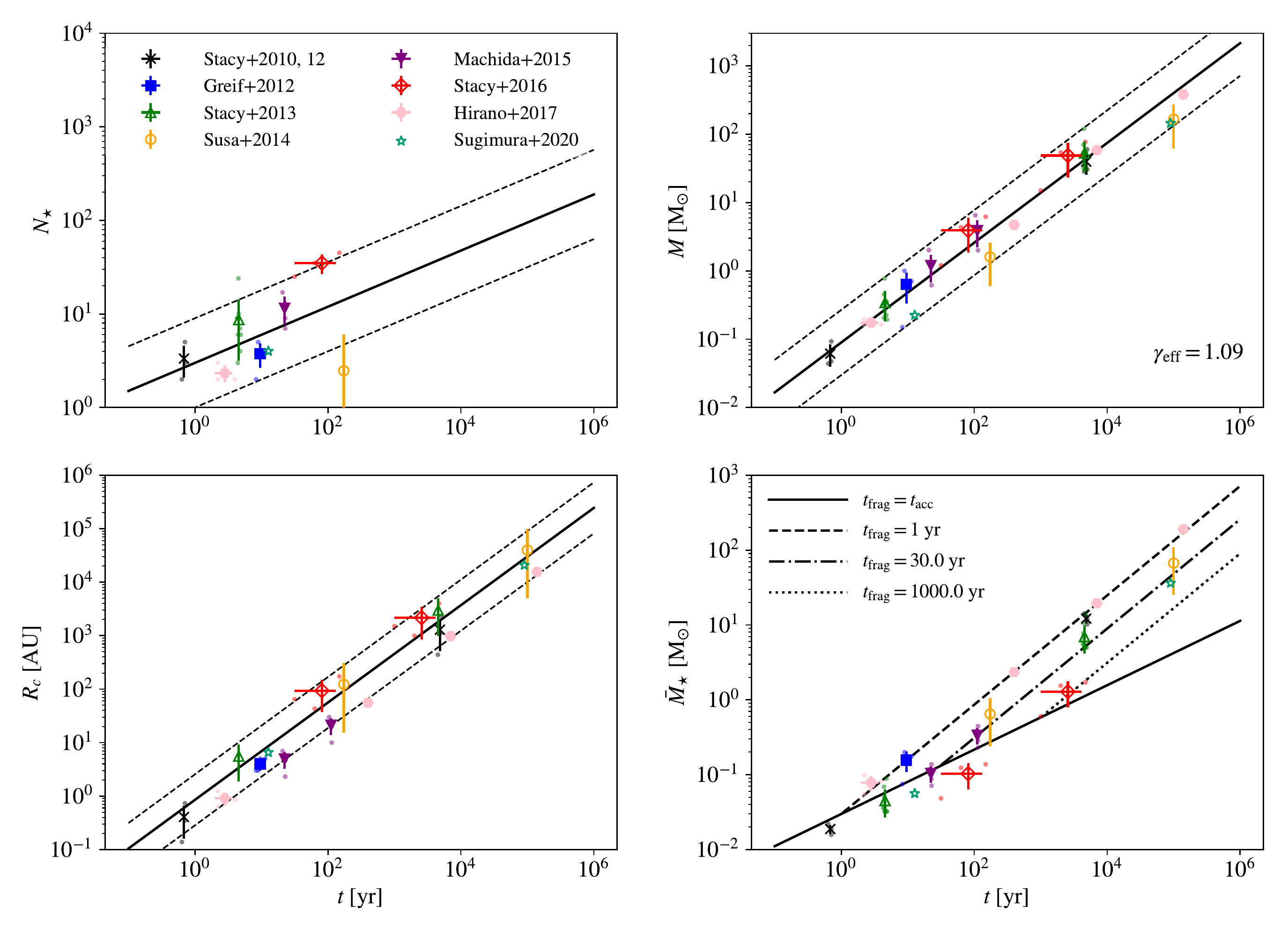}
\caption{Universal solution of Pop~III systems in Phase 1 of disk fragmentation and protostar accretion (Equ.~\ref{e2}-\ref{e4}) for the total number $N_{\star}$ and mass $M$ of (proto)stars, cluster size $R_{c}$, as well as the mean stellar mass $\bar{M}_{\star}\equiv M/N_{\star}$ (clockwise). For $N_{\star}$, $M$ and $R_{c}$, the solution itself is shown with solid lines, and the range of $\sim 0.5$~dex scatters with thin dashed lines. For $\bar{M}_{\star}$, we consider four cases with $t_{\rm frag}=t_{\rm acc}$ (solid), 1~yr (dashed), 30~yr (dashed-dotted) and 1000~yr (dotted), which generally capture the range of outcomes in the literature. 
The underlying simulation data from \citet{stacy2010first,stacy2012first,greif2012formation,stacy2013constraining,susa2014mass,machida2015accretion,stacy2016building,hirano2017formation,sugimura2020birth} are also shown, with filled (empty) marks corresponding to simulations without (with) sink particles. When multiple star-forming clouds are simulated, the mean values are shown with $1\sigma$ error bars, on top of the results for individual clouds denoted by dots, if available. For \citet{hirano2017formation}, data for three runs with different resolution are considered, for which only individual results are shown for the original physical quantities. }
\label{f1}
\end{figure*}

Now we fit the above scaling relations to representative simulation data,
as shown in Fig.~\ref{f1}. For $R_{c}$, $M$ and the mean stellar mass $\bar{M}_{\star}\equiv M/N_{\star}$, each simulation contributes two (sets of)\footnote{\citet{greif2012formation} only contributes one set of data points, as in their case $n_{\rm th}=n_{\rm ad}$.} data points to the plot, one for the original physical quantities ($Q_{\rm phs}$) in the simulation, and the other for the normalized quantities ($Q_{\rm norm}$), which satisfy $R_{c,\rm norm}=(n_{\rm ad}/n_{\rm th})^{\gamma_{\rm eff}/2-1} R_{\rm phs}$, $M_{\rm norm}=(n_{\rm ad}/n_{\rm th})^{3\gamma_{\rm eff}/2-2}M_{\rm phs}$ and $t_{\rm norm}=(n_{\rm ad}/n_{\rm th})^{-1/2}t_{\rm phs}$ according to Equations~(\ref{e1}), given $n_{\rm ad}=10^{19}\ \rm cm^{-3}$ as the critical density above which dense cores become adiabatic (optically-thick). For $N_{\star}$, on the other hand, only normalized quantities are shown. 
These simulations are consistent with the universal solution, within deviations of $\sim 0.5$~dex:
\begin{align}
	N_{\star}&\simeq 3(t/\mathrm{yr})^{0.3}\ ,\label{e2}\\
	R_{c}&\simeq \mathrm{AU}\ (t/ \mathrm{yr})^{2-\gamma_{\rm eff}}\ ,\label{e3}\\
	M&\simeq 400\ \mathrm{M_{\odot}}\ \left[t/(10^{5}\ \rm yr)\right]^{4-3\gamma_{\rm eff}}\ ,\label{e4}
\end{align}
which holds at the early stage of Phase 1 \textit{when the effect of feedback is still unimportant}. 

At later stages, radiative feedback from protostars heats up the gas, such that accretion onto the disk declines \citep{stacy2016building}, and fragmentation is suppressed. Eventually, the in-falling gas is evaporated by UV radiation, and accretion shut down. For simplicity, we model the feedback effect with the accretion and fragmentation timescales, $t_{\rm acc}$ and $t_{\rm frag}$. The former corresponds to the time in which the star cluster can grow (in both size and mass) by accretion before gas evaporation, i.e. the duration of Phase 1. While  the latter denotes the time elapsed before formation and removal (by mergers and ejection) of fragments reach equilibrium such that the total number of \textit{surviving} (proto)stars is saturated. Note that we are always concerned with the number of \textit{surviving} fragments, instead of all fragments that have formed. In general, fragmentation will continue as long as accretion of gas onto the disk is rapid enough to trigger gravitational instability, but the number of surviving fragments can be saturated before accretion is completely shut down. The rate of mergers and ejection of fragments increases with the number of surviving fragments, while the rate of forming new fragments is proportional to the accretion rate. Therefore, an equilibrium will be reached at some moment that is denoted as $t_{\rm frag}$ in our model. Now for the newly-born Pop~III star cluster at the end of Phase 1, we have $N_{\star}=N_{\star}(t=t_{\rm frag})$, $R_{0}=R_{c}(t=t_{\rm acc})$ and $M=M(t=t_{\rm acc})$ from Equations~(\ref{e2}-\ref{e4}).

\subsection{Newly-born Pop~III star clusters}
\label{s2.2}
Given the global properties $R_{0}$, $M$ and $N_{\star}$ of a newly-born cluster, we need to further determine the detailed distribution of stars in mass and phase space. The former is captured by the initial mass function (IMF), $n_{\star}\equiv dN/dm_{\star}$, which is an input element for our model. We assume $n_{\star}\propto m_{\star}^{-\alpha}$ for simplicity, in the mass range $M_{\rm min}-M_{\rm max}$, where $\alpha$ and $M_{\min}$ are treated as adjustable parameters, while $M_{\max}$ can be derived from them and the mean stellar mass $\bar{M}_{\star}\equiv M/N_{\star}$. To cover the wide range of results for the Pop~III IMF in the literature (e.g. \citealt{greif2011simulations,stacy2013constraining,susa2014mass,hirano2015primordial,chon2020supermassive}), we investigate several cases for $\alpha\sim -0.17-2$ and $M_{\rm min}\sim 1-10\ \rm M_{\odot}$ (see Sec.~\ref{s4.3}). By sampling $n_{\star}\equiv n_{\star}(m_{\star}|\alpha,M_{\min},\bar{M}_{\star})$, we distribute $M$ into $N\sim N_{\star}$ stars\footnote{We generate stellar masses until the sum of the generated masses is larger than $M$, and then set the mass of the last star $i$ to $M-\sum_{j=1}^{i-1} m_{\star,j}$. During this process we impose an upper limit of $500\ \rm M_{\odot}$ to the stellar masses. } of masses $m_{\star,i}$, $i=1,\ 2,\ \cdots,\ N$.

Given these $N$ stars with assigned masses, we construct the phase space distribution based on the following features of disk fragmentation and accretion \citep{susa2019merge}:
\begin{itemize}
    \item All stars are approximately in the same plane (inherited from the disk).
    \item Each star in the cluster is the final product of a surviving fragment (i.e. protostellar core). The higher the stellar mass, the earlier the fragment is formed.
    \item New fragments are formed in the accretion disks associated with individual pre-existing fragments, rather than being all associated with the central massive star (see fig.~6 of \citealt{susa2019merge}).
\end{itemize}

\begin{figure}
\includegraphics[width=1\columnwidth]{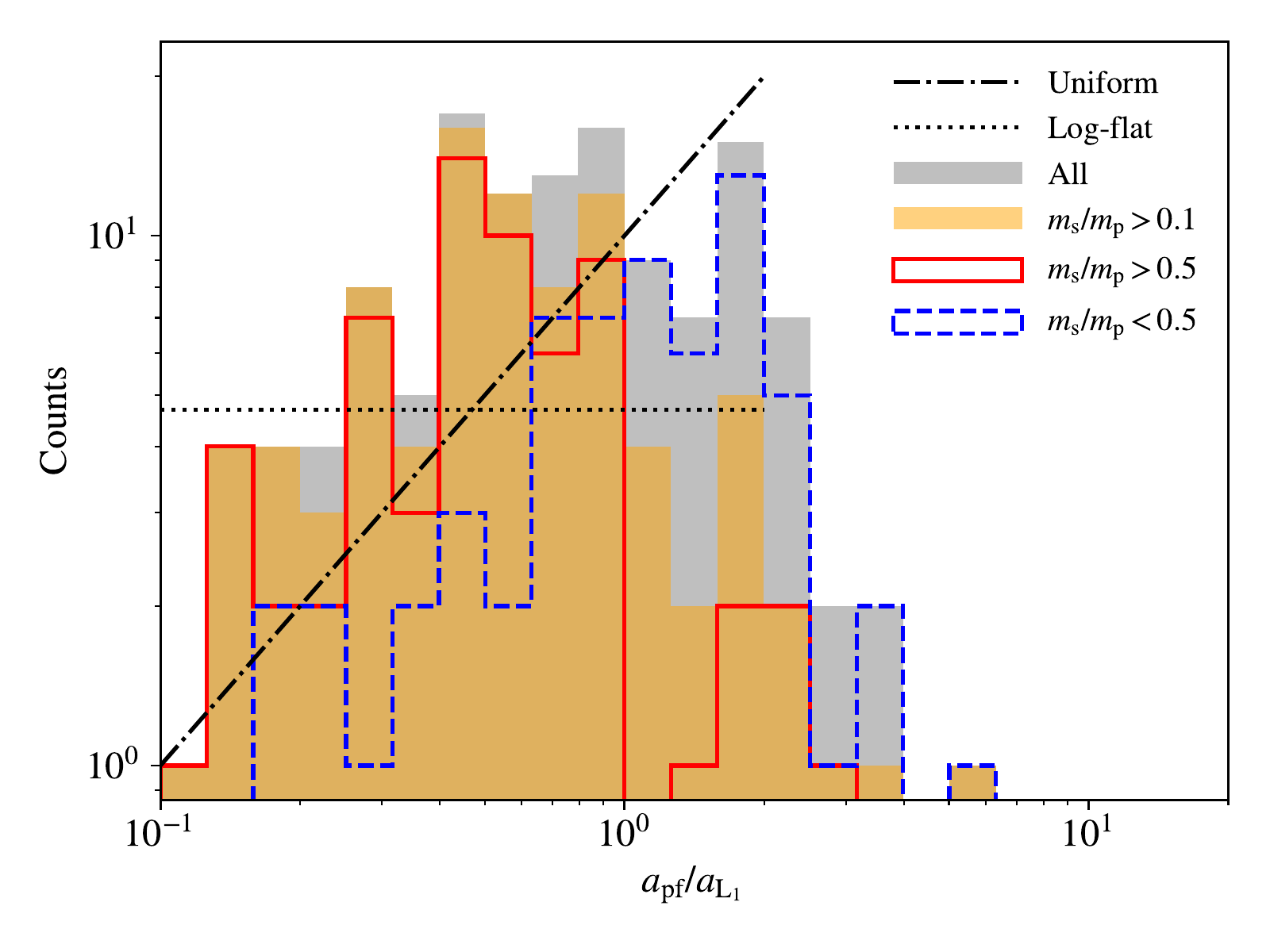}
\caption{Distribution of $a_{\rm pf}/a_{\rm L_{1}}$ from \citealt{susa2019merge} (see their fig.~7), where $a_{\rm pf}$ is the separation between the newly-formed fragment and its `primary' (of mass $m_{\rm p}$), and $a_{\rm L_{1}}$ is the distance from the `primary' to the L1 point in the binary system made of the `primary' and its pre-existing companion (of mass $m_{\rm s}$). }
\label{adis}
\end{figure}

The basic idea is to connect all stars in a hierarchy of pairs/binaries, which reflects the fragmentation history. We first sort the stars in descending order of mass (i.e. $m_{\star,i}>m_{\star,j}$ for $i<j$), which meanwhile establishes the formation sequence of the corresponding fragments in time\footnote{Note that our model considers \textit{surviving} fragments, which are defined as end products instead of specific usually short-lived fragments in the detailed fragmentation process. In reality, fragments emerge and disappear, such that earlier formed fragments do not necessarily survive longer or grow to larger masses \citep{oliva2020modeling}. However, the masses of surviving fragments do generally grow in time. The formation sequence of \textit{surviving} fragments by our definition actually reflects the formation sequence of the corresponding host disks/primaries.}. Then the stars are processed along the formation sequence, such that each star is attached to another star:
	\begin{itemize}
	\item The second most massive star is attached to the most massive star to form the \textit{root} binary (processed).
	\item When $i-1$ stars have been processed ($i>2$), the $i$th star will be attached to one of these $i-1$ stars with probabilities proportional to their masses\footnote{We assume that the rate of forming fragments is proportional to the disk mass, which is then proportional to the mass of the central (most massive) protostar in the disk.}. The chosen star $j$ and star $i$ are grouped into a binary system. This step is repeated until all stars are in binaries.
	\item Each time a binary is formed (with stars $i$ and $j$), the orbital eccentricity $e_{ij}$ is drawn from a thermal distribution $P(e)=2e$ for $e\in [0,1)$ \citep{duquennoy1991multiplicity}.
	\item The semi-major axis $a_{ij}$ is drawn from an uniform distribution 
	whose range is given by $a_{\max}=a_{\rm L_{1}}/(1+e_{ij})$ and $a_{\min}=\max[a_{\rm lobe}/(1-e_{ij}),0.1a_{\max}]$ for $i>2$ and $j>1$. Here $a_{\rm lobe}\approx R_{\star,\rm ZAMS, 1}[0.6q_{1}^{2/3}+\ln(1+q_{1}^{1/3})]/(0.49q_{1}^{1/3})$ \citep{eggleton1983aproximations} is the minimum separation for the primary to not fill its Roche lobe, given the mass ratio of the primary ($m_{1}=m_{\star,j}$) and the secondary ($m_{2}=m_{\star,i}$) $q_{1}=m_{1}/m_{2}$, and the radius of the primary $R_{\star,\rm ZAMS,1}$. $a_{\rm L_{1}}$ is the distance from $j$ to the L1 point of the \textit{parent} binary in which $j$ is the secondary. 
	To avoid too small clusters, inconsistent with the pre-determined size $R_{0}$, we set $a_{\max}=R$ for $j=1$ and $a_{\min}=R/3$ if meanwhile $i=2$ (i.e. the \textit{root} binary). For $j=1$ and $i>2$, $a_{\min}=\max[a_{\rm lobe}/(1-e_{ij}),0.1a_{\rm L_{1}}/(1+e_{ij})]$, where $a_{\rm L_{1}}$ denotes the distance from the most massive star to the L1 point of the \textit{root} binary. 
	\end{itemize}
A key element in this process is the distribution of semi-major axes. Our implementation is based on the fragment properties encountered in small-scale hydrodynamic simulations \citep{susa2019merge}. It is found that when a new fragment is formed around a pre-existing fragment (of mass $m_{\rm p}$, called `primary') that itself has a companion (of mass $m_{\rm s}$, called `secondary'), if the companion mass is not too small ($m_{\rm s}/m_{\rm p}\gtrsim 0.5$), the new fragment tends to reside in the influence sphere of the `primary'. This is illustrated in Fig.~\ref{adis} with the distribution of $a_{\rm pf}/a_{\rm L_{1}}$, where $a_{\rm pf}$ is the separation between the newly-formed fragment and its `primary', and $a_{\rm L_{1}}$ is the distance from the `primary' to the L1 point in the binary system made of the `primary' and its pre-existing companion. It turns out that for ($m_{\rm s}/m_{\rm p}\gtrsim 0.5$), $a_{\rm pf}\sim 0.1-1 a_{\rm L_{1}}$, and $a_{\rm pf}/a_{\rm L_{1}}$ approximately follows an uniform distribution, different from the widely-used log-flat distribution for binary separations \citep{abt1983normal}\footnote{The lower limit $0.1 a_{\rm L_{1}}$ is valid (at the end of Phase 1) only if (1) small-scale ($\lesssim 10$~AU) fragmentation around the protostar is suppressed and (2) binaries expand during accretion due to angular momentum conservation. The former is consistent with the analysis in \citet{liaow2019}, based on the optical depth of $\mathrm{H}_{2}$ emission, i.e. the main coolant in primordial star forming-disks. The latter is supported by simulations (see e.g. \citealt{sugimura2020birth}). However, it may emerge from limited resolution or missing physics (e.g. magnetic fields). Therefore, to explore the case in which $a_{ij}>0.1a_{\rm L_{1}}$ does not hold, we ran test simulations with $a_{\min}=a_{\rm lobe}$, as discussed in Sec.~\ref{s4.4}.}. In our case, for simplicity we only consider the effect of the most massive companion of the `primary'\footnote{In principle, one should consider all companions of the `primary' which are massive enough to influence fragmentation ($m_{\rm s}/m_{\rm p}\gtrsim 0.5$), or the one closest to the `primary' (which seems to be the definition in \citealt{susa2019merge}). This leads to smaller $a_{ij}$. We ran test simulations to evaluate the relevant effects (see Sec.~\ref{s4.4}).}, which means that $m_{\rm s}/m_{\rm p}\gtrsim 1$ always holds (according to the formation sequence) and justifies $a_{\max}=a_{\rm L_{1}}/(1+e_{ij})$.

\begin{figure}
\includegraphics[width=1\columnwidth]{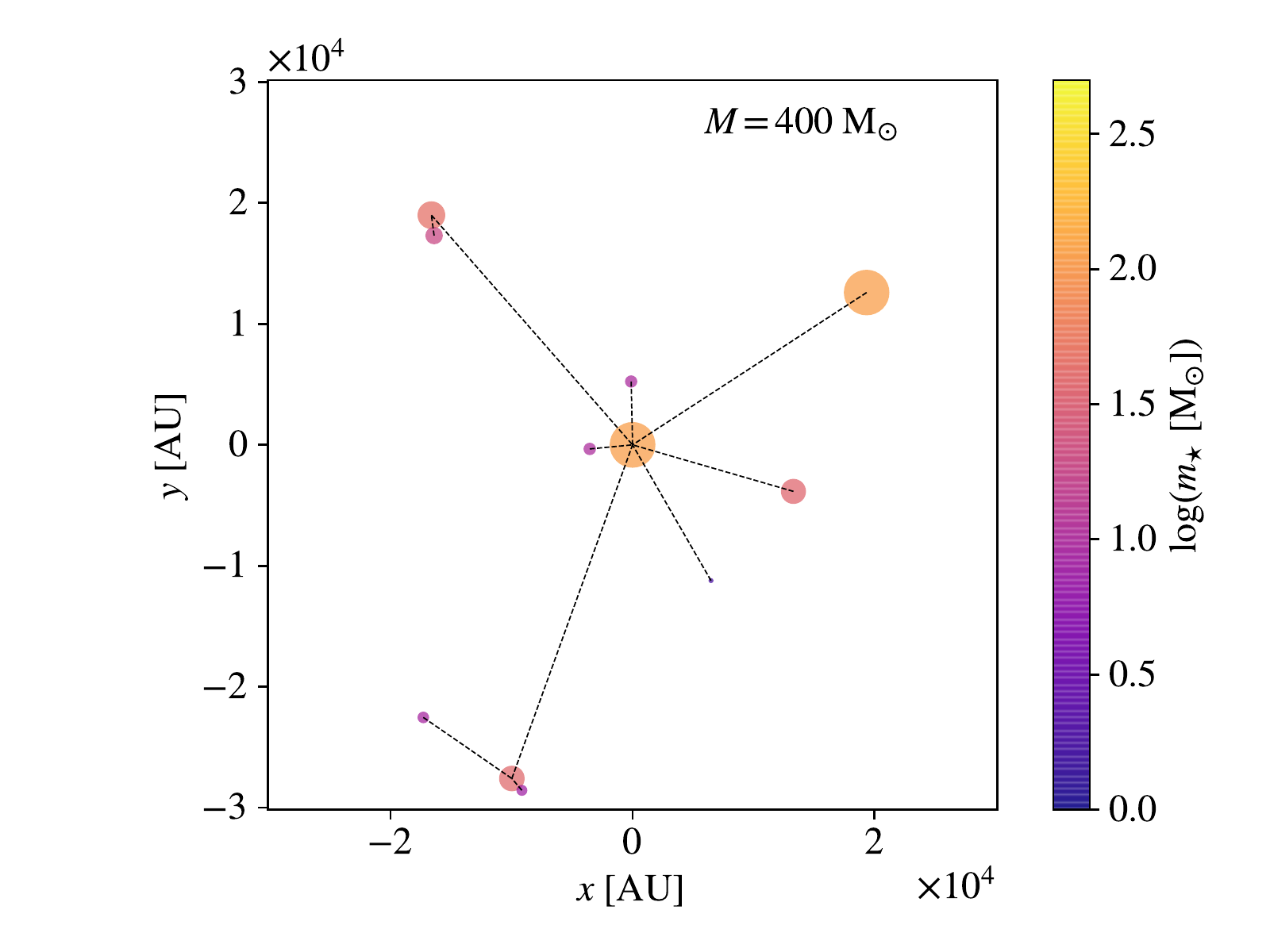}
\caption{Projected distribution of stars in the disk ($xy$) plane of a representative newly-born Pop~III cluster with $M=400\ \rm M_{\odot}$, $N_{\star}\approx 12$, $\alpha=1$ and $M_{\min}=1\ \rm M_{\odot}$. Stars are color coded by their masses, whose label sizes are also proportional to their physical zero-age main sequence (ZAMS) radii. The hierarchy of binaries is shown by connecting the two stars in each binary with thin dashed lines. }
\label{f2}
\end{figure}

Finally, given the hierarchy of binaries and the binary parameters, we allocate positions and velocities to stars, again following the formation sequence, such that the phase space location of star $i$ is determined relative to its primary $j$, according to the binary parameters $a_{ij}$ and $e_{ij}$, as follows.
	\begin{itemize}
	\item The most massive star ($i=1$) is in the disk plane. For $i>1$, the vertical scatter with respect to the disk plane follows an uniform distribution in the range of $[-0.05 a_{ij},0.05 a_{ij}]$.
	\item All binary orbits are initialized at apocenters.
	\item If $N_{j}$ star(s) is (are) attached to star $j$, the azimuthal angular position of the $n$th companion\footnote{If there are multiple companions, they are randomly shuffled, such that the azimuthal angular positions do not correlate with stellar masses.} is given by $\theta_{0}+2\pi (n+l)/N_{j}$, where $\theta_{0}$ is drawn from a uniform distribution in $[0,2\pi]$, and $l$ from a uniform distribution in $[0,\delta n]$. We have $\delta n=0.5$ by default. 
	\end{itemize}
Fig.~\ref{f2} shows an example of newly-born Pop~III clusters with $t_{\rm frag}=0.1$~kyr, $t_{\rm acc}=0.1$~Myr, $\alpha=1$ and $M_{\min}=1\ \rm M_{\odot}$, in terms of the projected distribution of 11 stars in the disk ($xy$) plane. Based on the range of $\bar{M}_{\star}(t)$ obtained in hydrodynamic simulations (see the bottom-right panel of Fig.~\ref{f1}), constraints on the total mass of Pop~III stars per minihalo from the 21-cm absorption signal \citep{schauer2019constraining}, $M\sim 500-1000\ \rm M_{\odot}$, we explore the parameter space defined by $t_{\rm frag}\sim 10-1000$~yr and $t_{\rm acc}\sim 0.1-1$~Myr (see Sec.~\ref{s4.1} and \ref{s4.2}).

\subsection{N-body simulations}
\label{s2.3}
With initial conditions generated by the above Phase 1 model, we run N-body simulations to evolve the newly-born Pop~III clusters during Phase 2. We adopt the 4th order pure N-body solver \texttt{ph4} from the Astrophysical MUltipurpose Software Environment (\textsc{AMUSE}, \citealt{zwart2009multiphysics,zwart2013multi,pelupessy2013astrophysical,portegies2018astrophysical,amuse})\footnote{\url{https://amuse.readthedocs.io/en/latest/index.html}.}. Stars are treated as point masses without gravitational softening. However, we do consider stellar collisions such that two stars will coalesce into one star\footnote{We do not include gravitational recoils for mergers, assuming that the gravitational energy released by mergers is carried away by radiation or winds under spherical symmetry.} if their relative distance is smaller than the sum of their radii. Here for simplicity we fix the stellar radii to the zero-age main sequence (ZAMS) radii, based on the fitting formulae for Pop~III stellar evolution at a typical metallicity $Z=10^{-6}\ \rm Z_{\odot}$ from \citealt{tanikawa2020fitting} (see their table A4). We also ignore other effects of stars having finite sizes and internal structures, such as (binary) stellar evolution and tidal circularization. For a typical Pop~III binary with $m_{1}=85\ \rm M_{\odot}$ and $m_{2}=45\ \rm M_{\odot}$ (see Sec.~\ref{s4}), given the radius $R_{\star,\rm G,1}\sim 10^{3}\ \rm R_{\odot}$ and luminosity $L_{1}\sim 10^{6.5}\ \rm L_{\odot}$ of the primary during the giant phase according to \citet{tanikawa2020fitting}, the tidal circularization timescale can be estimated as $t_{\rm circ}\sim 58\ {\rm Gyr}\cdot(a/100\ \rm AU)^{8}$ following \citet[][see their Sec. 2.3]{hurley2002evolution}. Here we have assumed a depth of $R_{\rm env}\sim 0.5R_{\star,\rm G,1}$ and a mass of $M_{\rm env}\sim 0.5m_{1}$ for the convective envelope, and adopted the numerical factor $f_{\rm conv}=1$, as the eddy turnover time-scale $\tau_{\rm conv}\sim 0.5$~yr is much smaller than the orbital period $P_{\rm orb}\gtrsim 10$~yr in our case. For $t_{\rm circ}$ to be comparable with the duration of the giant phase $\tau_{\rm G,1}\sim 0.3$~Myr, it is required that $a\lesssim 20$~AU. As shown in Sec.~\ref{s4}, for all the models that follow the universal solution (\ref{e2}-\ref{e4}), less than 1\% of binaries have such small separations, such that the negligence of tidal effects in our simulations has little impact on most of our binary statistics\footnote{However, if the cluster is initially much smaller, of only a few AU, close binaries with $a\lesssim 20$~AU can be dominant ($\gtrsim 50$\%, see Sec.~\ref{s4.4}). In that case, our eccentricity distribution may be biased to highly eccentric binaries if the simulation goes long enough. Nevertheless, such compact clusters will be dispersed with $\lesssim 1$~kyr, comparable to the tidal circularization timescale and much smaller than stellar lifetimes, such that tidal effects can be taken into account by post processing.} and does not change the conclusions of this paper. 

For simplicity, we do not consider dynamical friction of gas and the background potential from the dark matter halo during Phase 2, 
since their effects are unimportant, while scatterings among stars play the major role in binary formation, especially for close binaries (\citealt{ryu2016formation}, RT16 henceforth). Besides, \citet{stacy2013constraining} found that the self-gravity from the disk/cluster is stronger than that from the background potential regarding possible ejection of protostars. We have run test simulations which include the background potentials from a typical star-forming minihalo ($M_{\rm halo}\sim 10^{6}\ \rm M_{\odot}$) and atomic-cooling (AC) halo ($M_{\rm halo}\sim 2\times 10^{7}\ \rm M_{\odot}$) at $z=15$, with cored power-law density profiles\footnote{The cored power-law profile with a power-law index of -1 is a good approximation to the NFW profile, as low-mass haloes ($M_{\rm halo}\lesssim 10^{8}\ \rm M_{\odot}$) at high redshifts have very low concentrations ($c\lesssim 0.2$) according to the trend of halo concentration evolution seen in cosmological simulations \citep{dutton2014cold}.}, $\rho\propto r^{-1}$ for $r>r_{c}$, and $\rho=\rm const.$ for $r\le r_{c}$, where the core size is 1\% of the virial radius ($r_{c}=0.01 R_{\rm vir}$). It turns out that including the background potential only reduces the ejection fraction of stars by up to 30\%, and has little effects ($\lesssim 3$\%) on the overall evolution and binary statistics of the clusters. 
The quality of calculation is controlled by adjusting timestep parameters to ensure that energy conservation is maintained within 3\% of relative errors throughout the simulation\footnote{We have verified that imposing a more strict criterion $\delta E/E<0.1$\% leads to negligible (a few percent) changes in median binary properties, although it can enhance the rates of collision, X-ray binary and binary black hole mergers by up to a factor of 2. The differences may also be caused by the small sample size of such objects.}. 
The duration of the simulation, $t_{\rm end}$, extends over at least 10 relaxation timescales of the system, 
while $\lesssim 4$~Myr, shorter than the lifetime of stars that may end up as core-collapse supernovae (CCSNe), corresponding to $m_{\star}\lesssim 40\ \rm M_{\odot}$. Therefore, we do not consider supernova kicks, assuming that more massive stars ($m_{\star}\gtrsim 40\ \rm M_{\odot}$) either collapse completely into black holes (BHs) without kicks, or explode as pair-instability supernovae (PISNe) which leave no remnants\footnote{Stars that undergo PISNe will be removed from the simulation on-the-fly. We also use the fitting formulae in \citet{tanikawa2020fitting} to derive the stellar lifetimes.}. 

Actually, AMUSE is capable of coupling pure/direct N-body simulations with (binary) stellar evolution, and even (radiative) hydrodynamics of the ambient gas (see e.g. \citealt{boekholt2018formation,alister2020formation,wall2020modelling}), and thus can be used to design much more complex numerical experiments for formation and evolution of Pop~III clusters. We defer such sophisticated approaches to future work and focus on the dependence of binary statistics on the initial conditions of newly-born Pop~III clusters, as parameterized by our Phase 1 model. 

\section{Cluster evolution}
\label{s3}

Before investigating the binary statistics of Pop~III stars systematically under different initial conditions, in this section we present the general trends of cluster evolution in Phase 2 driven by N-body dynamics. We use the fiducial model \texttt{tf1e2ta1e5a1m1} with $t_{\rm frag}=0.1$~kyr, $t_{\rm acc}=0.1$~Myr, $M=400\ \rm M_{\odot}$, $N_{\star}\approx 12$, $R_{0}\simeq 3\times 10^{4}$~AU, $\alpha=1$ and $M_{\min}=1\ \rm M_{\odot}$, as an example. The results from other models are similar (see Table~\ref{t1} for a list of models considered). For this model, the (initial) dynamical and relaxation timescales of the system are $t_{\rm dyn}\simeq \sqrt{R_{\rm vir}^{3}/(GM)}\simeq 10$~kyr and $t_{\rm relax}\simeq 0.138N_{\star}/\ln(\gamma N_{\star})t_{\rm dyn}\simeq 6.3 t_{\rm dyn}\simeq 63$~kyr, where we estimate the virial radius as $R_{\rm vir}\sim 0.4 R_{0}$ and adopt $\gamma=0.11$ \citep{amuse}. We ran 1000 simulations for 4~Myr, such that at the end massive stars ($m_{\star}\gtrsim 40\ \rm M_{\odot}$) have died in BHs and PISNe.

\begin{figure*}
\includegraphics[width=2\columnwidth]{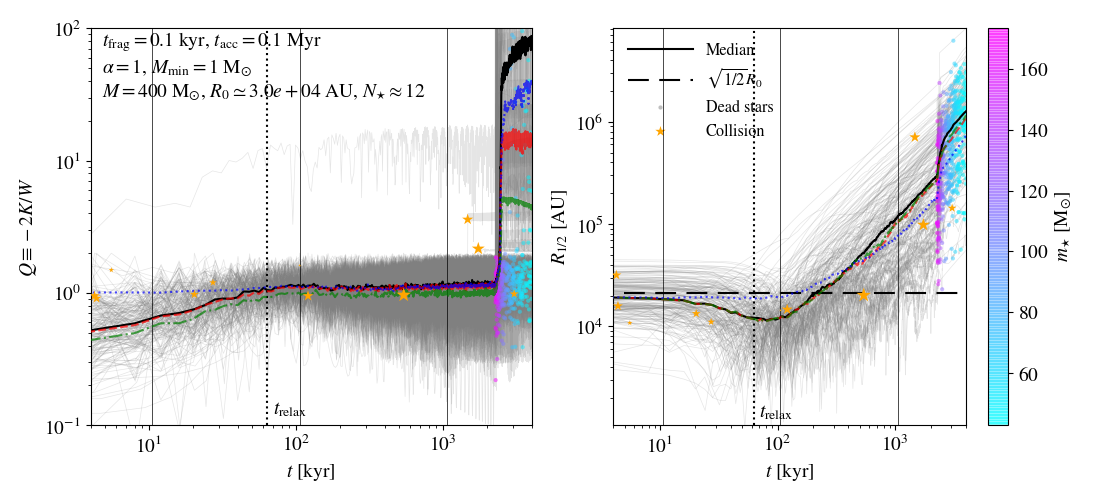}
\caption{Evolution of the virial parameter (left) and 50\% Lagrangian radius (right) from 200 runs for the fiducial model with $M=400\ \rm M_{\odot}$, $N_{\star}\approx 12$, $R_{0}\simeq 3\times 10^{4}$~AU, $\alpha=1$ and $M_{\min}=1\ \rm M_{\odot}$. The median tracks are shown with the solid curve on top of the evolutionary tracks for individual clusters, denoted by thin gray lines. For comparison, we also show the median tracks for two test simulations under the same conditions, but including the background potentials for typical star-forming minihaloes ($M_{\rm halo}\sim 10^{6}\ \rm M_{\odot}$, red dashed) and AC haloes ($M_{\rm halo}\sim 2\times 10^{7}\ \rm M_{\odot}$, green dashed-dotted), at $z=15$. The median tracks for \texttt{tf1e2ta1e5a1m1\_vir} where the initial velocities are re-scaled to have $Q=1$ are shown with the blue dotted curves. Collision events are labelled with stars whose sizes are proportional to the masses of the merger products. Stellar deaths (as BHs or PISNe) are shown with dots, color coded by progenitor masses. The thin vertical solid lines denote 1, 10 and 100 dynamical timescales $ t_{\rm dyn}$. The dashed vertical line shows the relaxation timescale $t_{\rm relax}$. In the right panel, the horizontal long-dashed line denotes $\sqrt{1/2}R_{0}$ as an estimation to the initial half-mass radius (from a $\rho\propto r^{-1}$ density profile, see Fig.~\ref{pro}). }
\label{qrt}
\end{figure*}

Fig.~\ref{qrt} shows the evolution of the virial parameter $Q\equiv 2K/W$ and 50\% Lagrangian radius (i.e. half-mass radius) from 200 runs, where $K$ and $W$ are the kinematic and potential energies. According to the median evolution track, the system contracts at $t\lesssim t_{\rm relax}\sim 10 t_{\rm dyn}\sim 100$~kyr and then re-expands gradually by relaxation, meanwhile reaching virial equilibrium ($Q\sim 1$). The virial equilibrium is broken by PISNe ($Q\gtrsim 4$) in some clusters at $t\sim 2$~Myr, which also facilitate cluster expansion/dispersion. This process is illustrated in Fig.~\ref{sketch}. The system has expanded by a factor of $\sim10$~(100) at $t\sim 1\ (4)$~Myr in terms of the half-mass radius. We verify that the background potential from the dark matter halo has little effect on the evolution, except for slightly slowing down the ejected companions of PISN progenitors, which somewhat delays cluster dispersion. As long as the input IMF includes a significant fraction of PISN progenitors, the host systems will experience strong disruption by PISNe at $t\sim 2$~Myr, which also affects binary statistics. Since it is interesting to take into account PISN progenitors in binaries, we focus on the binary statistics before PISNe in the next section. Actually, in most cases the majority of binary hardening by scatters happens at $t\lesssim 2$~Myr (see below).

\begin{figure*}
\includegraphics[width=2\columnwidth]{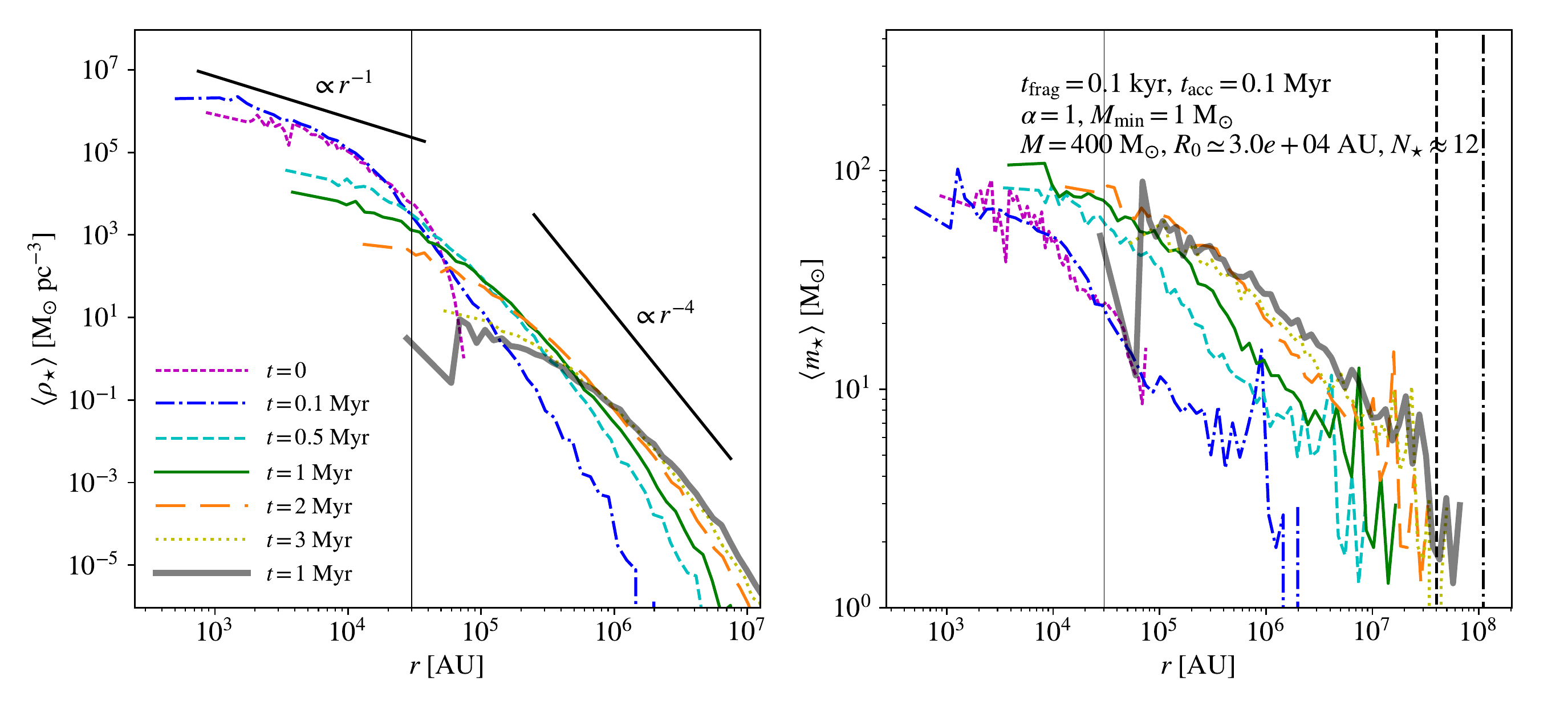}
\caption{Evolution of the mean density (left) and stellar mass (right) profiles from the stacked results of 1000 runs for the fiducial model with $M=400\ \rm M_{\odot}$, $N_{\star}\approx 12$, $R_{0}\simeq 3\times 10^{4}$~AU, $\alpha=1$ and $M_{\min}=1\ \rm M_{\odot}$. The initial cluster size $R_{0}$ and virial radius of typical minihaloes ($M_{\rm halo}\sim 10^{6}\ \rm M_{\odot}$) and AC haloes ($M_{\rm halo}\sim 2\times 10^{7}\ \rm M_{\odot}$) at redshift $z=15$ are shown with the (thin) solid, dashed and dashed-dotted vertical lines. }
\label{pro}
\end{figure*}

Beside dynamical properties, we also assess the evolution of internal structure in terms of mean density and stellar mass profiles from stacked results of 1000 runs, as shown in Fig.~\ref{pro}. The mean stellar mass profile embodies the mean stellar masses at different radial bins, which is a measurement of mass segregation. Similar to the trend shown in Fig.~\ref{qrt}, the system contracts at $t\lesssim 10t_{\rm dyn}\sim 100$~kyr, while the central stellar density remains as high as $\sim 10^{6}\ \rm M_{\odot}\ pc^{-3}$, with some low-mass stars ($m_{\star}\sim 1\ \rm M_{\odot}$) having already been scattered out to $r\sim 70 R_{0}\sim 10^{6}$~
AU. Subsequently, the cluster keeps expanding at $t\gtrsim 100$~kyr, and the central density of stars decreases with time, particularly after PISNe ($t\gtrsim 2$~Myr), reaching a few $\rm M_{\odot}\ pc^{-3}$ in the end. Throughout the evolution, the inner density profile approximately follows $\rho\propto r^{-1}$, while the density profile at the outskirt flattens with time, becoming $\rho\propto r^{-4}$ at $t\gtrsim 100t_{\rm dyn}\sim 1$~Myr. Mass segregation is present at the beginning and maintained to the end, such that the shape of the mean stellar mass profile does not change significantly. By the end of the simulation at $t=4$~Myr, some low-mass stars have reached the halo boundary ($R_{200}\sim 10^{8}$~AU for typical haloes), while the majority of massive stars ($m_{\star}\gtrsim 10\ \rm M_{\odot}$) remains well inside the halo ($r\lesssim 10^{7}\ \mathrm{AU}\lesssim 0.3 R_{200}$)\footnote{When the background potential is included, no stars can reach the halo boundary by $t=4$~Myr, although a fraction (a few to twenty percent) of stars are unbound to the halo.}. Actually, the profile of the scatters in stellar mass also decreases with radius, although less sharply compared with the mean stellar mass profile. This implies that low-mass stars can be everywhere in the cluster, while massive stars prefer the inner region.

\begin{figure}
\includegraphics[width=1\columnwidth]{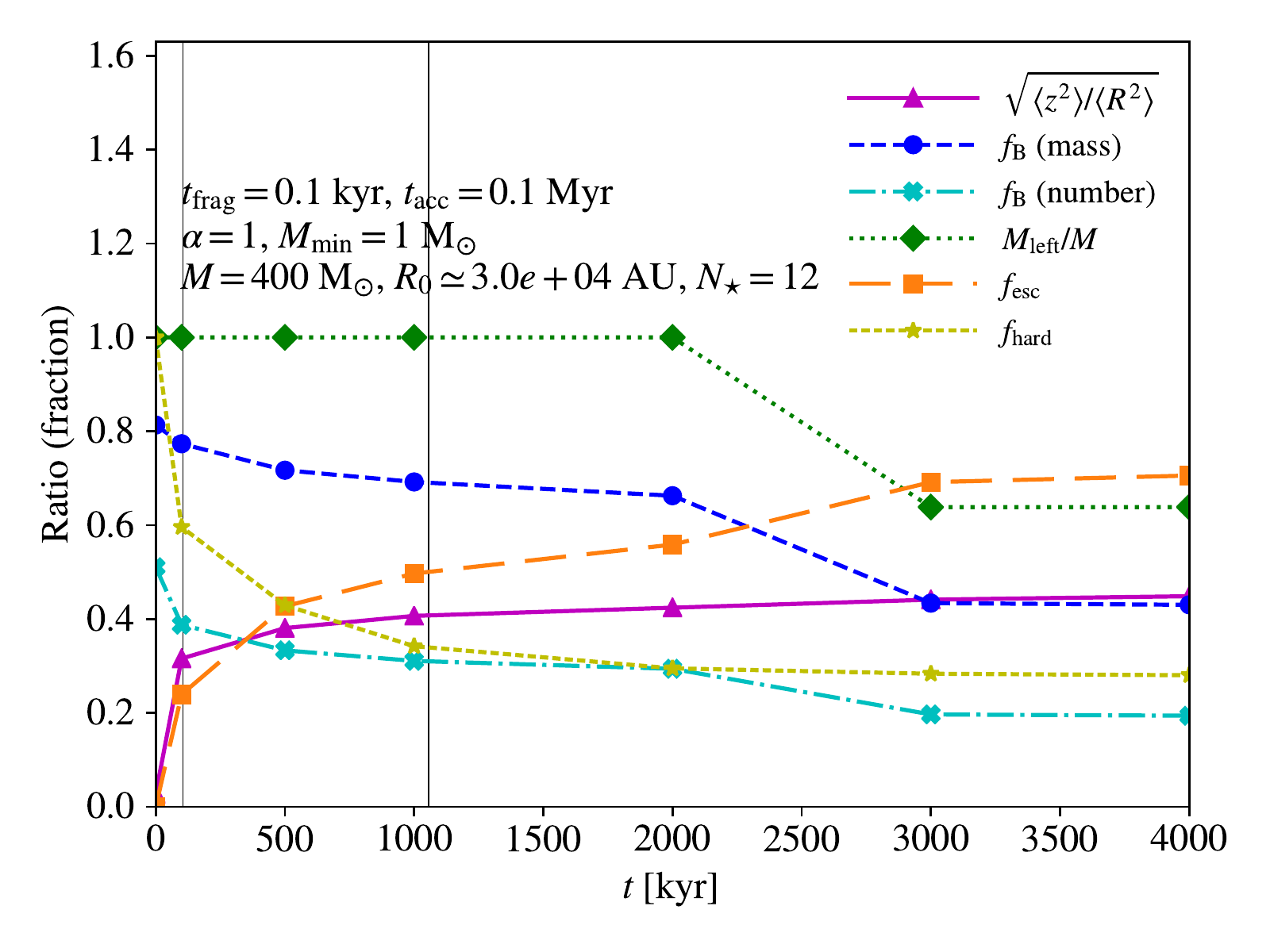}
\caption{Evolution of the disk thickness parameter $\sqrt{\langle z^{2}\rangle/\langle R^{2}\rangle}$ (solid, triangles), binary fraction $f_{\rm B}$ in terms of mass (dashed, circles) and number (dashed-dotted, crosses), ratio of leftover mass (alive stars + remnants) over initial mass $M_{\rm left}/M$ (dotted, diamonds), fraction of stars unbound from the cluster (long-dashed, squares), and binary hardening parameter $f_{\rm hard}\equiv\langle a\rangle_{\rm med}/\langle a(t=0)\rangle_{\rm med}$ (short-dashed, stars), from 1000 runs for the fiducial model with $M=400\ \rm M_{\odot}$, $N_{\star}\approx 12$, $R_{0}\simeq 3\times 10^{4}$~AU, $\alpha=1$ and $M_{\min}=1\ \rm M_{\odot}$. The vertical lines denote 10 and 100 dynamical timescales.}
\label{evo}
\end{figure}

\begin{figure}
\includegraphics[width=1\columnwidth]{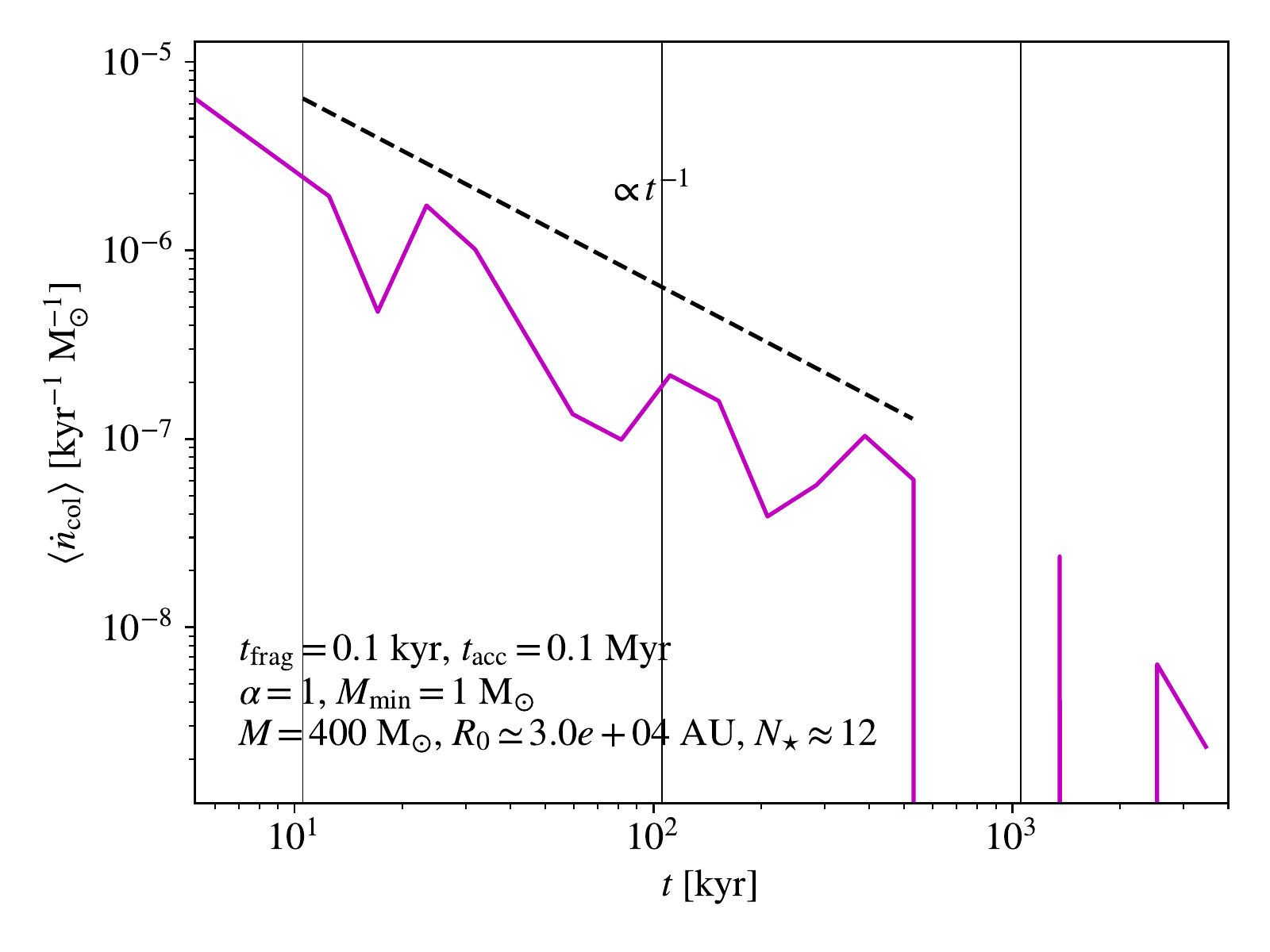}
\caption{Average collision rate of stars (per initial stellar mass) as a function of time, from 1000 runs for the fiducial model with $M=400\ \rm M_{\odot}$, $N_{\star}\approx 12$, $R_{0}\simeq 3\times 10^{4}$~AU, $\alpha=1$ and $M_{\min}=1\ \rm M_{\odot}$. The vertical lines denote 1, 10 and 100 dynamical timescales. The collision rate decreases almost linearly with time at $t\lesssim 50 t_{\rm dyn}$.}
\label{colrate}
\end{figure}

Finally, we look into the effects of scatters among stars on cluster shape, binary properties and ejection of stars, as shown in Fig.~\ref{evo}. We define the disk thickness parameter as $\sqrt{\langle z^{2}\rangle/\langle R^{2}\rangle}$, where the brackets denote mass-weighted average over all stars from stacked results of 1000 runs. The closer this parameter is to $1/\sqrt{2}\sim 0.7$, the closer the system is to spherical symmetry. We further define the binary hardening parameter as the ratio of current over initial median separations in binaries, $f_{\rm hard}\equiv\langle a\rangle_{\rm med}/\langle a(t=0)\rangle_{\rm med}$. Starting from a thin disk of $\sqrt{\langle z^{2}\rangle/\langle R^{2}\rangle}\sim 0.03$, the system is significantly puffed up at $t\lesssim 10t_{\rm dyn}$, thereafter ($t\gtrsim10 t_{\rm dyn}\sim 0.1$~Myr) the system is still oblate, with almost constant thickness parameter of $\sim 0.4<1/\sqrt{2}$. The fraction of stars in binaries decreases (but not significantly) with time, especially for the initial contraction ($t\lesssim 0.1$~Myr) and late stage ($t\gtrsim 2$~Myr) when massive stars ($m_{\star}\sim 120-240\ \rm M_{\odot}$) start to explode as PISNe. In general, the majority of binary hardening and ejection of stars (by scatters) happens at $t\lesssim 5-10 t_{\rm relax}$ ($\sim 50-100 t_{\rm dyn}$ for \texttt{tf1e2ta1e5a1m1}), if not considering PISNe, which boost the ejection fraction by 0.13 at $t\sim 2$~Myr. By $t\sim 100 t_{\rm dyn}\sim 1$~Myr, the ejection fraction is as high as $\sim 50$ (22) \% in terms of number (mass) of stars, and binaries are hardened by a factor of $\sim 3$. 
We also derive the average collision rate of stars (per initial stellar mass)  as a function of time (Fig.~\ref{colrate}), which decreases almost linearly with time for $t\lesssim 50 t_{\rm dyn}$, beyond which our sample of clusters is too small to see the trend. For the fiducial model, the collision rate is very low ($\sim 10^{-7}-10^{-5}\ \rm kyr^{-1}\ M_{\odot}^{-1}$), such that less than 1 percent of stars undergo collisions. However, if the cluster is initially much smaller than predicted by the universal solution~(Equ.~\ref{e2}-\ref{e4}), the collision rate can be significantly enhanced (see Sec.~\ref{s4.4})

When virial equilibrium is imposed on the initial conditions (\texttt{tf1e2ta1e5a1m1\_vir}), there is no initial contraction, but the system still significantly expands at $t\gtrsim t_{\rm relax}\sim 10 t_{\rm dyn}\sim 100$~kyr, however at a slower rate compared with the fiducial case (Fig.~\ref{qrt}). Furthermore, the collision frequency and ejection fraction of stars are reduced by 54 (52)\% and 21 (34)\% in terms of number (mass). The binary hardening parameter is also reduced (by a factor of $\sim 2$), followed by significant reduction of hard binaries (HDBs), by up to a factor of $\sim 4$, and so is the disk thickness parameter. Here HDBs are defined by $a[(100\ \mathrm{M_{\odot}})^{2}/(m_{1}m_{2})](1+e)/(1-e)\lesssim 10^{4}\ {\rm AU}\sim 0.05\ \rm pc$, such that they will be hardened (even at apocenters) by interactions with typical surrounding low-mass Pop~II/I stars with a stellar mass $m_{\star}\sim 1\ \rm M_{\odot}$ and a velocity dispersion of $\sigma\sim 30\ \rm km\ s^{-1}$. Nevertheless, the overall fraction of stars in binaries is slightly increased. These trends (see Table~\ref{t2} for details) are caused by the fact that without the initial contraction, interactions among stars are less violent compared with the fiducial case. They are consistent with the picture that interactions among stars destroy soft binaries and harden hard binaries. 

\section{Binary statistics}
\label{s4}

\begin{table*}
    \centering
    \caption{Phase 1 model parameters. We simulate 1000 random realizations of Pop~III clusters for each model other than the fiducial one (\texttt{tf1e2ta1e5a1m1}), for which we run 10000 simulations. The `standard' initial distribution of $a$ is adopted for most simulations as an uniform distribution with $a_{\min}\sim 0.1 a_{\rm L_{1}}$ and $a_{\max}\sim a_{\rm L_{1}}$, given $a_{\rm L_{1}}$ as the distance from the primary to the L1 point defined by its most massive companion. While for \texttt{tf1e2ta1e5a1m1\_tight}, we have $a_{\min}=a_{\rm lobe}$. For \texttt{tf1e2ta1e5a1m1\_close}, all massive companions of the primary are considered, such that $a_{\max}\sim \min(a_{\mathrm{ L_{1}},n})$, for $m_{\mathrm{s},n}/m_{\rm p}>0.5$. In addition to the models shown here, we further consider a variant of \texttt{tf1e2ta1e5a1m1\_greif} called \texttt{tf1e2ta1e5a1m1\_greif\_ncol}, where the stellar radii are all set to $10^{-4}\ \mathrm{AU}\ll R_{\star,\rm ZAMS}$ to suppress stellar collision, with other conditions unchanged. Moreover, we also explore the case in which the initial velocities of stars are re-scaled to meet virial equilibrium, for the fiducial model, denoted by \texttt{tf1e2ta1e5a1m1\_vir}.}
    \begin{tabular}{ccccccccccc}
    \hline
         Model & $t_{\rm frag}$ & $t_{\rm acc}$ & $\alpha$ & $M_{\min}$ & $N_{\star}$ & $M$ & $R_{0}$ & $a_{\min}$ & $a_{\max}$ & $t_{\rm end}$\\
         & [kyr] & [Myr] & & $[\rm M_{\odot}]$ & & $[\rm M_{\odot}]$ & [AU] & & & [Myr]\\
         \hline 
         \texttt{tf1e2ta1e5a1m1} & 0.1 & 0.1 & 1 & 1 & 12 & 400 & $3\times 10^{4}$ & $\sim 0.1a_{\rm L_{1}}$ & $\sim a_{\rm L_{1}}$ & 1\\
         \texttt{tf1e2ta1e5a1m10} & 0.1 & 0.1 & 1 & 10 & 12 & 400 & $3\times 10^{4}$ & $\sim 0.1a_{\rm L_{1}}$ & $\sim a_{\rm L_{1}}$ & 1\\
         \texttt{tf1e2ta1e5a017m1} & 0.1 & 0.1 & -0.17 & 1 & 12 & 400 & $3\times 10^{4}$ & $\sim 0.1a_{\rm L_{1}}$ & $\sim a_{\rm L_{1}}$ & 1\\
         \texttt{tf1e2ta1e5a0m1} & 0.1 & 0.1 & 0 & 1 & 12 & 400 & $3\times 10^{4}$ & $\sim 0.1a_{\rm L_{1}}$ & $\sim a_{\rm L_{1}}$ & 1\\
         \texttt{tf1e2ta1e5a05m1} & 0.1 & 0.1 & 0.5 & 1 & 12 & 400 & $3\times 10^{4}$ & $\sim 0.1a_{\rm L_{1}}$ & $\sim a_{\rm L_{1}}$ & 1\\
         \texttt{tf1e2ta1e5a15m1} & 0.1 & 0.1 & 1.5 & 1 & 12 & 400 & $3\times 10^{4}$ & $\sim 0.1a_{\rm L_{1}}$ & $\sim a_{\rm L_{1}}$ & 1\\
         \texttt{tf1e2ta1e5a2m10} & 0.1 & 0.1 & 2 & 10 & 12 & 400 & $3\times 10^{4}$ & $\sim 0.1a_{\rm L_{1}}$ & $\sim a_{\rm L_{1}}$ & 1\\
         \hline
         \texttt{tf1e1ta1e5a1m1} & 0.01 & 0.1 & 1 & 1 & 6 & 400 & $3\times 10^{4}$ & $\sim 0.1a_{\rm L_{1}}$ & $\sim a_{\rm L_{1}}$ & 1\\
         \texttt{tf1e3ta1e5a1m1} & 1 & 0.1 & 1 & 1 & 24 & 400 & $3\times 10^{4}$ & $\sim 0.1a_{\rm L_{1}}$ & $\sim a_{\rm L_{1}}$ & 1\\
         \hline
         \texttt{tf1e2ta3e5a1m1} & 0.1 & 0.3 & 1 & 1 & 12 & 892 & $8.2\times 10^{4}$ & $\sim 0.1a_{\rm L_{1}}$ & $\sim a_{\rm L_{1}}$ & 2\\
         \texttt{tf1e3ta1e6a1m1} & 1 & 1 & 1 & 1 & 24 & 2148 & $2.5\times 10^{5}$ & $\sim 0.1a_{\rm L_{1}}$ & $\sim a_{\rm L_{1}}$ & 4\\
         \hline
         \texttt{tf1e2ta1e5a1m1\_small} & 0.1 & 0.1 & 1 & 1 & 12 & 400 & $10^{4}$ & $\sim 0.1a_{\rm L_{1}}$ & $\sim a_{\rm L_{1}}$ & 0.2\\
         \texttt{tf1e2ta1e5a1m1\_comp} & 0.1 & 0.1 & 1 & 1 & 12 & 400 & $2\times 10^{3}$ & $\sim 0.1a_{\rm L_{1}}$ & $\sim a_{\rm L_{1}}$ & 0.2\\
         \texttt{tf1e2ta1e5a1m1\_greif} & 0.1 & 0.1 & 1 & 1 & 12 & 400 & 7 & $\sim 0.1a_{\rm L_{1}}$ & $\sim a_{\rm L_{1}}$ & 0.001\\
         \texttt{tf1e2ta1e5a1m1\_tight} & 0.1 & 0.1 & 1 & 1 & 12 & 400 & $3\times 10^{4}$ & $ a_{\rm lobe}$ & $\sim a_{\rm L_{1}}$ & 1\\
         \texttt{tf1e2ta1e5a1m1\_close} & 0.1 & 0.1 & 1 & 1 & 12 & 400 & $3\times 10^{4}$ & $\sim 0.1a_{\max}$ & $\sim \min(a_{\mathrm{ L_{1}},n})$ & 1\\
         \hline
    \end{tabular}
    \label{t1}
\end{table*}

We now explore the dependence of binary statistics on initial conditions set by Phase 1 models. Throughout this work, binaries and multiple systems are identified with the following procedure. We process the stars in descending order of mass. For each star $i$, we calculate the binding energies of its parent system $p_{i}$ and the parent systems $p_{j}$ of stars less massive than it. When there are negative binding energies, we identify a binary/multiple system for $p_{i}$ and the $p_{j}$ that gives the smallest (negative) binding energy. Here the parent system $p_{i}$ denotes the highest-level pre-existing binary/multiple system that contains star $i$. If star $i$ has not been included in any systems, $p_{i}$ is just star $i$ itself. We only consider multiple systems with up to 4 stars. If not specified, below we show the results for all binaries including those in multiple systems.

We investigate 11+5+2 models that cover a physically motivated region in the parameter space, i.e. $t_{\rm frag}\sim 10-10^{3}$~yr, $t_{\rm acc}\sim 0.1-1$~Myr, $\alpha\sim -0.17-2$ and $M_{\min}\sim 1-10\ \rm M_{\odot}$, and possible deviations from our `standard' Phase 1 picture, as summarized in Table~\ref{t1}. For the fiducial model \texttt{tf1e2ta1e5a1m1} with $t_{\rm frag}=0.1$~kyr, $t_{\rm acc}=0.1$~Myr, $M=400\ \rm M_{\odot}$, $N_{\star}\approx 12$, $R_{0}\simeq 3\times 10^{4}$~AU, $\alpha=1$ and $M_{\min}=1\ \rm M_{\odot}$, we follow the Phase 2 evolution of 10000 clusters randomly generated from the procedure described in Sec.~\ref{s2.2}. While for other models, we only simulate 1000 clusters each. The binary properties are calculated at $t\sim 10 t_{\rm relax}$. Before looking into different Phase 1 parameters, we first focus on the results from our fiducial case (at $t=1$~Myr), which are compared with the Pop~III binary statistics widely used in previous studies (e.g. \citealt{greif2012formation,stacy2013constraining,kinugawa2014possible,hartwig2016,
ryu2016formation,belczynski2017likelihood,kinugawa2020chirp}). 

\begin{figure*}
\includegraphics[width=1.7\columnwidth]{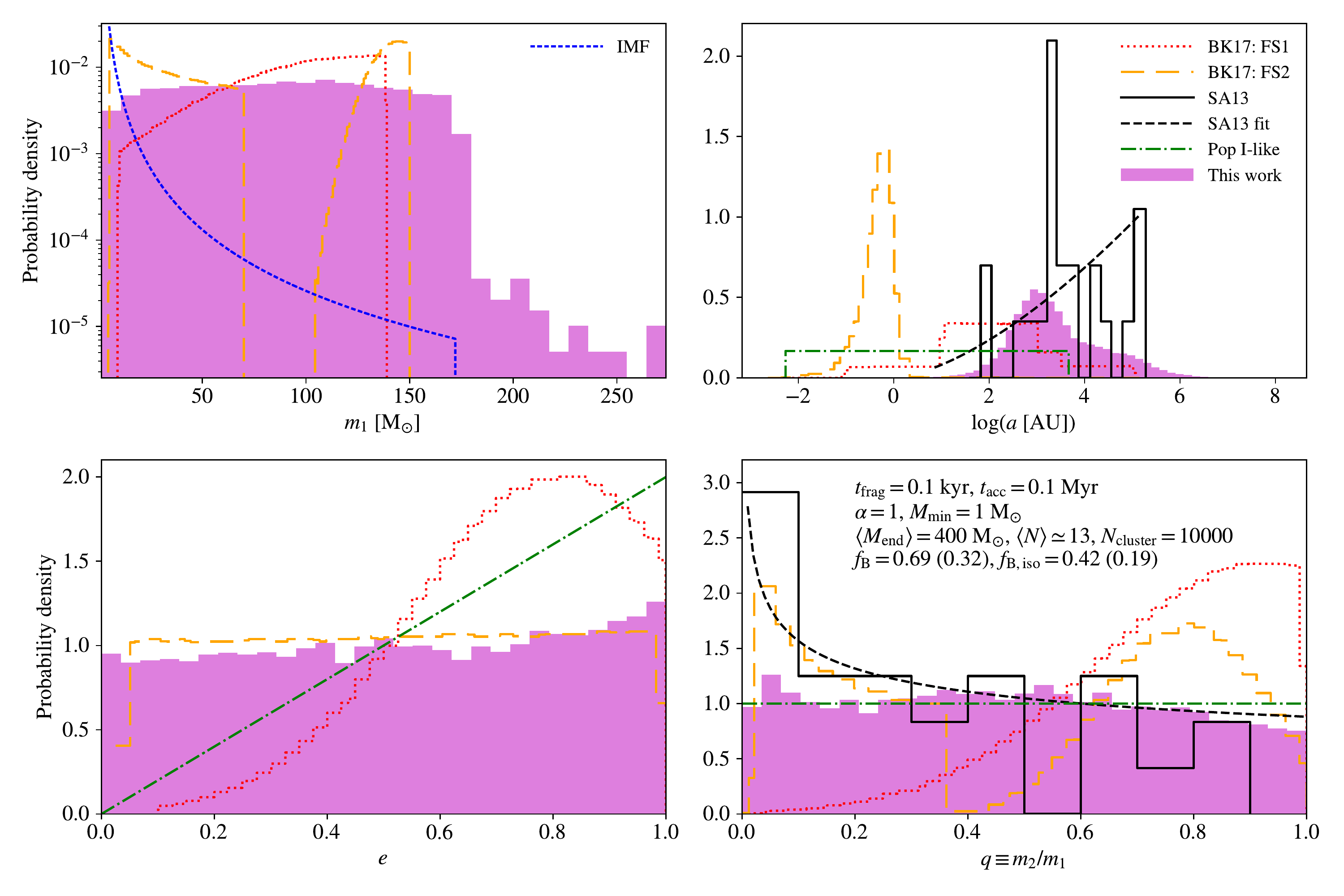}
\caption{Binary statistics from 10000 runs for the fiducial model \texttt{tf1e2ta1e5a1m1} (at $t=1$~Myr, purple histograms), in terms of the distributions of primary mass $m_{1}$, semi-major axis (separation) $a$, (secondary to primary) mass ratio $q\equiv m_{2}/m_{1}$ and orbital eccentricity $e$ (clockwise). In the annotation of the bottom-right panel, $f_{\rm B}$ ($f_{\rm B,iso}$) denotes the fraction of stars in (isolated) binaries, where the values in (out of) the brackets are derived with respect to number (mass) of stars. For comparison, we show the properties of binary \textit{proto}stars in \citealt{stacy2013constraining} (SA13, see their fig.~8 and 9) with solid contours and dashed curves (for power-law fits), assuming a typical total mass of protostar binaries $m_{1}+m_{2}= 5\rm \ M_{\odot}$. We also show the binary statistics used in \citealt{belczynski2017likelihood} (Bk17, see their fig.~6-9), which are obtained from N-body simulations in RT16. They consider two models, FS1 (dotted contours) and FS2 (long-dashed contours), whose initial conditions are based on the small-scale simulations by SA13 and \citet{greif2012formation}, respectively. We also present the trends in Pop~I like binaries with the dashed-dotted contours, some of which are often applied to Pop~III binaries as zeroth-order approximations (e.g. \citealt{kinugawa2014possible,hartwig2016}). In the top-left panel for $m_{1}$, the input power-law IMF (of a slope $\alpha=1$) is shown with the short-dashed curve.}
\label{f3}
\end{figure*}

As shown in Fig.~\ref{f3}, there are significant differences between our fiducial model and literature results, especially for the distributions of primary mass $m_{1}$ and separation $a$. The distribution of primary mass is highly sensitive to the IMF. In general, if the IMF follows a power law, the primary mass distribution (PMD) appears to be another power law flattened (more top-heavy) compared with the IMF, as more massive stars are more likely to form binaries\footnote{In other words, massive binaries are more likely to survive disruptions from scatters during cluster relaxation. This phenomenon is clearly shown in Fig.~\ref{bfrac}. Therefore, it is unphysical to assume that the PMD is identical to the IMF when stars are born in clusters whose dynamical evolution favors massive binaries.}. For our fiducial model and the FS1 model used in \citealt{belczynski2017likelihood} (BK17, henceforth), based on RT16, the differences in power-law slope for the PMD and IMF are approximately 1. While for the FS2 model in BK17, since the IMF is no longer a simple power law, the connection between PMD and IMF is more complex, but they still have similar shapes. In conclusion, the PMD discrepancies for different studies can be understood with the differences in the input IMFs. 

For the distribution of $a$, our fiducial model produces more wide binaries than the FS1 and FS2 models in BK17, as well as the Pop~I-like case. Particularly, close binaries with $a\lesssim 100$~AU are almost absent in our fiducial model, but make up significant fractions of the binary populations in other studies. The difference is caused by the different initial conditions, especially cluster sizes, which will be discussed in details below (Sec.~\ref{s4.4}) and in Appendix~\ref{a2}. Generally speaking, the FS1 and FS2 models in BK17 adopt the sizes of disks in \citet[][SA13 henceforth]{stacy2013constraining} and \citet{greif2012formation}, when protostars have grown to only a few $\rm M_{\odot}$, but artificially turn the protostars into stars by boosting their masses by a factor of a few hundred. In reality, according to the universal solution (see Equ.~\ref{e2}-\ref{e4}), growth of the protostars by accretion also leads to expansion of the system due to angular momentum conservation. As a result, with similar total cluster masses, our fiducial model exhibits much larger initial cluster sizes, which in turn result in wider binaries. This difference exists as long as the cluster size is set by the universal solution, and leads to interesting consequences for Pop~III X-ray binaries (XRBs) as well as binary black hole (BBH) mergers (see Sec.~\ref{s5} and Appendix~\ref{a1}). 

When comparing the distribution of $a$ for protostars in SA13 with our results, we find that the difference is rather small, with our distribution slightly biased towards smaller separations. Actually, SA13 describe the state of the system in the middle of Phase 1, when gravitational interactions between (proto)stars are not the only agent. This has two effects that tend to cancel each other, such that the distributions of $a$ in our fiducial model and SA13 are similar: First, the disk is smaller than what is adopted in our model at the \textit{end} of Phase 1 as the initial condition for Phase 2. Second, the hardening of binaries by scatters of stars during Phase 2 is missing in SA13.

In our fiducial model, both the distributions of orbital eccentricity $e$ and (secondary to primary) mass ratio $q\equiv m_{2}/m_{1}$ $q$ are almost uniform in the range of [0, 1], consistent with the FS2 model in BK17 for $e$ and the Pop-I like case for $q$. Compared with SA13, the fraction of binaries with $q\lesssim 0.1$ is smaller by a factor of $\sim 3$. This could be caused by biased weighting in our Phase 1 scheme (Sec.~\ref{s2.2}) for the assignment of newly formed fragments to existing fragments, or scatters of stars in Phase 2, during which massive stars tend to replace the low-mass companions of massive primaries. 
Both FS1 and FS2 from BK17 under-predict the fraction of binaries with small mass ratios compared with our model. This may be caused by different procedures of identifying binaries, especially for wide binaries. Interestingly, our fiducial model, as well as the FS1 model in BK17, does not reproduce/maintain the thermal distribution of $e$ seen in present-day binaries \citep{duquennoy1991multiplicity} and adopted for the initial conditions. Actually, in all the 18 models explored here, the distribution of $e$ is always approximately uniform. This implies that the dynamics of Pop~III stars formed in minihaloes is fundamentally different from that of present-day stars formed in molecular clouds embedded in galactic disks. 

Next, in the following subsections, we evaluate the effects of five key Phase 1 parameters one by one: fragmentation timescale (Sec.~\ref{s4.1}), accretion timescale (Sec.\ref{s4.2}), initial mass function (Sec.~\ref{s4.3}), cluster size and initial distribution of binary separation (Sec.~\ref{s4.4}). It turns out that binary statistics is particularly sensitive to the last two parameters. A more detailed summary of the overall properties and trends for the 18 models is provided in Appendix~\ref{a1}, in terms of key cluster evolution and binary statistical parameters and outcomes (see Table~\ref{t2}). 

\subsection{Fragmentation timescale}
\label{s4.1}
\begin{figure*}
\includegraphics[width=1.7\columnwidth]{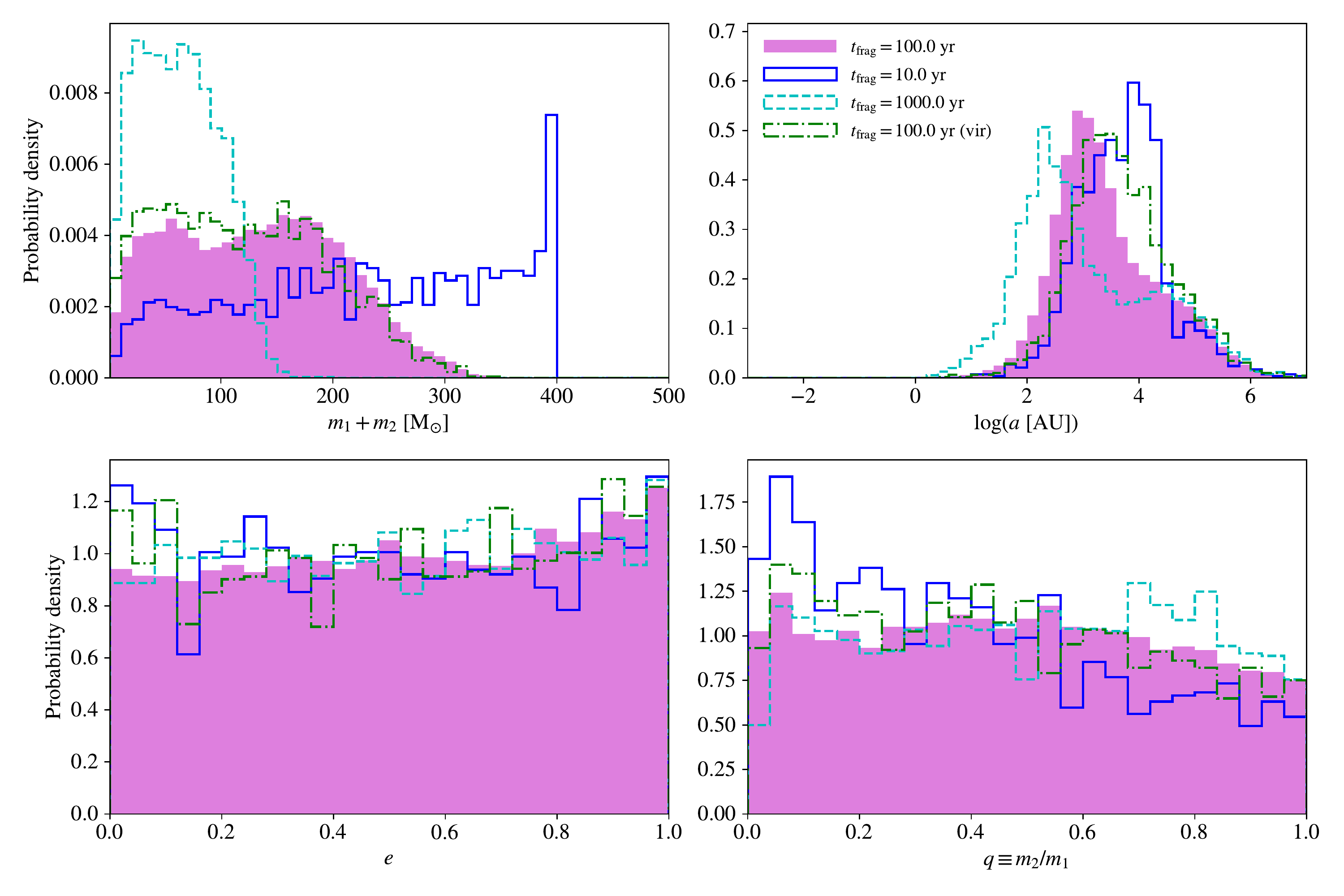}
\caption{Binary statistics for three models with different fragmentation timescales (i.e. numbers of stars), \texttt{tf1e2ta1e5a1m1} (purple histograms), \texttt{tf1e1ta1e5a1m1} (blue solid contours) and \texttt{tf1e3ta1e5a1m1} (cyan dashed contours), and the variant of the fiducial model with imposed virial equilibrium in the initial conditions, \texttt{tf1e2ta1e5a1m1\_vir} (green dashed-dotted contours), in terms of the distributions of total mass $m_{1}+m_{2}$, semi-major axis (separation) $a$, (secondary to primary) mass ratio $q\equiv m_{2}/m_{1}$ and orbital eccentricity $e$ (clockwise). }
\label{comp0}
\end{figure*}

Fig.~\ref{comp0} shows the binary statistics for three models with different fragmentation timescales $t_{\rm frag}$ and other parameters fixed to the fiducial values, as well as the variant of the fiducial model in which initial velocities of stars are re-scaled to establish virial equilibrium (see Table~\ref{t1}). As described in Sec.~\ref{s2.1}, the fragmentation timescale $t_{\rm frag}$ determines the typical number of stars in a cluster $N_{\star}$, and therefore, the upper bound of the IMF, $M_{\max}$, since the lower bound and slope of the IMF, as well as the total cluster mass are fixed. Larger $t_{\rm frag}$ leads to larger $N_{\star}$ and smaller $M_{\rm max}$. The effects of $t_{\rm frag}$ are most salient for the distributions of total binary mass and separation. With different $t_{\rm frag}$, the total binary mass distributions (TBMDs) have different mass cut-offs, but similar shapes. The reason is that the TBMD is also closely related to the IMF, such that the IMF upper bound $M_{\rm max}$ determines the cut-off mass of TBMD. The difference in TBMD simply reflects the difference in $M_{\rm max}$. 

For the distribution of separation, it turns out that a larger $N_{\rm star}$ (i.e. larger $t_{\rm frag}$) leads to more frequent scatters of stars, and therefore, stronger hardening of binaries (see Table~\ref{t2}) and more close binaries. Meanwhile, the distribution of mass ratio tends to be biased towards the small end with a smaller number of stars. This can also be explained by the frequency of scatters determined by $N_{\star}$, as more frequent scatters tend to destroy more binaries with small mass ratios by exchanges. Similarly, the effects of imposing virial equilibrium at the beginning of Phase 2 show up as a shift towards wide binaries in the distribution of separation, and a mass ratio distribution slightly biased to the small end. The reason is that the initial virial equilibrium suppress scatters of stars by avoiding the initial contraction, as shown in Sec.~\ref{s3}. 

\subsection{Accretion timescale}
\label{s4.2}
\begin{figure*}
\includegraphics[width=1.7\columnwidth]{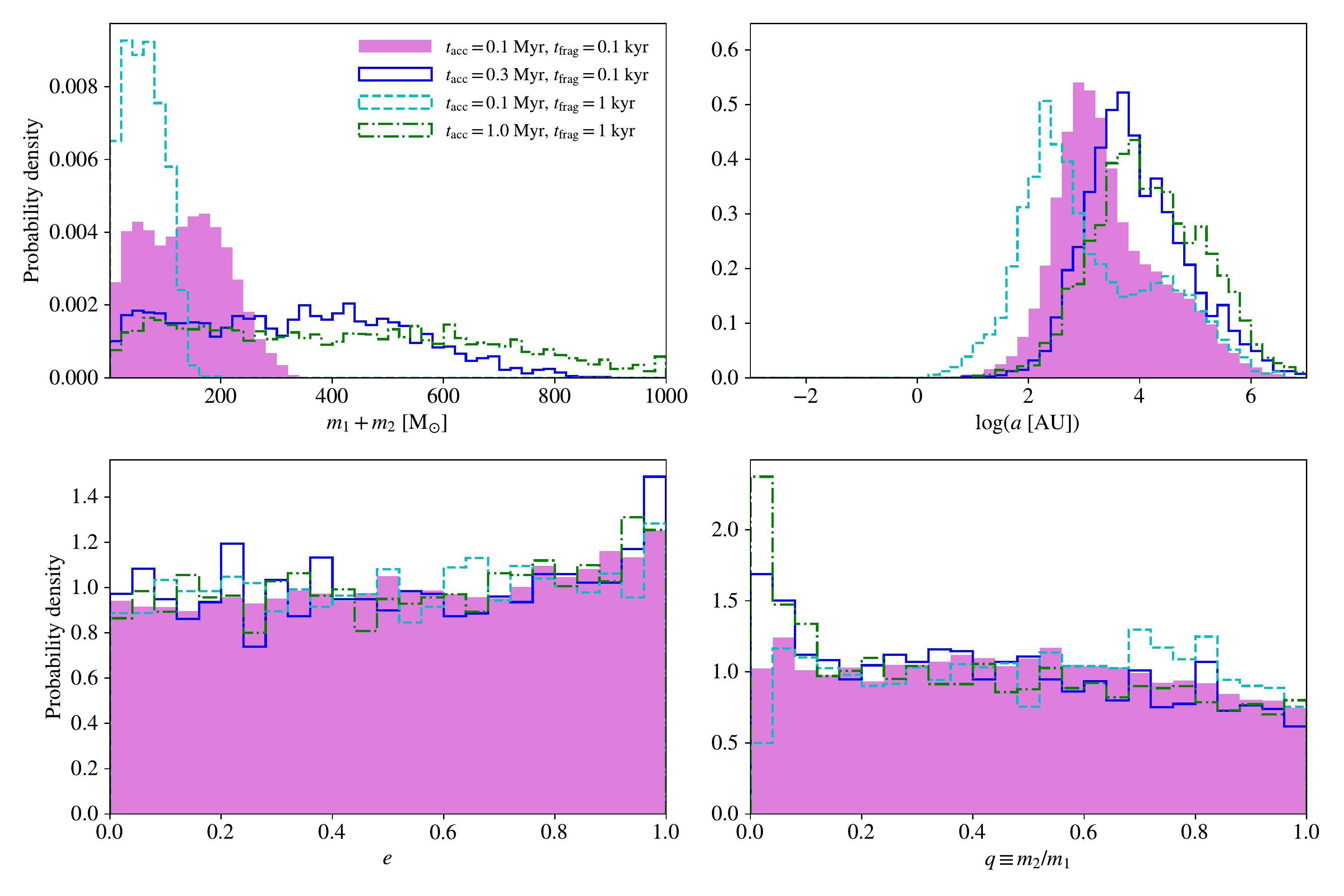}
\caption{Binary statistics for two groups of models with different accretion timescales (i.e. total masses of stars and cluster sizes), \texttt{tf1e2ta1e5a1m1} (purple histograms) and \texttt{tf1e2ta3e5a1m1} (blue solid contours), \texttt{tf1e3ta1e5a1m1} (cyan dashed contours) and \texttt{tf1e3ta1e6a1m1} (green dashed-dotted contours), in terms of the distributions of total mass $m_{1}+m_{2}$, semi-major axis (separation) $a$, (secondary to primary) mass ratio $q\equiv m_{2}/m_{1}$ and orbital eccentricity $e$ (clockwise). Here for \texttt{tf1e3ta1e6a1m1}, we show the results at $t=2$~Myr to include the binaries that contain PISN progenitors.}
\label{comp1}
\end{figure*}

Fig.~\ref{comp1} shows the binary statistics for four models with different accretion timescales $t_{\rm acc}$ (see Table~\ref{t1}). Again, the effects of $t_{\rm acc}$ mainly show up in the TBMD and distribution of separation. Note that both the total cluster mass $M$ and size $R_{c}$ increase quasi-linearly with $t_{\rm acc}$ according to the universal solution~(Equ.\ref{e2}-\ref{e4}). Similar to the case of $t_{\rm frag}$, $t_{\rm acc}$ also affects the TBMD cut-off mass, as it sets the upper bound of the IMF, $M_{\rm max}$, via $M$, whereas for the distribution of separation, it turns out that a larger $t_{\rm acc}$ results in less close binaries. This trend is caused by two mechanisms: First, the initial distribution of separation in the binary hierarchy itself scales with the cluster size, such that a larger $t_{\rm acc}$ biases the initial distribution to wide binaries. Second, the density of stars in the cluster actually decreases with $t_{\rm acc}$, which leads to less scatters of stars and weaker hardening of binaries (see Table~\ref{t2}). Besides, the suppression of scatters by increasing $t_{\rm acc}$ also increases the fraction of binaries with small mass ratios.

\subsection{Initial mass function}
\label{s4.3}
\begin{figure*}
\includegraphics[width=1.7\columnwidth]{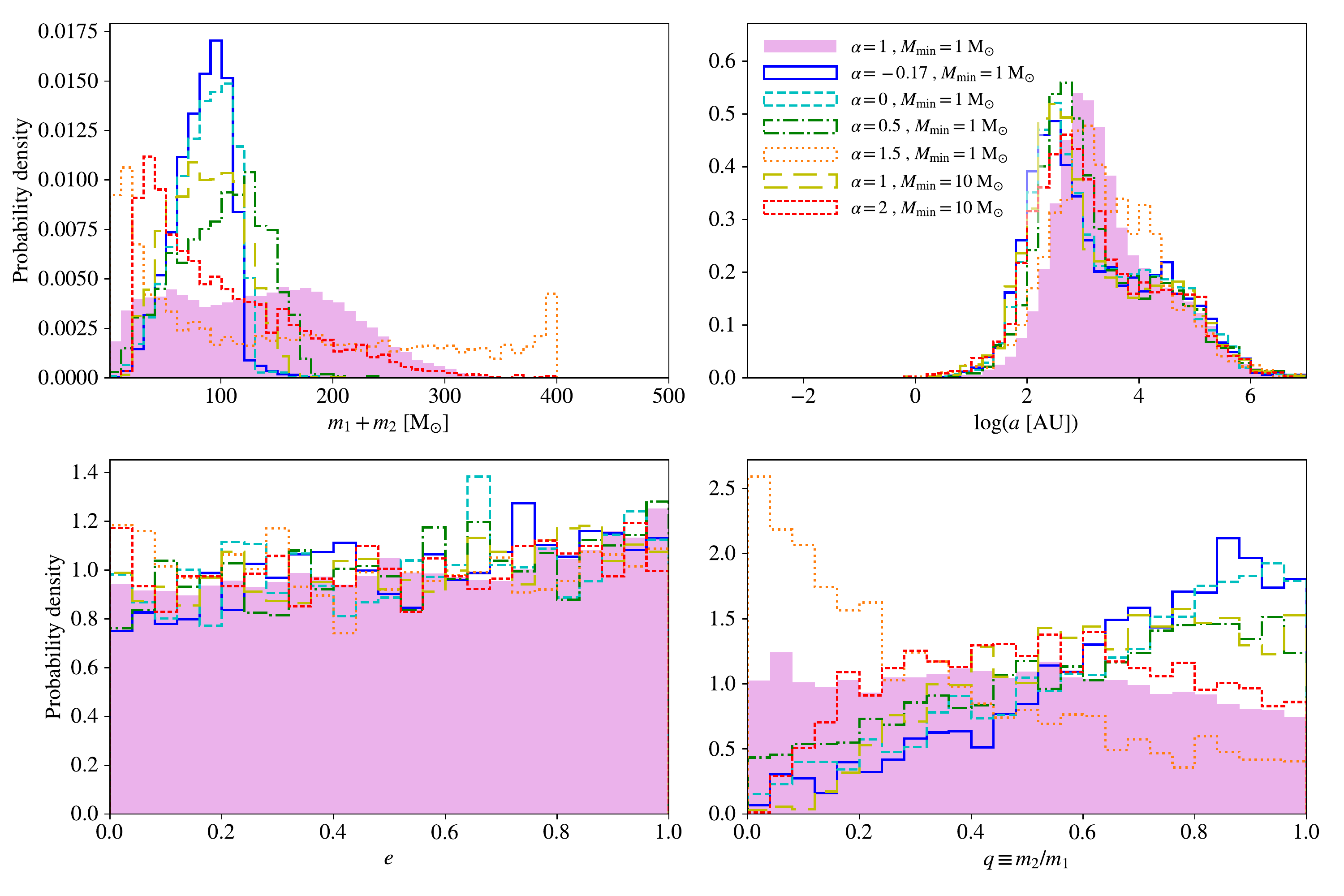}
\caption{Binary statistics for seven models with different IMFs (parameterized by the slope $\alpha$ and lower bound $M_{\min}$), \texttt{tf1e2ta1e5a1m1} (purple histograms), \texttt{tf1e2ta1e5a017m1} (blue solid), \texttt{tf1e2ta1e5a0m1} (cyan dashed), \texttt{tf1e2ta1e5a05m1} (green dashed-dotted), \texttt{tf1e2ta1e5a15m1} (orange dotted), \texttt{tf1e2ta1e5a1m10} (yellow long-dashed) and \texttt{tf1e2ta1e5a2m10} (red short-dashed), in terms of the distributions of total mass $m_{1}+m_{2}$, semi-major axis (separation) $a$, (secondary to primary) mass ratio $q\equiv m_{2}/m_{1}$ and orbital eccentricity $e$ (clockwise).}
\label{comp2}
\end{figure*}

Fig.~\ref{comp2} shows the binary statistics for seven models with different IMFs (parameterized by the slope $\alpha$ and lower bound $M_{\min}$) and other parameters fixed to the fiducial values (see Table~\ref{t1}). Larger $\alpha$ and smaller $M_{\min}$ correspond to a less top-heavy IMF. Again, the different TBMDs reflect the differences in IMF, since the TBMD is also a more top-heavy version of the IMF, similar to the case of the PMD. Interestingly, the effect of IMF on the distribution of separation is weak, with up to $\sim$25\% variations in the binary hardening parameter (see Table~\ref{t2}). Nevertheless, the distribution of mass ratio is sensitive to the IMF, such that a more top-heavy IMF biases the distribution towards the large end. This is consistent with intuition that reducing the fraction of low-mass stars will decrease the number of binaries with small mass ratios, by simply reducing the chance of forming binaries with low-mass stars and also enhancing the chance of kicking low-mass companions out of their systems. 

\subsection{Cluster size and initial distribution of binary separation}
\label{s4.4}
\begin{figure*}
\includegraphics[width=1.7\columnwidth]{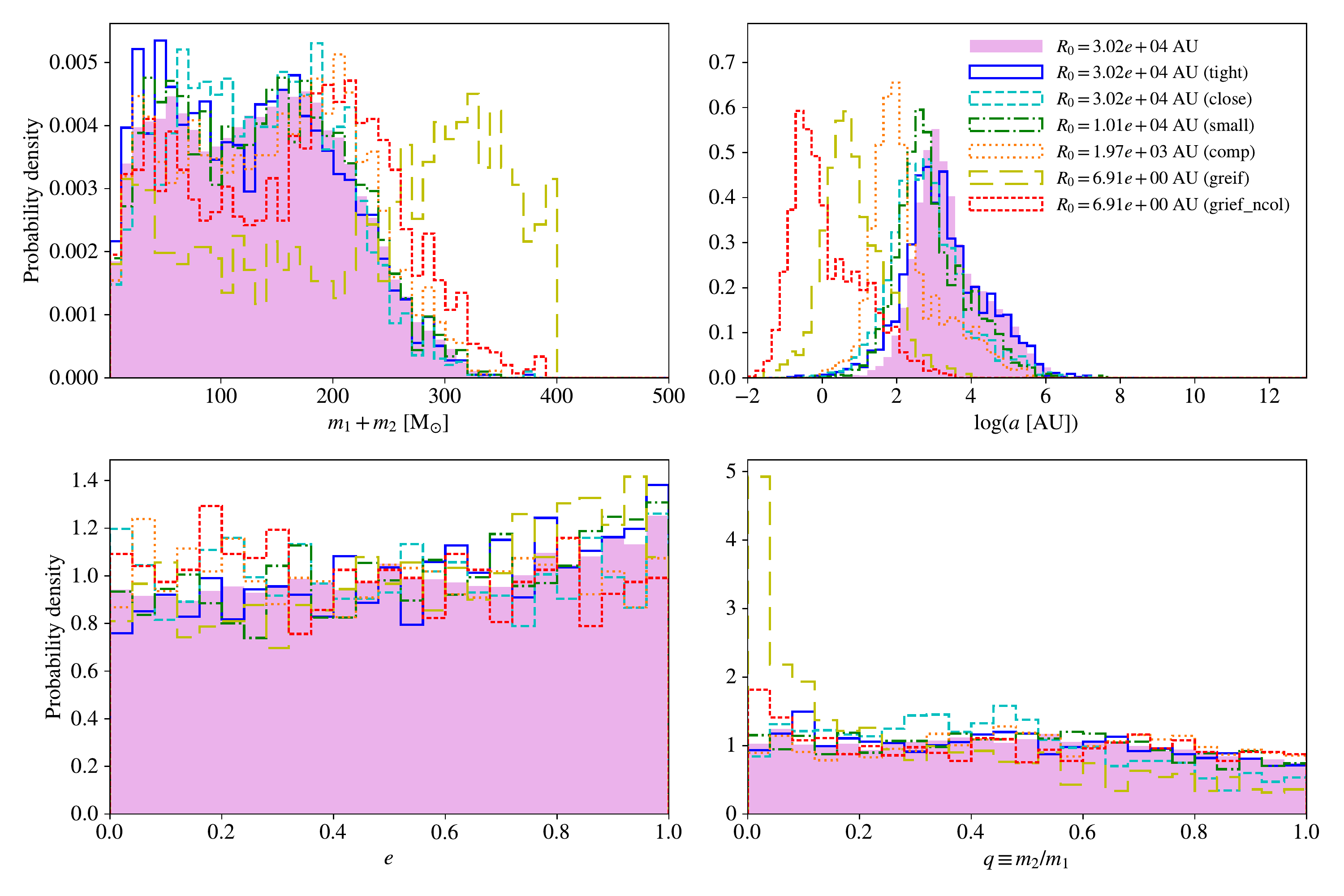}
\caption{Binary statistics for seven models with different cluster sizes, initial distributions of binary separation and treatments of stellar collision, \texttt{tf1e2ta1e5a1m1} (purple histograms), \texttt{tf1e2ta1e5a1m1\_tight} (blue solid), \texttt{tf1e2ta1e5a1m1\_close} (cyan dashed), \texttt{tf1e2ta1e5a1m1\_small} (green dashed-dotted), \texttt{tf1e2ta1e5a1m1\_comp} (orange dotted), \texttt{tf1e2ta1e5a1m1\_greif} (yellow long-dashed) and \texttt{tf1e2ta1e5a1m1\_greif\_ncol} (red short-dashed), in terms of the distributions of total mass $m_{1}+m_{2}$, semi-major axis (separation) $a$, (secondary to primary) mass ratio $q\equiv m_{2}/m_{1}$ and orbital eccentricity $e$ (clockwise).}
\label{comp3}
\end{figure*}

Finally, we relax some assumptions in the Phase 1 model to consider the effects of cluster size $R_{0}$ and initial distribution of binary separation on binary statistics, by comparing seven different models with the total cluster mass, number of stars and IMF fixed, as summarized in Fig.~\ref{comp3}. Although the universal solution~(Equ.~\ref{e2}-\ref{e4}) is consistent with many small-scale simulations, the sample of minihaloes covered by these simulations is still too small to capture all types of environments for Pop~III star formation during early structure formation. In light of this, we consider a series of models with smaller cluster sizes than the fiducial case, to mimic special conditions of strong convergence flows (e.g. during halo mergers) and lower than normal specific angular momentum carried by the collapsing cloud. In \texttt{tf1e2ta1e5a1m1\_small}, the cluster size is reduced by a factor of 3, which reflects the scatters in cluster size from existing small-scale simulations (see Fig.~\ref{f1}). In \texttt{tf1e2ta1e5a1m1\_comp} and \texttt{tf1e2ta1e5a1m1\_greif}, we set the cluster sizes to those of the star-forming disks in SA13 and \citet{greif2012formation}, respectively, to mimic the FS1 and FS2 models in BK17. At last, we further consider a variant of \texttt{tf1e2ta1e5a1m1\_greif} called \texttt{tf1e2ta1e5a1m1\_greif\_ncol}, where stellar collision is suppressed by setting all the stellar radii to $10^{-4}\ \mathrm{AU}\ll R_{\star,\rm ZAMS}$. It turns out that the distribution of separation after Phase 2 evolution is very sensitive to the (initial) cluster size (at the end of Phase 1). When stellar collisions are unimportant (i.e. for $R\gtrsim 10^{3}$~AU and \texttt{tf1e2ta1e5a1m1\_greif\_ncol}), the median separation $\langle a\rangle_{\rm med}$ is approximately proportional to $R_{0}$ (see Fig.~\ref{size}). In the extreme case of \texttt{tf1e2ta1e5a1m1\_greif}, with $R_{0}\sim 7$~AU and $R_{\star,\rm ZAMS}$ adopted to identify collisions, almost each star will experience one collision event on average (see Table~\ref{t2}), such that a significant fraction of close binaries ($a\lesssim 1$~AU), especially with large mass ratios ($q\gtrsim 0.1$), will be disrupted, and there is a higher fraction of massive binaries ($m\gtrsim 300\ \rm M_{\odot}$) made of massive stars formed in collisions. 

\begin{figure}
\includegraphics[width=1\columnwidth]{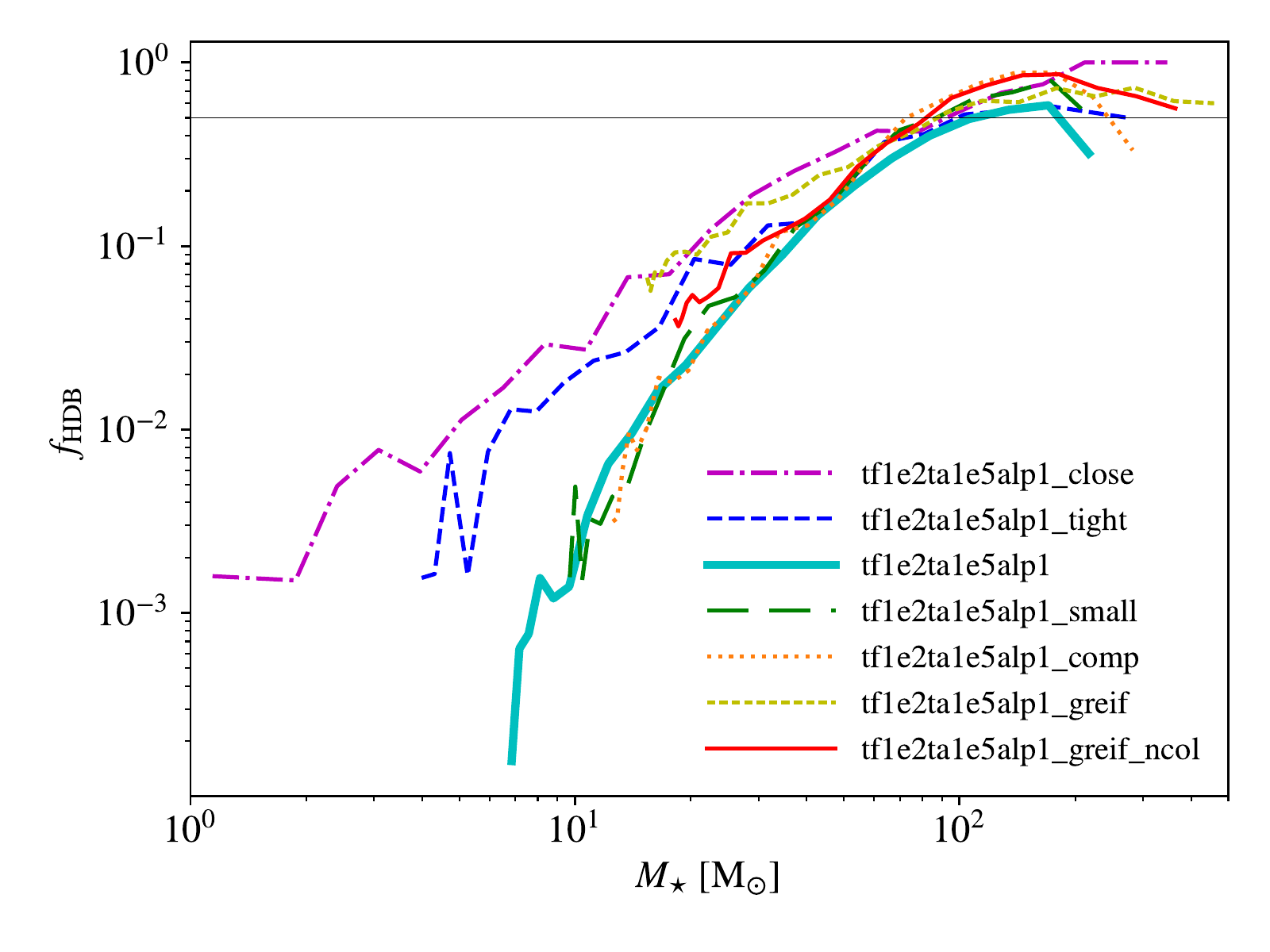}
\caption{Fraction of stars in HDBs as a function of stellar mass for seven models with different cluster sizes, initial distributions of binary separation and treatments of stellar collision (see Table~\ref{t1} for definitions). The thin horizontal line corresponds to 50\%.} 
\label{bfrac}
\end{figure}

Beside cluster size, there are also uncertainties in the fragment properties encountered in small-scale simulations,  
due to limited resolution and simulation time. Therefore, in addition to the default treatment described in Sec.~\ref{s2.2}, we further consider two other ways of initializing binary separations, that lead to more close binaries at the end of Phase 1: For \texttt{tf1e2ta1e5a1m1\_tight}, we have $a_{\min}=a_{\rm lobe}$, where $a_{\rm lobe}$ is the minimum separation for the primary to not fill its Roche lobe. For \texttt{tf1e2ta1e5a1m1\_tight}, all massive companions of the primary are considered to define the influence sphere of the primary, i.e. $a_{\max}\sim \min(a_{\mathrm{ L_{1}},n})$, for $m_{\mathrm{s},n}/m_{\rm p}>0.5$. Not surprisingly, these two models produce more close binaries compared with the fiducial case, but not significantly so, such that the fraction of HDBs is increased by 3.4\% and 28\%, respectively (see Table~\ref{t2}). Particularly, the fraction of low-mass stars ($m_{\star}\lesssim 10\ \rm M_{\odot}$) in HDBs is increased from 0 to $\sim 10^{-3}-0.02$, as shown in Fig.~\ref{bfrac}. Interestingly, the binary hardening parameter actually becomes larger, indicating that the increase in close binaries is inherited from the initial condition rather than driven by stronger dynamical hardening. These outcomes imply that HDBs involving low-mass stars must have very small separations at the end of Phase 1 to survive the following Phase 2 evolution. Besides, such low-mass (close) binaries tend to be extremely rare as long as the total mass of stars is high enough (a few $100\ \rm M_{\odot}$) to have a significant fraction of massive stars (with a top-heavy IMF), which can disrupt them. 

\section{Implications and discussion}
\label{s5}
In this section, we discuss the implications of the Pop~III binary statistics derived from our models, focusing on the X-ray binaries (XRBs, Sec.~\ref{s5.1}) and binary black holes (BBHs, Sec~\ref{s5.2}). For conciseness and simplicity, we do not model binary stellar evolution (i.e. mass transfer, common envelope phases, tidal circularization), when characterizing these two populations of binaries involving BHs. Instead, we adopt optimistic approximations to establish upper limits. Nevertheless, we show below that even with optimistic assumptions, our models generally predict that the signals of XRBs and BBHs originated from Pop~III stars are likely quite weak.

\subsection{X-ray binaries}
\label{s5.1}
To derive the population of XRBs from a given catalog of binary stars, we use a highly simplified model for Eddington accretion via stable mass transfer triggered by Roche lobe overflow. Detailed modelling of X-ray binary evolution is beyond the scope of this study. 

We again adopt the fitting formulae for Pop~III stellar evolution (at a typical metallicity $Z=10^{-6}\ \rm Z_{\odot}$) from \citet{tanikawa2020fitting} to calculate the lifetimes and stellar radii during the main sequence (MS) and giant phases ($\tau_{k}$ and $R_{\star, k}$, $k=\rm MS,\ G$), as well as remnant masses ($m_{\rm rem}$). Once the primary becomes a BH with $m_{\rm BH}=M_{\rm rem,1}>0$ after $\tau_{1}\approx \tau_{\rm MS,1}+\tau_{\rm G, 1}$, we follow the evolution of the secondary. If the secondary is still on the MS, i.e. $\tau_{\rm MS,2}>\tau_{1}$, we compare the pericenter separation $a(1-e)$ to the threshold for the secondary to fill its Roche lobe, $a_{\rm lobe,2}\approx R_{\star,\rm MS, 2}[0.6q^{2/3}+\ln(1+q^{1/3})]/(0.49q^{1/3})$ \citep{eggleton1983aproximations}. If the secondary is in the giant phase, we do the same evaluation by replacing $R_{\star,\rm MS, 2}$ with $R_{\star,\rm G, 2}$. As long as there is one phase in which $a<a_{\rm lobe,2}$, we identify the binary as the progenitor of a XRB. We set the (asymptotic) lifetime of the XRB to $t_{\rm XRB}=\tau_{\rm MS,2}+\tau_{\rm G,2}-\tau_{1}$ if $a<a_{\rm lobe,2}$ is achieved during MS. While if $a<a_{\rm lobe,2}$ is only found in the giant phase, we have $t_{\rm XRB}=\min(\tau_{\rm MS,2}+\tau_{\rm G,2}-\tau_{1},\tau_{\rm G,2})$. Here, as optimistic estimations, we set the MS (giant) radius to that at the end of the MS (giant) phase. Besides, by using the \textit{pericenter} separation, we have also adopted an optimistic assumption that accretion at the pericenter will circularize the orbit on a short timescale to insure continuous accretion over the entire evolution.

Once $t_{\rm XRB}$ is known, we can estimate the total mass accreted onto the XRB as $m_{\rm acc}=\min\{ m_{2},\ m_{\rm BH}[\exp(t_{\rm XRB}/\tau_{\rm Edd})-1]\}$, where $\tau_{\rm Edd}\sim 40$~Myr is the Eddington accretion timescale with an efficiency $\epsilon=0.125$. Note that if the secondary mass $m_{2}$ can be exhausted, the true lifetime is smaller than the asymptotic estimation. For simplicity, we have ignored the effect of mass loss on the evolution of the secondary. We also ignore possible common envelope phase that may rise from unstable mass transfer (e.g. when the secondary has a convective envelope), which can shrink the binary orbit and even lead to merger. 
With $m_{\rm acc,i}$ for each individual XRB $i$, we define $f_{\rm acc}\equiv M_{\rm acc}/M_{\rm tot}$ as the efficiency of accretion onto XRBs, given the total accreted mass $M_{\rm acc}=\sum_{i}m_{\rm acc, i}$ and mass of stars $M_{\rm tot}=M N_{\rm sim}$ from $N_{\rm sim}$ simulations. Then we can derive the accreted mass density of Pop~III XRBs from $f_{\rm acc}$ and any given Pop~III star formation history, as
\begin{align}
\begin{split}
    \rho_{\rm acc}(z)&=f_{\rm acc}\rho_{\star,\rm PopIII}(z)\ ,\\
    \rho_{\star,\rm PopIII}(z)&=\int_{\infty}^{z}\dot{\rho}_{\star,\rm PopIII}(dt/dz')dz'\ ,
\end{split}\label{e5}
\end{align}
where $\dot{\rho}_{\star,\rm PopIII}$ is the Pop~III star formation rate density (SFRD).

It turns out that the efficiency of forming XRBs, $n_{\rm XRB}\equiv N_{\rm XRB}/M_{\rm tot}$, is very low in our models, given the total number of XRBs obtained in the simulations. For the fiducial model \texttt{tf1e2ta1e5a1m1}, we have $n_{\rm XRB}=1.45\times 10^{-5}\ \rm M_{\odot}^{-1}$, and most of these XRBs only have accretion turned on during the giant phase of the secondary. The median BH mass, XRB lifetime and accreted mass are $\langle m_{\rm BH}\rangle_{\rm med}=45\ \rm M_{\odot}$, $\langle t_{\rm XRB}\rangle_{\rm med}=0.3$~Myr and $\langle m_{\rm acc}\rangle_{\rm med}=0.37\ \rm M_{\odot}$, from typical progenitors with $\langle m_{1}\rangle_{\rm med}=91\ \rm M_{\odot}$, $\langle m_{2}\rangle_{\rm med}=63\ \rm M_{\odot}$, $\langle a\rangle_{\rm med} = 481$~AU and $\langle e\rangle_{\rm med}=0.987$. As the giant phase is typically very short, leading to low accreted mass, the accretion efficiency is also very low: $f_{\rm acc}=6.79\times 10^{-6}$. As shown in Sec.~\ref{s4.1}-\ref{s4.3}, changing the IMF, fragmentation and accretion timescales will not significantly affect the population of close binaries, such that we always have $n_{\rm XRB}\lesssim 5.5\times 10^{-5}\ \rm M_{\odot}$ and $f_{\rm acc}\lesssim 1.8\times 10^{-5}$ (see Table~\ref{t2}).

\begin{figure}
\includegraphics[width=1\columnwidth]{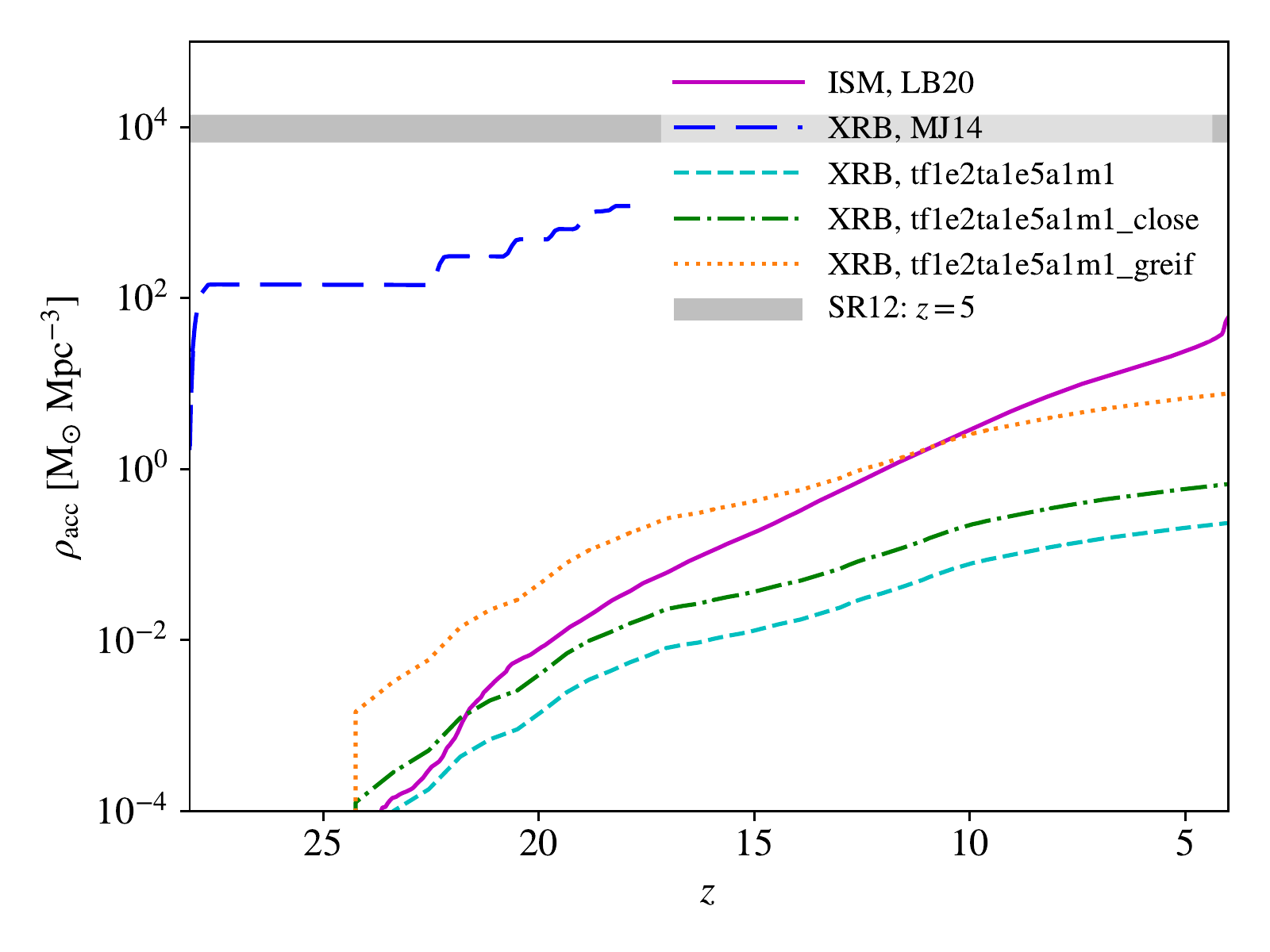}
\caption{Total (co-moving) accreted mass density of Pop~III BHs $\rho_{\rm acc}$ vs. redshift $z$, for accretion from the interstellar medium (ISM) obtained with the cosmological simulation \texttt{FDbox\_Lseed} in \citealt{boyuan2020} (LB20, solid) and X-ray binaries (XRBs), in the simulation of \citealt{jeon2014radiative} (MJ14, long-dashed), and inferred from three models of this work: \texttt{tf1e2ta1e5a1m1} (dashed), \texttt{tf1e2ta1e5a1m1\_close} (dashed-dotted) and \texttt{tf1e2ta1e5a1m1\_greif} (dotted). The shaded region shows the upper limit at $z=5$, placed by the unresolved cosmic x-ray background, from \citealt{salvaterra2012limits} (SR12). } 
\label{rhoacc}
\end{figure}

\begin{figure}
\includegraphics[width=1\columnwidth]{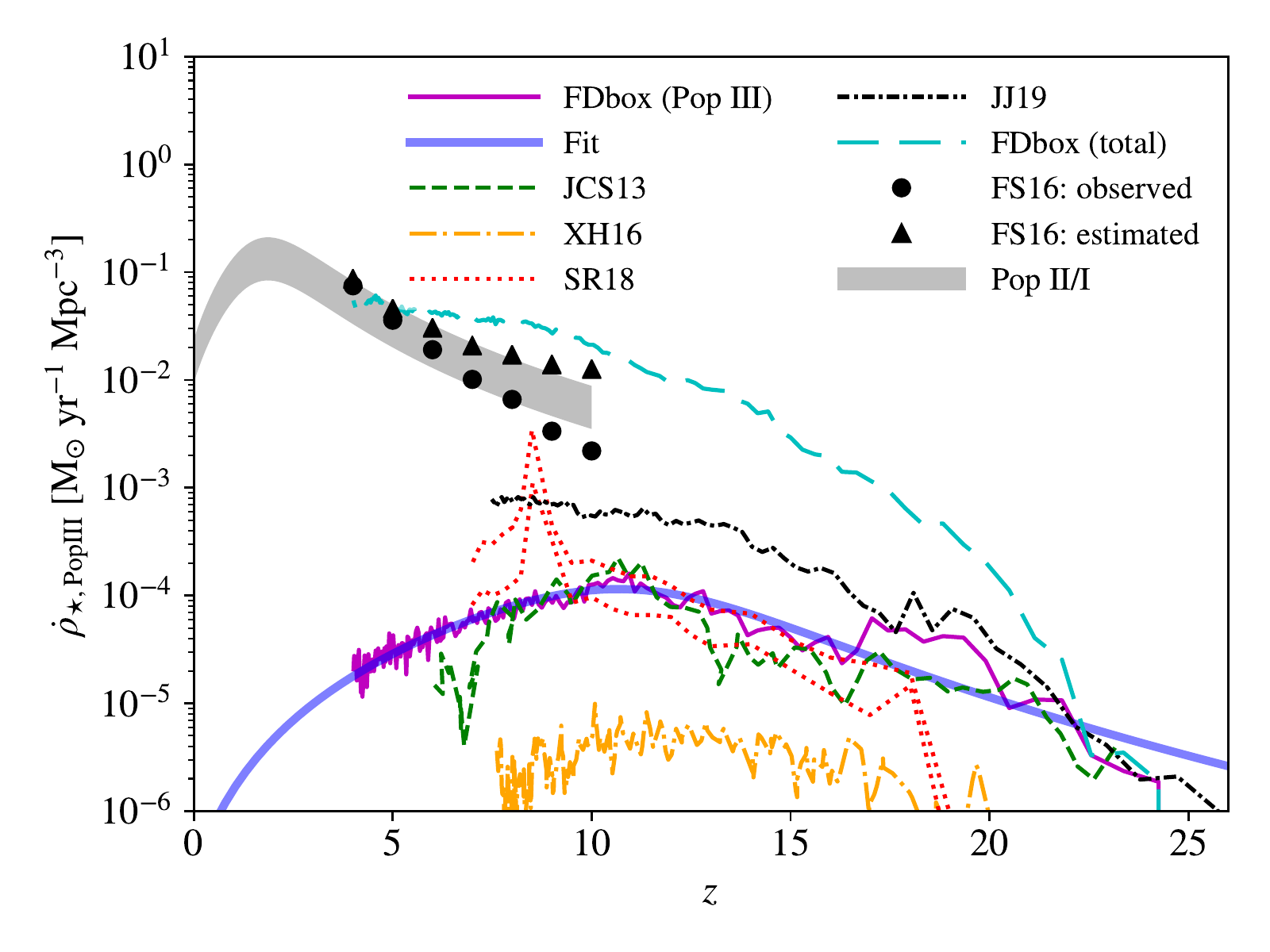}
\vspace{-20pt}
\caption{Co-moving Pop~III SFRD (Fig. 5 of \citealt{liu2020did}). The simulation data from \texttt{FDbox\_Lseed} in \citet{boyuan2020} and the corresponding fit (see their Equ. 6) are shown with thin and thick solid curves. We further plot the results from other cosmological simulations in \citealt{johnson2013first} (JCS13; with Lyman-Werner feedback), \citealt{xu2016late} (XH16), \citealt{sarmento2018following} (SR18) and \citealt{jaacks2019legacy} (JJ19), with the dashed, dashed-dotted, dotted and densely dashed-dotted curves, which demonstrate the range of Pop~III star formation histories in current models. Note that the XH16 results are based on a zoom-in simulation for a low-density region ($\langle\delta\rangle=-0.26$ at $z=8$), which should be regarded as lower limits. The SR18 results include two cases with (upper) and without (lower) unresolved inefficient metal mixing. For comparison, we plot the (extrapolated) total SFRD (with 0.2 dex scatters) from \citealt{madau2014araa} (shaded region), inferred by UV and IR galaxy surveys, such as \citealt{finkelstein2016observational} (FS16; data points). The corresponding simulated total SFRD is shown with the long-dashed curve.}
\label{sfrd}
\end{figure}

For more extreme models with some key assumptions for Phase~1 relaxed (see Sec.~\ref{s4.4}), the efficiency of XRBs can be significantly enhanced by up to two orders of magnitude (see Table~\ref{t2} and Fig.~\ref{size}). However, the contribution from Pop~III XRBs to the cosmic (unresolved) X-ray background seems to be negligible, even in these extreme cases. For instance, Fig.~\ref{rhoacc} shows the (co-moving) accreted mass density of Pop~III XRBs derived from three typical models, \texttt{tf1e2ta1e5a1m1}, \texttt{tf1e2ta1e5a1m1\_close} and \texttt{tf1e2ta1e5a1m1\_greif}, in comparison with the result from \citet{jeon2014radiative}, and the accreted mass density for ISM accretion from the cosmological simulation \texttt{FDbox\_Lseed} in \citet{boyuan2020}. Here we adopt the Pop~III SFRD from \citet{boyuan2020} to be self-consistent, shown in Fig.~\ref{sfrd}. As discussed in \citet{liu2020did}, our Pop~III SFRD is approximately the median value among various simulation results \citep{tornatore2007population,wise2011birth,johnson2013first,xu2016late,sarmento2018following}. Integrating $\dot{\rho}_{\star,\mathrm{PopIII}}$ across cosmic history gives a density of all Pop~III stars ever formed, $\sim 10^{5}\ \rm M_{\odot}\ Mpc^{-3}$, which is lower by up to one order of magnitude than the upper limit set by the recent \textit{Planck} measurement of the optical depth to electron scattering \citep{visbal2015,inayoshi2016gravitational}. Therefore, we may have underestimated cosmic Pop~III star formation by up to one order of magnitude. 
Note that the same simulation also reproduces the total SFRD inferred from observation \citep{madau2014araa,finkelstein2016observational} within a factor of 2. We refer the reader to Section 3 of \citet{boyuan2020} for more detailed detailed descriptions of the simulation results.  %

The accreted mass density from XRBs can only be larger than that of ISM accretion at very early times ($z\gtrsim 12-22$), and remains below $10^{-3}$ of the value required to explain the cosmic X-ray background observed at $z=5$ \citep{salvaterra2012limits}. Taking into account possible underestimation of the Pop~III SFRD can boost the accreted mass density by up to a factor of 10, still far below the observational upper limit. The efficiencies of forming Pop~III XRBs and their accretion predicted in our models are much lower (by factors of $\sim 10-10^{4}$) than the values used in previous studies focusing on the feedback effects from Pop~III XRBs (e.g. \citealt{jeon2014radiative,hummel2015first,ryu2016formation}, see Fig.~\ref{size} for details). This implies that the feedback from Pop~III XRBs is unimportant, at least for the global background. 

\subsection{Binary black holes and gravitational waves}
\label{s5.2}
For simplicity, we do not consider binary stellar evolution and supernova kicks during the formation of BBHs, such that the population of BBHs is constructed simply by changing the stellar masses to remnant masses (also with the stellar evolution model in \citealt{tanikawa2020fitting}), with the binary orbital parameters ($a$ and $e$) fixed\footnote{Here our simplified method is valid for massive Pop~III wide binaries, which do not experience close binary interactions and tend to collapse directly into BHs (with no natal kicks). We model the BBH mergers from close binaries separately.}. As minihaloes tend to merge into larger structures in a short timescale $\sim 100$~Myr, during the lifetimes of most BBHs, they will be surrounded by low-mass Pop~II/I stars formed in larger structures, such that only hard binaries can survive the 3-body interactions with surrounding stars. In light of this, we have restricted our analysis to hard BBHs, defined as $a[(100\ \mathrm{M_{\odot}})^{2}/(m_{\rm BH,1}m_{\rm BH,2})](1+e)/(1-e)\lesssim 10^{4}\ {\rm AU}\sim 0.05\ \rm pc$, which will be hardened rather than disrupted by interactions with typical surrounding low-mass Pop~II/I stars with a stellar mass $m_{\star}\sim 1\ \rm M_{\odot}$ and a velocity dispersion of $\sigma\sim 30\ \rm km\ s^{-1}$. 

We then calculate the time taken for each (hard) BBH to merge, called delay time $t_{\rm delay}$, under three conditions. (i) We consider isolated evolution purely driven by emission of gravitational waves, where $t_{\rm delay}\propto a^{4}$. (ii) We further include 3-body hardening by surrounding stars, following the semi-analytical model described in \citet{boyuan2020}, in turn based on extrapolations of observed BH-host scaling relations. The reader is referred to \citet{boyuan2020} for details. In general, this model captures the strength of 3-body hardening with a parameter $\gamma$, which is the slope of the density profile of stars around the BBH. We consider $\gamma\sim 0.5-1.5$, where the value $1.5$ implies a stellar density $\rho_{\star}\sim 10^{4}\ \rm M_{\odot}\ pc^{-3}$, typical at the centers of globular and nuclear star clusters. Lastly, it is well known that binary stellar evolution is very important for Pop~III BBHs (e.g. \citealt{kinugawa2014possible,belczynski2017likelihood,tanikawa2020merger}). In light of this, we adopt a simple optimistic assumption to estimate its effect: (iii) All XRBs identified in Sec.~\ref{s5.1} with Roche lobe overflow of the secondary, will experience common envelope evolution, which reduces the separation to cause the BBH to merge within a Hubble time. We defer more complete calculations, combining all the three mechanisms above, to future work. Note that we do not include natal kicks for the primary to become a BH. Although massive Pop III stars ($\gtrsim 40~\rm \ M_{\odot}$) are usually expected to collapse directly into BHs without natal kicks (e.g. \citealt{heger2003massive}), such kicks, if present, can enhance the chance of common envelope evolution by shrinking the binary orbit, especially for BHs in the pulsational pair instability gap of $M_{\rm BH}\sim 55-130\ \rm M_{\odot}$ \citep{tanikawa2020merger}. Therefore, we may have underestimated the efficiency of BBHs in the common envelope channel.

Once $t_{\rm delay}$ is known, we can derive the efficiency of BBH mergers as $f_{\rm BBH}\equiv N_{\rm BBH}/M_{\rm tot}$, where $N_{\rm BBH}$ is the total number of BBHs that can merge within a Hubble time ($t_{\rm delay}<t_{\rm H}\sim 13.7$~Gyr). The values of $f_{\rm BBH}$ for all the 18 models explored in this work are listed in Table~\ref{t2}, and the effects of changing our Phase 1 assumptions are shown in Fig.~\ref{size}.

It turns out that under the `standard' Phase 1 picture, no BBHs can merge within a Hubble time with purely isolated evolution (i), regardless of the IMF, accretion and fragmentation timescales. With 3-body hardening (ii), only the extreme case with $\gamma=1.5$ can have non-zero efficiencies up to $\sim 2\times 10^{-3}\ \rm M_{\odot}^{-1}$. The question then becomes which fraction of BBHs can fall into such dense star clusters (with $\rho_{\star}\gtrsim 10^{4}\ \rm M_{\odot}\ pc^{-3}$) to be significantly hardened. Here we roughly estimate the fraction of Pop~III BHs in nuclear star clusters $f_{\rm BH,NSC}$. According to \citet{boco2020growth}, dynamical friction of the gaseous disks in early-type galaxies can drive BHs of $\sim 100\ \rm M_{\odot}$ into the center within $\lesssim 1$~Gyr, which are initially $\lesssim 300$~pc away, when gas is mainly supported by rotation. We assume that BHs within $r\lesssim 300$~pc from galaxy centers at $z\lesssim 6$ (corresponding to a cosmic age of $t\gtrsim 1$~Gyr) tend to fall into nuclear star clusters (with $\rho_{\star}\sim 10^{5}\ \rm M_{\odot}\ pc^{-3}$), which are common in high-$z$ AC cooling haloes with $M_{\rm halo}\gtrsim 10^{8}\ \rm M_{\odot}$ (e.g. \citealt{devecchi2009formation,devecchi2010high,devecchi2012high}). We calculate the fraction of Pop~III BHs within 60 (300)~pc of galaxy centers at $z=4$ as $\sim 0.03$ (0.3), from the cosmological simulation \texttt{FD\_box\_Lseed} in \citealt{boyuan2020}. Here 60~pc is the gravitational softening length of gas particles in the simulation, beyond which dynamical friction by gas is unresolved. Therefore, we have $f_{\rm BH, NSC}\sim 0.03-0.3$ and the true efficiency for Pop~III BBH mergers driven by 3-body hardening in nuclear star clusters is $\hat{f}_{\rm BBH}=f_{\rm BBH,NSC}f_{\rm BBH}\sim 7\times 10^{-6}-6\times 10^{-4}\ \rm M_{\odot}^{-1}$. When considering optimistic common envelope evolution (iii), we have $f_{\rm BBH}\sim n_{\rm XRB}\sim 1.45-5.5\times 10^{-5}\ \rm M_{\odot}^{-1}$. 

However, if the initial cluster size can be as small as a few AU (for \texttt{tf1e2ta1e5a1m1\_greif} and \texttt{tf1e2ta1e5a1m1\_greif\_ncol}, see Sec.~\ref{s4.4}), even isolated evolution (i) results in $f_{\rm BBH}\sim 5.5\times 10^{-5}-6.4\times 10^{-4}\ \rm M_{\odot}^{-1}$, similar to the cases including 3-body hardening (ii). The common envelope evolution model (iii), on the other hand, predicts $f_{\rm BBH}\sim  10^{-5}-10^{-3}\ \rm M_{\odot}^{-1}$. In conclusion, our fiducial model \texttt{tf1e2ta1e5a1m1} predicts $f_{\rm BBH}\sim 10^{-5}-10^{-4}\ \rm M_{\odot}^{-1}$, for which both 3-body hardening (ii) and common envelope evolution (iii) are important for facilitating in-spirals of BBHs to result in mergers within a Hubble time. 

\section{Summary and Conclusions}
\label{s6}
We use N-body simulations to study the dynamical evolution of Population~III (Pop~III) star clusters and the resulting binary statistics. Our simulations are based on a physically motivated framework for the initial conditions of newly-formed Pop~III star clusters (Sec.~\ref{s2}), which takes into account the features of hierarchical fragmentation and scale-free nature of disk evolution during Pop~III star formation, found in small-scale hydrodynamic simulations (e.g. \citealt{beuther2019high,susa2019merge,clark2020emergent,oliva2020modeling}). The key concept of this framework is a universal solution of protostellar disk evolution (see Equ.~\ref{e2}-\ref{e4} and Fig.~\ref{f2}), associating the global properties of Pop~III clusters (i.e. mass, size and number of stars) with the fragmentation and accretion timescales, $t_{\rm frag}$ and $t_{\rm acc}$, which is obtained by fitting simulation data. By exploring 18 models with different initial condition parameters, such as $t_{\rm frag}$ and $t_{\rm acc}$, initial mass function (IMF), distribution of (initial) binary separation $a$, cluster size and stellar collision scheme (see Table~\ref{t1} and \ref{t2}), we obtain the general trends of Pop~III cluster evolution (Sec.~\ref{s3}), and evaluate the dependence of Pop~III statistics on initial conditions (Sec.~\ref{s4} and Appendix~\ref{a1}), as well as the relevant implications for Pop~III X-ray binaries (XRBs, Sec.~\ref{s5.1}) and binary black hole (BBH) mergers (Sec.~\ref{s5.2}). Our main findings and conclusions are summarized below.
\begin{itemize}
    \item After gas removal by (proto)stellar feedback, Pop~III star clusters first undergo a contraction phase for approximately one relaxation timescale ($\sim$ a few dynamical timescales) during which virial equilibrium is established and the size of the system is reduced by a factor of $\sim 2$. Then the system re-expands gradually due to two-body relaxation, and the majority of binary evolution (i.e. formation, disruption and hardening of binaries), driven by gravitational scatterings among stars, occurs within a few relaxation timescales ($\sim$ a few Myr), accompanied by puffing up of the disk, collisions and ejections of stars. In most cases, by the end of the simulation ($\sim$ 10 relaxation timescales), about half of the stars are unbound from the cluster (and host minihalo of $M_{\rm halo}\sim 10^{6}\ \rm M_{\odot}$), corresponding to $\sim 0.1-0.3$ of the total stellar mass. Stellar collisions are not important (with a few events in every $10^{4}\ \rm M_{\odot}$), as long as the initial cluster size is much larger than a few AU. Pair-instability supernovae, if present under an IMF that is sufficiently top-heavy, can facilitate cluster dispersion after $\sim 2$~Myr, boosting the fraction of ejected stars by $\sim 75$\%.
    \item Our models predict much fewer close binaries ($a\lesssim 100$~AU) than in previous studies with simplified treatments of Pop~III binary statistics (e.g. \citealt{kinugawa2014possible,ryu2016formation,belczynski2017likelihood}). The reason is that previous studies have underestimated the cluster size by a factor of $\sim 15-10^{4}$ when turning protostars directly into stars without taking into account expansion of the system by angular-momentum conservation during accretion. We are able to roughly reproduce their results by reducing the cluster size accordingly (Appendix~\ref{a2}), confirming that (initial) cluster size is the most important initial condition parameter for Pop~III binary statistics (see Sec.~\ref{s4.4}).
    \item Other than the cluster size, binary statistics is also affected by the two characteristic timescales, $t_{\rm frag}$ and $t_{\rm acc}$, which are positively correlated with the total number and mass of stars in the cluster, respectively. In general, more close binaries are produced with larger $t_{\rm frag}$ (i.e. more stars) and smaller $t_{\rm acc}$ (i.e. smaller cluster sizes and lower total masses). However, the population of close binaries is insensitive to the IMF (with up to 25\% variations in binary hardening), although the IMF leaves strong imprints on the distributions of total binary mass and mass ratio. 
    \item We find that binaries involving low-mass stars ($\lesssim 10\ \rm M_{\odot}$) are very rare (with a binary fraction $\lesssim 0.02$) due to disruptions from massive stars during scatters. This holds as long as the cluster is massive enough (a few $100\ \rm M_{\odot}$) to have a significant fraction of massive stars. The only way to efficiently form such low-mass binaries, which are candidate progenitors of carbon-enhanced-metal-pool (CEMP) stars with AGB winds \citep{izzard2009population,abate2013wind,hansen2016}, is to shut down accretion early on or reduce the accretion rate significantly by protostellar feedback, the so-called low-mass mode of Pop~III star formation \citep{stacy2014first}. This situation deviates from the universal solution, and it remains uncertain how likely it occurs, probably only in certain types of host minihaloes. Nevertheless, CEMP stars may be the second generation of stars formed in clouds enriched by (low-energy) supernovae of Pop~III stars (e.g. \citealt{ji2015preserving,cooke2014carbon}) or by spinstars, i.e. fast rotating very metal-poor massive stars (e.g. \citealt{meynet2010c,maeder2015first,choplin2017some}).
    \item Due to the lack of close binaries in our models, we predict significantly (by a factor of $10-10^{3}$) lower efficiencies of forming Pop~III XRBs (only a few in every $10^{5}\ \rm M_{\odot}$) than assumed in previous studies (e.g. \citealt{jeon2014radiative,hummel2015first,ryu2016formation}). Moreover, most of these XRBs experience accretion only during the giant phase of the secondary, over $\lesssim 1$~Myr, such that the efficiency of accretion is even lower (by up to a factor of $10^{4}$). As a result, the contribution from Pop~III XRBs to the cosmic (unresolved) X-ray background \citep{salvaterra2012limits} is negligible ($\lesssim 10^{-3}$) even in extreme cases with very small (a few AU) clusters and enhanced close binary formation. It is actually overwhelmed by the contribution from (smooth) accretion onto Pop~III BHs from the interstellar medium inferred from cosmological simulations \citep{boyuan2020}, except for the very early stage ($z\gtrsim 12-22$). We conclude that the feedback from Pop~III XRBs is unimportant for the global evolution of the intergalactic medium.
    \item Again, without many close binaries, the estimated efficiencies\footnote{Note that construction of the population of Pop~III BBH mergers involves many astrophysical aspects which deserve detailed modelling, and the (initial) binary statistics is just the starting point. Our results based on simplified calculations are only for illustrative purposes. } of forming BBH mergers from Pop~III stars via common evolution of close binaries in our models, $f_{\rm BBH}\sim 10^{-5}\ \rm M_{\odot}$, are at the low end of the range obtained in previous studies (e.g. \citealt{kinugawa2014possible,hartwig2016,belczynski2017likelihood,boyuan2020}). For classical binary stellar evolution, the efficiency required to explain the total merger rate of BBHs observed by LIGO-Virgo \citep{abbott2019gwtc}, $f_{\rm BBH}\sim 10^{-4}\ \rm M_{\odot}$ \citep{kinugawa2014possible}, can only be achieved with much (a factor of $\gtrsim 15$) smaller cluster sizes. Nevertheless, this efficiency can also be reached by dynamical hardening of initially wide binaries in dense star clusters (e.g. \citealt{antonini2016merging,leigh2018rate,bl2020gw190521}), if a significant fraction ($\gtrsim 0.1$) of Pop~III BHs fall into dense star clusters. 
\end{itemize}

The main purpose of this paper is to provide an improved set of Pop~III binary statistics, which can be applied to many fields involving Pop~III binaries. To promote such broad follow-up studies, we make our data \href{https://drive.google.com/drive/folders/1JHBjhSBDPT3jdipdQtmr2IJ5TV6Xe6Id?usp=sharing}{public}. Our fiducial model and its variants following the universal solution can be regarded as the typical/average states of Pop~III star clusters formed in high-$z$ minihaloes. More extreme models with smaller cluster sizes are outliers from the universal solution, but can still contribute significantly to the observational signatures of Pop~III stars, depending on how likely such situations occur in early cosmic structure formation, which is still unknown.

To further improve our knowledge of Pop~III binaries, future theoretical efforts should systematically explore the final stage of Pop~III star formation, such as the growth of protostars and removal of gas by feedback, in a representative sample of star-forming clouds in a realistic cosmological context. Only in this way can we understand the deviations around the universal solution, and thus construct models closer to the complex reality, which should be a mixture of different modes of Pop~III star formation. Due to current limitation in computational power, a promising approach is hybrid simulations of N-body dynamics for protostars, sub-grid feedback schemes (based on sink particles) and radiative hydrodynamics for the surrounding gas (e.g. \citealt{wall2020modelling}), with advanced tools such as \textsc{AMUSE} \citep{zwart2009multiphysics,zwart2013multi,pelupessy2013astrophysical,portegies2018astrophysical,amuse}. On the observational side, more advanced gravitational wave instruments will come into operation over the next decades (e.g. \citealt{punturo2010einstein,gair2011imbhet,abbott2017exploring,abbott2018prospects,abbott2019search,dechihertz,tiango,robson2019construction,tianqin}). They will provide a greatly enhanced view of the demographics of BBH mergers at $z\lesssim 10$, where it will be possible to identify sub-populations dominated by Pop~III progenitors, whose properties (e.g. redshift evolution of merger rate, distributions of mass and effective spin) will place constraints on Pop~III binaries combined with theoretical predictions \citep{kinugawa2014possible,hartwig2016,belczynski2017likelihood,kinugawa2020chirp,boyuan2020,tanikawa2020merger,bl2020gw190521}. More broadly, considering how dynamical processes
shape the evolution of the first stellar systems promises a more complete understanding of early cosmic history.

\section*{Acknowledgments}
The authors acknowledge the Texas Advanced Computing Center (TACC) for providing HPC resources under XSEDE allocation TG-AST120024. Georges Meynet has received funding from the European Research Council (ERC) under the European Union's Horizon 2020 research and innovation program (grant agreement No 833925, project STAREX).


\section*{Data availability}
The data underlying this article are available at \href{https://drive.google.com/drive/folders/1JHBjhSBDPT3jdipdQtmr2IJ5TV6Xe6Id?usp=sharing}{this Google Drive folder}. The codes that generate/process the data will be shared on reasonable request to the corresponding authors.

\bibliographystyle{mnras}
\bibliography{ref} 

\begin{thebibliography}{}
\makeatletter
\relax
\def\mn@urlcharsother{\let\do\@makeother \do\$\do\&\do\#\do\^\do\_\do\%\do\~}
\def\mn@doi{\begingroup\mn@urlcharsother \@ifnextchar [ {\mn@doi@}
  {\mn@doi@[]}}
\def\mn@doi@[#1]#2{\def\@tempa{#1}\ifx\@tempa\@empty \href
  {http://dx.doi.org/#2} {doi:#2}\else \href {http://dx.doi.org/#2} {#1}\fi
  \endgroup}
\def\mn@eprint#1#2{\mn@eprint@#1:#2::\@nil}
\def\mn@eprint@arXiv#1{\href {http://arxiv.org/abs/#1} {{\tt arXiv:#1}}}
\def\mn@eprint@dblp#1{\href {http://dblp.uni-trier.de/rec/bibtex/#1.xml}
  {dblp:#1}}
\def\mn@eprint@#1:#2:#3:#4\@nil{\def\@tempa {#1}\def\@tempb {#2}\def\@tempc
  {#3}\ifx \@tempc \@empty \let \@tempc \@tempb \let \@tempb \@tempa \fi \ifx
  \@tempb \@empty \def\@tempb {arXiv}\fi \@ifundefined
  {mn@eprint@\@tempb}{\@tempb:\@tempc}{\expandafter \expandafter \csname
  mn@eprint@\@tempb\endcsname \expandafter{\@tempc}}}

\bibitem[\protect\citeauthoryear{Abate, Pols, Izzard, Mohamed  \&
  De~Mink}{Abate et~al.}{2013}]{abate2013wind}
Abate C.,  Pols O.,  Izzard R.,  Mohamed S.,   De~Mink S.,  2013, \aap, 552,
  A26

\bibitem[\protect\citeauthoryear{Abbott et~al.,}{Abbott
  et~al.}{2017}]{abbott2017exploring}
Abbott B.~P.,  et~al., 2017, Classical and Quantum Gravity, 34, 044001

\bibitem[\protect\citeauthoryear{Abbott et~al.,}{Abbott
  et~al.}{2018}]{abbott2018prospects}
Abbott B.~P.,  et~al., 2018, Living Reviews in Relativity, 21, 3

\bibitem[\protect\citeauthoryear{Abbott et~al.,}{Abbott
  et~al.}{2019a}]{abbott2019gwtc}
Abbott B.,  et~al., 2019a, Phys. Rev. X, 9, 031040

\bibitem[\protect\citeauthoryear{Abbott et~al.,}{Abbott
  et~al.}{2019b}]{abbott2019search}
Abbott B.,  et~al., 2019b, \prd, 100, 064064

\bibitem[\protect\citeauthoryear{Abbott et~al.,}{Abbott
  et~al.}{2020a}]{abbott2020gw190521}
Abbott R.,  et~al., 2020a, \prl, 125, 101102

\bibitem[\protect\citeauthoryear{Abbott et~al.,}{Abbott
  et~al.}{2020b}]{abbott2020properties}
Abbott R.,  et~al., 2020b, \apj, 900, L13

\bibitem[\protect\citeauthoryear{Abt}{Abt}{1983}]{abt1983normal}
Abt H.~A.,  1983, \araa, 21, 343

\bibitem[\protect\citeauthoryear{Alister~Seguel, Schleicher, Boekholt,
  Fellhauer  \& Klessen}{Alister~Seguel et~al.}{2020}]{alister2020formation}
Alister~Seguel P.,  Schleicher D.,  Boekholt T.,  Fellhauer M.,   Klessen R.,
  2020, \mnras, 493, 2352

\bibitem[\protect\citeauthoryear{{Andr{\'e} Oliva} \& {Kuiper}}{{Andr{\'e}
  Oliva} \& {Kuiper}}{2020}]{oliva2020modeling}
{Andr{\'e} Oliva} G.,  {Kuiper} R.,  2020, arXiv e-prints, \href
  {https://ui.adsabs.harvard.edu/abs/2020arXiv200813653A} {p. arXiv:2008.13653}

\bibitem[\protect\citeauthoryear{Antonini \& Rasio}{Antonini \&
  Rasio}{2016}]{antonini2016merging}
Antonini F.,  Rasio F.~A.,  2016, \apj, 831, 187

\bibitem[\protect\citeauthoryear{{Arca Sedda} et~al.,}{{Arca Sedda}
  et~al.}{2019}]{dechihertz}
{Arca Sedda} M.,  et~al., 2019, arXiv e-prints, \href
  {https://ui.adsabs.harvard.edu/abs/2019arXiv190811375A} {p. arXiv:1908.11375}

\bibitem[\protect\citeauthoryear{Bavera et~al.,}{Bavera
  et~al.}{2020}]{bavera2020origin}
Bavera S.~S.,  et~al., 2020, \aap, 635, A97

\bibitem[\protect\citeauthoryear{Belczynski, Holz, Bulik  \&
  O'Shaughnessy}{Belczynski et~al.}{2016}]{belczynski2016first}
Belczynski K.,  Holz D.~E.,  Bulik T.,   O'Shaughnessy R.,  2016, \nat, 534,
  512

\bibitem[\protect\citeauthoryear{Belczynski, Ryu, Perna, Berti, Tanaka  \&
  Bulik}{Belczynski et~al.}{2017}]{belczynski2017likelihood}
Belczynski K.,  Ryu T.,  Perna R.,  Berti E.,  Tanaka T.~L.,   Bulik T.,  2017,
  \mnras, 471, 4702

\bibitem[\protect\citeauthoryear{Beuther et~al.,}{Beuther
  et~al.}{2019}]{beuther2019high}
Beuther H.,  et~al., 2019, \aap, 621, A122

\bibitem[\protect\citeauthoryear{Boco, Lapi  \& Danese}{Boco
  et~al.}{2020}]{boco2020growth}
Boco L.,  Lapi A.,   Danese L.,  2020, \apj, 891, 94

\bibitem[\protect\citeauthoryear{Boekholt, Schleicher, Fellhauer, Klessen,
  Reinoso, Stutz  \& Haemmerl{\'e}}{Boekholt
  et~al.}{2018}]{boekholt2018formation}
Boekholt T.,  Schleicher D.,  Fellhauer M.,  Klessen R.,  Reinoso B.,  Stutz
  A.,   Haemmerl{\'e} L.,  2018, \mnras, 476, 366

\bibitem[\protect\citeauthoryear{Bromm}{Bromm}{2013}]{bromm2013}
Bromm V.,  2013, Rep. Prog. Phys., 76, 112901

\bibitem[\protect\citeauthoryear{Bromm \& Yoshida}{Bromm \&
  Yoshida}{2011}]{bromm2011first}
Bromm V.,  Yoshida N.,  2011, \araa, 49, 373

\bibitem[\protect\citeauthoryear{{Chiaki} \& {Yoshida}}{{Chiaki} \&
  {Yoshida}}{2020}]{chiaki2020}
{Chiaki} G.,  {Yoshida} N.,  2020, arXiv e-prints, \href
  {https://ui.adsabs.harvard.edu/abs/2020arXiv200806107C} {p. arXiv:2008.06107}

\bibitem[\protect\citeauthoryear{Chon \& Omukai}{Chon \&
  Omukai}{2020}]{chon2020supermassive}
Chon S.,  Omukai K.,  2020, \mnras, 494, 2851

\bibitem[\protect\citeauthoryear{Choplin, Hirschi, Meynet  \&
  Ekstr{\"o}m}{Choplin et~al.}{2017}]{choplin2017some}
Choplin A.,  Hirschi R.,  Meynet G.,   Ekstr{\"o}m S.,  2017, \aap, 607, L3

\bibitem[\protect\citeauthoryear{Clark \& Whitworth}{Clark \&
  Whitworth}{2020}]{clark2020emergent}
Clark P.~C.,  Whitworth A.~P.,  2020, arXiv preprint arXiv:2008.09808

\bibitem[\protect\citeauthoryear{Cooke \& Madau}{Cooke \&
  Madau}{2014}]{cooke2014carbon}
Cooke R.~J.,  Madau P.,  2014, \apj, 791, 116

\bibitem[\protect\citeauthoryear{{Dayal} \& {Ferrara}}{{Dayal} \&
  {Ferrara}}{2018}]{dayal2018}
{Dayal} P.,  {Ferrara} A.,  2018, \mn@doi [\physrep]
  {10.1016/j.physrep.2018.10.002}, \href
  {https://ui.adsabs.harvard.edu/\#abs/2018PhR...780....1D} {780, 1}

\bibitem[\protect\citeauthoryear{Devecchi \& Volonteri}{Devecchi \&
  Volonteri}{2009}]{devecchi2009formation}
Devecchi B.,  Volonteri M.,  2009, \apj, 694, 302

\bibitem[\protect\citeauthoryear{Devecchi, Volonteri, Colpi  \&
  Haardt}{Devecchi et~al.}{2010}]{devecchi2010high}
Devecchi B.,  Volonteri M.,  Colpi M.,   Haardt F.,  2010, \mnras, 409, 1057

\bibitem[\protect\citeauthoryear{Devecchi, Volonteri, Rossi, Colpi  \&
  Portegies~Zwart}{Devecchi et~al.}{2012}]{devecchi2012high}
Devecchi B.,  Volonteri M.,  Rossi E.,  Colpi M.,   Portegies~Zwart S.,  2012,
  \mnras, 421, 1465

\bibitem[\protect\citeauthoryear{Duquennoy, Mayor  \& Halbwachs}{Duquennoy
  et~al.}{1991}]{duquennoy1991multiplicity}
Duquennoy A.,  Mayor M.,   Halbwachs J.-L.,  1991, \aaps, 88, 281

\bibitem[\protect\citeauthoryear{Dutton \& Maccio}{Dutton \&
  Maccio}{2014}]{dutton2014cold}
Dutton A.~A.,  Maccio A.~V.,  2014, \mnras, 441, 3359

\bibitem[\protect\citeauthoryear{Dvorkin, Vangioni, Silk, Uzan  \&
  Olive}{Dvorkin et~al.}{2016}]{dvorkin2016metallicity}
Dvorkin I.,  Vangioni E.,  Silk J.,  Uzan J.-P.,   Olive K.~A.,  2016, \mnras,
  461, 3877

\bibitem[\protect\citeauthoryear{Eggleton}{Eggleton}{1983}]{eggleton1983aproximations}
Eggleton P.~P.,  1983, \apj, 268, 368

\bibitem[\protect\citeauthoryear{Farrell, Groh, Hirschi, Murphy, Kaiser,
  Ekstr{\"o}m, Georgy  \& Meynet}{Farrell et~al.}{2020}]{farrell2020gw190521}
Farrell E.~J.,  Groh J.~H.,  Hirschi R.,  Murphy L.,  Kaiser E.,  Ekstr{\"o}m
  S.,  Georgy C.,   Meynet G.,  2020, arXiv preprint arXiv:2009.06585

\bibitem[\protect\citeauthoryear{{Feng}, {Wang}, {Hu}, {Hu}  \& {Wang}}{{Feng}
  et~al.}{2019}]{tianqin}
{Feng} W.-F.,  {Wang} H.-T.,  {Hu} X.-C.,  {Hu} Y.-M.,   {Wang} Y.,  2019,
  \mn@doi [\prd] {10.1103/PhysRevD.99.123002}, \href
  {https://ui.adsabs.harvard.edu/abs/2019PhRvD..99l3002F} {99, 123002}

\bibitem[\protect\citeauthoryear{Fialkov \& Barkana}{Fialkov \&
  Barkana}{2014}]{fialkov2014rich}
Fialkov A.,  Barkana R.,  2014, \mnras, 445, 213

\bibitem[\protect\citeauthoryear{Fialkov, Barkana, Pinhas  \& Visbal}{Fialkov
  et~al.}{2013}]{fialkov2013complete}
Fialkov A.,  Barkana R.,  Pinhas A.,   Visbal E.,  2013, \mnras, 437, L36

\bibitem[\protect\citeauthoryear{Fialkov, Cohen, Barkana  \& Silk}{Fialkov
  et~al.}{2017}]{fialkov2017constraining}
Fialkov A.,  Cohen A.,  Barkana R.,   Silk J.,  2017, \mnras, 464, 3498

\bibitem[\protect\citeauthoryear{Finkelstein}{Finkelstein}{2016}]{finkelstein2016observational}
Finkelstein S.~L.,  2016, \pasa, 33

\bibitem[\protect\citeauthoryear{{Fragos} et~al.,}{{Fragos}
  et~al.}{2013}]{fragos2013xray}
{Fragos} T.,  et~al., 2013, \mn@doi [\apj] {10.1088/0004-637X/764/1/41}, \href
  {https://ui.adsabs.harvard.edu/abs/2013ApJ...764...41F} {764, 41}

\bibitem[\protect\citeauthoryear{Frebel \& Norris}{Frebel \&
  Norris}{2015}]{frebel2015near}
Frebel A.,  Norris J.~E.,  2015, \araa, 53, 631

\bibitem[\protect\citeauthoryear{{Gair}, {Mandel}, {Miller}  \&
  {Volonteri}}{{Gair} et~al.}{2011}]{gair2011imbhet}
{Gair} J.~R.,  {Mandel} I.,  {Miller} M.~C.,   {Volonteri} M.,  2011, \mn@doi
  [General Relativity and Gravitation] {10.1007/s10714-010-1104-3}, \href
  {https://ui.adsabs.harvard.edu/abs/2011GReGr..43..485G} {43, 485}

\bibitem[\protect\citeauthoryear{G{\"o}tberg, de Mink, McQuinn, Zapartas, Groh
  \& Norman}{G{\"o}tberg et~al.}{2020}]{gotberg2020contribution}
G{\"o}tberg Y.,  de Mink S.,  McQuinn M.,  Zapartas E.,  Groh J.,   Norman C.,
  2020, \aap, 634, A134

\bibitem[\protect\citeauthoryear{Greif, Springel, White, Glover, Clark, Smith,
  Klessen  \& Bromm}{Greif et~al.}{2011}]{greif2011simulations}
Greif T.~H.,  Springel V.,  White S.~D.,  Glover S.~C.,  Clark P.~C.,  Smith
  R.~J.,  Klessen R.~S.,   Bromm V.,  2011, \apj, 737, 75

\bibitem[\protect\citeauthoryear{Greif, Bromm, Clark, Glover, Smith, Klessen,
  Yoshida  \& Springel}{Greif et~al.}{2012}]{greif2012formation}
Greif T.~H.,  Bromm V.,  Clark P.~C.,  Glover S.~C.,  Smith R.~J.,  Klessen
  R.~S.,  Yoshida N.,   Springel V.,  2012, \mnras, 424, 399

\bibitem[\protect\citeauthoryear{Han, Podsiadlowski  \& Lynas-Gray}{Han
  et~al.}{2007}]{han2007binary}
Han Z.,  Podsiadlowski P.,   Lynas-Gray A.,  2007, \mnras, 380, 1098

\bibitem[\protect\citeauthoryear{Han, Ge, Chen  \& Chen}{Han
  et~al.}{2020}]{han2020binary}
Han Z.,  Ge H.,  Chen X.,   Chen H.,  2020, arXiv preprint arXiv:2009.08611

\bibitem[\protect\citeauthoryear{{Hansen}, {Andersen}, {Nordstr{\"o}m},
  {Beers}, {Placco}, {Yoon}  \& {Buchhave}}{{Hansen} et~al.}{2016}]{hansen2016}
{Hansen} T.~T.,  {Andersen} J.,  {Nordstr{\"o}m} B.,  {Beers} T.~C.,  {Placco}
  V.~M.,  {Yoon} J.,   {Buchhave} L.~A.,  2016, \mn@doi [\aap]
  {10.1051/0004-6361/201527409}, \href
  {https://ui.adsabs.harvard.edu/abs/2016A&A...588A...3H} {588, A3}

\bibitem[\protect\citeauthoryear{{Hartwig}, {Volonteri}, {Bromm}, {Klessen},
  {Barausse}, {Magg}  \& {Stacy}}{{Hartwig} et~al.}{2016}]{hartwig2016}
{Hartwig} T.,  {Volonteri} M.,  {Bromm} V.,  {Klessen} R.~S.,  {Barausse} E.,
  {Magg} M.,   {Stacy} A.,  2016, \mn@doi [\mnras] {10.1093/mnrasl/slw074},
  \href {https://ui.adsabs.harvard.edu/abs/2016MNRAS.460L..74H} {460, L74}

\bibitem[\protect\citeauthoryear{Heger, Fryer, Woosley, Langer  \&
  Hartmann}{Heger et~al.}{2003}]{heger2003massive}
Heger A.,  Fryer C.,  Woosley S.,  Langer N.,   Hartmann D.~H.,  2003, \apj,
  591, 288

\bibitem[\protect\citeauthoryear{Hirano \& Bromm}{Hirano \&
  Bromm}{2017}]{hirano2017formation}
Hirano S.,  Bromm V.,  2017, \mnras, 470, 898

\bibitem[\protect\citeauthoryear{Hirano, Hosokawa, Yoshida, Omukai  \&
  Yorke}{Hirano et~al.}{2015}]{hirano2015primordial}
Hirano S.,  Hosokawa T.,  Yoshida N.,  Omukai K.,   Yorke H.~W.,  2015, \mnras,
  448, 568

\bibitem[\protect\citeauthoryear{Hummel, Stacy, Jeon, Oliveri  \& Bromm}{Hummel
  et~al.}{2015}]{hummel2015first}
Hummel J.~A.,  Stacy A.,  Jeon M.,  Oliveri A.,   Bromm V.,  2015, \mnras, 453,
  4136

\bibitem[\protect\citeauthoryear{Hurley, Tout  \& Pols}{Hurley
  et~al.}{2002}]{hurley2002evolution}
Hurley J.~R.,  Tout C.~A.,   Pols O.~R.,  2002, \mnras, 329, 897

\bibitem[\protect\citeauthoryear{Inayoshi, Kashiyama, Visbal  \&
  Haiman}{Inayoshi et~al.}{2016}]{inayoshi2016gravitational}
Inayoshi K.,  Kashiyama K.,  Visbal E.,   Haiman Z.,  2016, \mnras, 461, 2722

\bibitem[\protect\citeauthoryear{Izzard, Glebbeek, Stancliffe  \& Pols}{Izzard
  et~al.}{2009}]{izzard2009population}
Izzard R.~G.,  Glebbeek E.,  Stancliffe R.~J.,   Pols O.~R.,  2009, \aap, 508,
  1359

\bibitem[\protect\citeauthoryear{Jaacks, Finkelstein  \& Bromm}{Jaacks
  et~al.}{2019}]{jaacks2019legacy}
Jaacks J.,  Finkelstein S.~L.,   Bromm V.,  2019, \mnras, 488, 2202

\bibitem[\protect\citeauthoryear{Jeon, Pawlik, Bromm  \&
  Milosavljevi{\'c}}{Jeon et~al.}{2014}]{jeon2014radiative}
Jeon M.,  Pawlik A.~H.,  Bromm V.,   Milosavljevi{\'c} M.,  2014, \mnras, 440,
  3778

\bibitem[\protect\citeauthoryear{Ji, Frebel  \& Bromm}{Ji
  et~al.}{2015}]{ji2015preserving}
Ji A.~P.,  Frebel A.,   Bromm V.,  2015, \mnras, 454, 659

\bibitem[\protect\citeauthoryear{Johnson, Dalla  \& Khochfar}{Johnson
  et~al.}{2013}]{johnson2013first}
Johnson J.~L.,  Dalla V.~C.,   Khochfar S.,  2013, \mnras, 428, 1857

\bibitem[\protect\citeauthoryear{Kinugawa, Inayoshi, Hotokezaka, Nakauchi  \&
  Nakamura}{Kinugawa et~al.}{2014}]{kinugawa2014possible}
Kinugawa T.,  Inayoshi K.,  Hotokezaka K.,  Nakauchi D.,   Nakamura T.,  2014,
  \mnras, 442, 2963

\bibitem[\protect\citeauthoryear{Kinugawa, Nakamura  \& Nakano}{Kinugawa
  et~al.}{2020a}]{kinugawa2020chirp}
Kinugawa T.,  Nakamura T.,   Nakano H.,  2020a, arXiv preprint arXiv:2005.09795

\bibitem[\protect\citeauthoryear{Kinugawa, Nakamura  \& Nakano}{Kinugawa
  et~al.}{2020b}]{kinugawa2020formation}
Kinugawa T.,  Nakamura T.,   Nakano H.,  2020b, arXiv preprint arXiv:2009.06922

\bibitem[\protect\citeauthoryear{Kruckow, Tauris, Langer, Kramer  \&
  Izzard}{Kruckow et~al.}{2018}]{kruckow2018progenitors}
Kruckow M.~U.,  Tauris T.~M.,  Langer N.,  Kramer M.,   Izzard R.~G.,  2018,
  \mnras, 481, 1908

\bibitem[\protect\citeauthoryear{{Kuns}, {Yu}, {Chen}  \& {Adhikari}}{{Kuns}
  et~al.}{2019}]{tiango}
{Kuns} K.~A.,  {Yu} H.,  {Chen} Y.,   {Adhikari} R.~X.,  2019, arXiv e-prints,
  \href {https://ui.adsabs.harvard.edu/abs/2019arXiv190806004K} {p.
  arXiv:1908.06004}

\bibitem[\protect\citeauthoryear{Leigh et~al.,}{Leigh
  et~al.}{2018}]{leigh2018rate}
Leigh N.~W.,  et~al., 2018, \mnras, 474, 5672

\bibitem[\protect\citeauthoryear{{Liao}}{{Liao}}{2020}]{liaow2019}
{Liao} W.,  2020, in American Astronomical Society Meeting Abstracts \#235. p.
  443.03

\bibitem[\protect\citeauthoryear{{Liu} \& {Bromm}}{{Liu} \&
  {Bromm}}{2020a}]{bl2020gw190521}
{Liu} B.,  {Bromm} V.,  2020a, arXiv e-prints, \href
  {https://ui.adsabs.harvard.edu/abs/2020arXiv200911447L} {p. arXiv:2009.11447}

\bibitem[\protect\citeauthoryear{{Liu} \& {Bromm}}{{Liu} \&
  {Bromm}}{2020b}]{boyuan2020}
{Liu} B.,  {Bromm} V.,  2020b, \mn@doi [\mnras] {10.1093/mnras/staa1362}, \href
  {https://ui.adsabs.harvard.edu/abs/2020MNRAS.495.2475L} {495, 2475}

\bibitem[\protect\citeauthoryear{Liu \& Bromm}{Liu \&
  Bromm}{2020c}]{liu2020did}
Liu B.,  Bromm V.,  2020c, \mnras, 497, 2839

\bibitem[\protect\citeauthoryear{Liu \& Li}{Liu \&
  Li}{2006}]{liu2006population}
Liu X.-W.,  Li X.-D.,  2006, \aap, 449, 135

\bibitem[\protect\citeauthoryear{Machida \& Nakamura}{Machida \&
  Nakamura}{2015}]{machida2015accretion}
Machida M.~N.,  Nakamura T.,  2015, \mnras, 448, 1405

\bibitem[\protect\citeauthoryear{{Madau} \& {Dickinson}}{{Madau} \&
  {Dickinson}}{2014}]{madau2014araa}
{Madau} P.,  {Dickinson} M.,  2014, \mn@doi [\araa]
  {10.1146/annurev-astro-081811-125615}, \href
  {https://ui.adsabs.harvard.edu/abs/2014ARA&A..52..415M} {52, 415}

\bibitem[\protect\citeauthoryear{Maeder, Meynet  \& Chiappini}{Maeder
  et~al.}{2015}]{maeder2015first}
Maeder A.,  Meynet G.,   Chiappini C.,  2015, \aap, 576, A56

\bibitem[\protect\citeauthoryear{Mapelli, Giacobbo, Santoliquido  \&
  Artale}{Mapelli et~al.}{2019}]{mapelli2019properties}
Mapelli M.,  Giacobbo N.,  Santoliquido F.,   Artale M.~C.,  2019, \mnras, 487,
  2

\bibitem[\protect\citeauthoryear{Meng, Chen  \& Han}{Meng
  et~al.}{2009}]{meng2009single}
Meng X.,  Chen X.,   Han Z.,  2009, \mnras, 395, 2103

\bibitem[\protect\citeauthoryear{Meynet, Hirschi, Ekstrom, Maeder, Georgy,
  Eggenberger  \& Chiappini}{Meynet et~al.}{2010}]{meynet2010c}
Meynet G.,  Hirschi R.,  Ekstrom S.,  Maeder A.,  Georgy C.,  Eggenberger P.,
  Chiappini C.,  2010, Astronomy \& Astrophysics, 521, A30

\bibitem[\protect\citeauthoryear{{Mirabel}, {Dijkstra}, {Laurent}, {Loeb}  \&
  {Pritchard}}{{Mirabel} et~al.}{2011}]{mirabel2011}
{Mirabel} I.~F.,  {Dijkstra} M.,  {Laurent} P.,  {Loeb} A.,   {Pritchard}
  J.~R.,  2011, \mn@doi [\aap] {10.1051/0004-6361/201016357}, \href
  {https://ui.adsabs.harvard.edu/abs/2011A&A...528A.149M} {528, A149}

\bibitem[\protect\citeauthoryear{{Mirocha} \& {Furlanetto}}{{Mirocha} \&
  {Furlanetto}}{2019}]{Mirocha2019}
{Mirocha} J.,  {Furlanetto} S.~R.,  2019, \mn@doi [\mnras]
  {10.1093/mnras/sty3260}, \href
  {https://ui.adsabs.harvard.edu/\#abs/2019MNRAS.483.1980M} {483, 1980}

\bibitem[\protect\citeauthoryear{{Omukai} \& {Nishi}}{{Omukai} \&
  {Nishi}}{1998}]{omukai1998}
{Omukai} K.,  {Nishi} R.,  1998, \mn@doi [\apj] {10.1086/306395}, \href
  {https://ui.adsabs.harvard.edu/abs/1998ApJ...508..141O} {508, 141}

\bibitem[\protect\citeauthoryear{Pelupessy, van Elteren, de Vries, McMillan,
  Drost  \& Zwart}{Pelupessy et~al.}{2013}]{pelupessy2013astrophysical}
Pelupessy F.~I.,  van Elteren A.,  de Vries N.,  McMillan S.,  Drost N.,
  Zwart S.~P.,  2013, \aap, 557, A84

\bibitem[\protect\citeauthoryear{Portegies~Zwart \& McMillan}{Portegies~Zwart
  \& McMillan}{2018b}]{amuse}
Portegies~Zwart S.,  McMillan S.,  2018b, Astrophysical Recipes.
2514-3433, IOP Publishing

\bibitem[\protect\citeauthoryear{Portegies~Zwart \& McMillan}{Portegies~Zwart
  \& McMillan}{2018a}]{portegies2018astrophysical}
Portegies~Zwart S.,  McMillan S.,  2018a, \araa

\bibitem[\protect\citeauthoryear{Punturo et~al.,}{Punturo
  et~al.}{2010}]{punturo2010einstein}
Punturo M.,  et~al., 2010, Classical and Quantum Gravity, 27, 194002

\bibitem[\protect\citeauthoryear{{Qin}, {Mesinger}, {Park}, {Greig}  \&
  {Mu{\~n}oz}}{{Qin} et~al.}{2020}]{21cm2020}
{Qin} Y.,  {Mesinger} A.,  {Park} J.,  {Greig} B.,   {Mu{\~n}oz} J.~B.,  2020,
  \mn@doi [\mnras] {10.1093/mnras/staa1131}, \href
  {https://ui.adsabs.harvard.edu/abs/2020MNRAS.495..123Q} {495, 123}

\bibitem[\protect\citeauthoryear{Robson, Cornish  \& Liug}{Robson
  et~al.}{2019}]{robson2019construction}
Robson T.,  Cornish N.~J.,   Liug C.,  2019, Classical and Quantum Gravity, 36,
  105011

\bibitem[\protect\citeauthoryear{Ryu, Tanaka  \& Perna}{Ryu
  et~al.}{2016}]{ryu2016formation}
Ryu T.,  Tanaka T.~L.,   Perna R.,  2016, \mnras, 456, 223

\bibitem[\protect\citeauthoryear{Salvaterra, Haardt, Volonteri  \&
  Moretti}{Salvaterra et~al.}{2012}]{salvaterra2012limits}
Salvaterra R.,  Haardt F.,  Volonteri M.,   Moretti A.,  2012, \aap, 545, L6

\bibitem[\protect\citeauthoryear{Sana et~al.,}{Sana
  et~al.}{2012}]{sana2012binary}
Sana H.,  et~al., 2012, \sci, 337, 444

\bibitem[\protect\citeauthoryear{{Sana} et~al.,}{{Sana}
  et~al.}{2013}]{sana2013}
{Sana} H.,  et~al., 2013, \mn@doi [\aap] {10.1051/0004-6361/201219621}, \href
  {https://ui.adsabs.harvard.edu/abs/2013A&A...550A.107S} {550, A107}

\bibitem[\protect\citeauthoryear{Sarmento, Scannapieco  \& Cohen}{Sarmento
  et~al.}{2018}]{sarmento2018following}
Sarmento R.,  Scannapieco E.,   Cohen S.,  2018, \apj, 854, 75

\bibitem[\protect\citeauthoryear{Schauer, Liu  \& Bromm}{Schauer
  et~al.}{2019}]{schauer2019constraining}
Schauer A.~T.,  Liu B.,   Bromm V.,  2019, \apj, 877, L5

\bibitem[\protect\citeauthoryear{Secunda, Cen, Kimm, G{\"o}tberg  \& de
  Mink}{Secunda et~al.}{2020}]{secunda2020delayed}
Secunda A.,  Cen R.,  Kimm T.,  G{\"o}tberg Y.,   de Mink S.~E.,  2020, \apj,
  901, 72

\bibitem[\protect\citeauthoryear{Shao \& Li}{Shao \&
  Li}{2015}]{shao2015population}
Shao Y.,  Li X.-D.,  2015, \apj, 802, 131

\bibitem[\protect\citeauthoryear{Stacy \& Bromm}{Stacy \&
  Bromm}{2013}]{stacy2013constraining}
Stacy A.,  Bromm V.,  2013, \mnras, 433, 1094

\bibitem[\protect\citeauthoryear{Stacy \& Bromm}{Stacy \&
  Bromm}{2014}]{stacy2014first}
Stacy A.,  Bromm V.,  2014, \apj, 785, 73

\bibitem[\protect\citeauthoryear{Stacy, Greif  \& Bromm}{Stacy
  et~al.}{2010}]{stacy2010first}
Stacy A.,  Greif T.~H.,   Bromm V.,  2010, \mnras, 403, 45

\bibitem[\protect\citeauthoryear{Stacy, Greif  \& Bromm}{Stacy
  et~al.}{2012}]{stacy2012first}
Stacy A.,  Greif T.~H.,   Bromm V.,  2012, \mnras, 422, 290

\bibitem[\protect\citeauthoryear{Stacy, Bromm  \& Lee}{Stacy
  et~al.}{2016}]{stacy2016building}
Stacy A.,  Bromm V.,   Lee A.~T.,  2016, \mnras, 462, 1307

\bibitem[\protect\citeauthoryear{Stanway, Eldridge  \& Becker}{Stanway
  et~al.}{2016}]{stanway2016stellar}
Stanway E.~R.,  Eldridge J.,   Becker G.~D.,  2016, \mnras, 456, 485

\bibitem[\protect\citeauthoryear{Sugimura, Matsumoto, Hosokawa, Hirano  \&
  Omukai}{Sugimura et~al.}{2020}]{sugimura2020birth}
Sugimura K.,  Matsumoto T.,  Hosokawa T.,  Hirano S.,   Omukai K.,  2020, \apj,
  892, L14

\bibitem[\protect\citeauthoryear{Susa}{Susa}{2019}]{susa2019merge}
Susa H.,  2019, \apj, 877, 99

\bibitem[\protect\citeauthoryear{Susa, Hasegawa  \& Tominaga}{Susa
  et~al.}{2014}]{susa2014mass}
Susa H.,  Hasegawa K.,   Tominaga N.,  2014, \apj, 792, 32

\bibitem[\protect\citeauthoryear{Tanikawa, Susa, Yoshida, Trani  \&
  Kinugawa}{Tanikawa et~al.}{2020a}]{tanikawa2020merger}
Tanikawa A.,  Susa H.,  Yoshida T.,  Trani A.~A.,   Kinugawa T.,  2020a, arXiv
  preprint arXiv:2008.01890

\bibitem[\protect\citeauthoryear{Tanikawa, Yoshida, Kinugawa, Takahashi  \&
  Umeda}{Tanikawa et~al.}{2020b}]{tanikawa2020fitting}
Tanikawa A.,  Yoshida T.,  Kinugawa T.,  Takahashi K.,   Umeda H.,  2020b,
  \mnras, 495, 4170

\bibitem[\protect\citeauthoryear{Toonen, Nelemans  \& Zwart}{Toonen
  et~al.}{2012}]{toonen2012supernova}
Toonen S.,  Nelemans G.,   Zwart S.~P.,  2012, \aap, 546, A70

\bibitem[\protect\citeauthoryear{Tornatore, Ferrara  \& Schneider}{Tornatore
  et~al.}{2007}]{tornatore2007population}
Tornatore L.,  Ferrara A.,   Schneider R.,  2007, \mnras, 382, 945

\bibitem[\protect\citeauthoryear{{Visbal}, {Haiman}  \& {Bryan}}{{Visbal}
  et~al.}{2015}]{visbal2015}
{Visbal} E.,  {Haiman} Z.,   {Bryan} G.~L.,  2015, \mn@doi [\mnras]
  {10.1093/mnras/stv1941}, \href
  {https://ui.adsabs.harvard.edu/abs/2015MNRAS.453.4456V} {453, 4456}

\bibitem[\protect\citeauthoryear{Wall, Mac~Low, McMillan, Klessen, Zwart  \&
  Pellegrino}{Wall et~al.}{2020}]{wall2020modelling}
Wall J.~E.,  Mac~Low M.-M.,  McMillan S.~L.,  Klessen R.~S.,  Zwart S.~P.,
  Pellegrino A.,  2020, arXiv preprint arXiv:2003.09011

\bibitem[\protect\citeauthoryear{Wise, Turk, Norman  \& Abel}{Wise
  et~al.}{2011}]{wise2011birth}
Wise J.~H.,  Turk M.~J.,  Norman M.~L.,   Abel T.,  2011, \apj, 745, 50

\bibitem[\protect\citeauthoryear{Xu, Ahn, Wise, Norman  \& O'Shea}{Xu
  et~al.}{2014}]{xu2014heating}
Xu H.,  Ahn K.,  Wise J.~H.,  Norman M.~L.,   O'Shea B.~W.,  2014, \apj, 791,
  110

\bibitem[\protect\citeauthoryear{Xu, Norman, O’Shea  \& Wise}{Xu
  et~al.}{2016a}]{xu2016late}
Xu H.,  Norman M.~L.,  O’Shea B.~W.,   Wise J.~H.,  2016a, \apj, 823, 140

\bibitem[\protect\citeauthoryear{Xu, Ahn, Norman, Wise  \& O’Shea}{Xu
  et~al.}{2016b}]{xu2016x}
Xu H.,  Ahn K.,  Norman M.~L.,  Wise J.~H.,   O’Shea B.~W.,  2016b, \apj,
  832, L5

\bibitem[\protect\citeauthoryear{Zwart et~al.,}{Zwart
  et~al.}{2009}]{zwart2009multiphysics}
Zwart S.~P.,  et~al., 2009, \na, 14, 369

\bibitem[\protect\citeauthoryear{Zwart, McMillan, van Elteren, Pelupessy  \& de
  Vries}{Zwart et~al.}{2013}]{zwart2013multi}
Zwart S. F.~P.,  McMillan S.~L.,  van Elteren A.,  Pelupessy F.~I.,   de Vries
  N.,  2013, Computer Physics Communications, 184, 456

\makeatother
\end{thebibliography}

\appendix
\section{Key cluster and binary parameters}
\label{a1}
Table~\ref{t2} summarizes the key cluster evolution and statistical parameters for the 18 initial condition models explored in Sec.~\ref{s4}, which are defined in Table~\ref{t1}. In general, the fraction of stars unbound to the cluster (i.e. escape fraction), stellar collision rate, efficiencies from X-ray binaries (XRBs) and binary black hole (BBH) mergers increase with larger fragmentation timescale $t_{\rm frag}$ (i.e. larger number of stars), smaller accretion timescale $t_{\rm acc}$ (i.e. lower total mass and smaller cluster size) and more top-heavy IMF. While the fraction of stellar mass in (hard) binaries, binary hardening parameter and median binary separation follow the opposite trend. The fraction of stars in hard binaries decreases with $t_{\rm frag}$, $t_{\rm acc}$ and less top-heavy IMF, in contrary to the trend for the fraction of binaries in multiple systems. These outcomes can be explained by the following facts: 
\begin{itemize}
    \item Close binaries are formed by scatters between stars, which will be enhanced with larger number of stars and smaller cluster size.
    \item More frequent scatters and more top-heavy IMF enhance dispersion of multiple systems, resulting in binary populations more biased to massive stars, since both multiple systems and low-mass companions are vulnerable to disruption from massive stars.
    \item Both XRBs and BBH mergers are descendants of close binaries.
\end{itemize}
The reader is referred to Sec.~\ref{s4} for detailed interpretations.

\begin{figure*}
\includegraphics[width=1.7\columnwidth]{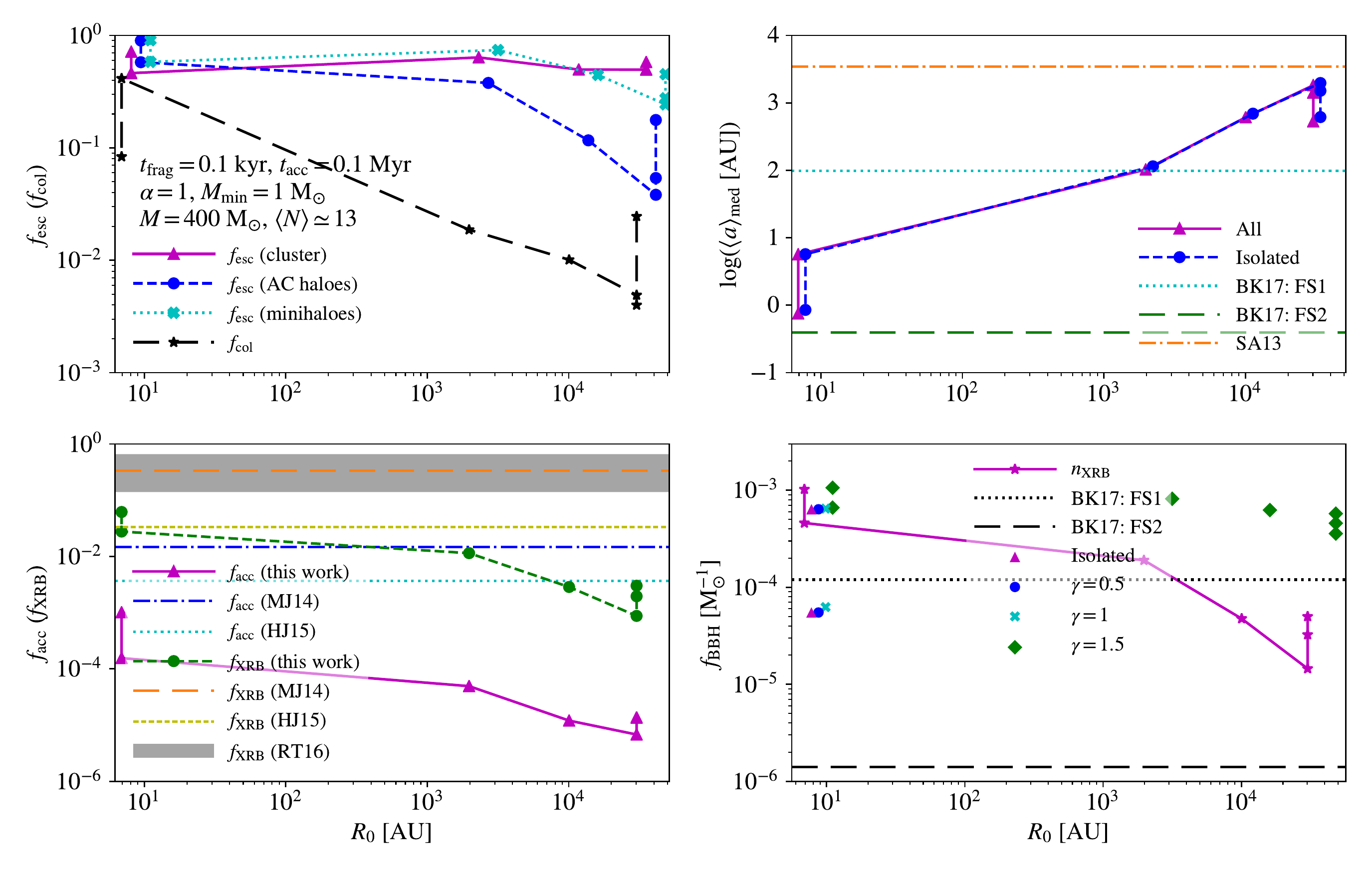}
\caption{Dependence of the escape and collision fractions, median separation of binaries, efficiency of X-ray binaries in terms of accreted mass and number of stars ($f_{\rm XRB}\equiv 2N_{\rm XRB}/N_{\rm tot}$, given the total numbers of XRBs and stars, $N_{\rm XRB}$ and $N_{\rm tot}$), as well as efficiency of BBH mergers, on cluster size and initial distribution of $a$, from the seven models explored in Sec.~\ref{s4.4} (see Table~\ref{t1}). At the large size end, from high (low) to low (high) $\langle a\rangle_{\rm med}$ ($f_{\rm esc}/$/$f_{\rm col}$/$f_{\rm acc}$/$f_{\rm XRB}$/$f_{\rm BBH}$) values, we show the cases for the fiducial distribution of $a$, $a_{\min}=a_{\rm lobe}$ (close) and $a_{\max}\sim \min(a_{\mathrm{L_{1}}, n})$ (tight). 
Similarly, at the small size end, we have two cases with (higher $\langle a\rangle_{\rm med}$) and without suppressed collision. In the former case, the stellar radii are all set to $10^{-4}\ \mathrm{AU}\ll R_{\star,\rm ZAMS}$ with other conditions unchanged. We have applied offsets in $R$ values to avoid overlapping of the data points. We also plot the number of XRBs per unit stellar mass $n_{\rm XRB}$ on top of $f_{\rm BBH}$, as it is an approximation to $f_{\rm BBH}$ with optimal isolated common envelope evolution. For comparison, we show the median separation from \citealt{stacy2013constraining} (SA13) for protostars, the results for $\langle a\rangle_{\rm med}$ and $f_{\rm BBH}$ from the FS1 and FS2 models used in \citealt{belczynski2017likelihood} (BK17), the efficiency of XRBs from \citealt{jeon2014radiative} (JM14), \citealt{hummel2015first} (HJ15), \citealt{ryu2016formation} (RT16, for their small-scale model).}
\label{size}
\end{figure*}

Fig.~\ref{size} illustrates the dependence of several key parameters on cluster size and initial distribution of binary separation, with the 7 models compared in Sec.~\ref{s4.4}.
It is shown that our `standard' Phase 1 picture introduced in Sec.~\ref{s2.1} and \ref{s2.2} predicts much less close binaries than literature results, such that the signals of Pop~III X-ray binaries (XRBs) and binary black hole (BBH) mergers are lower by up to three orders of magnitude. 

\begin{table*}
    \centering
    \caption{Key cluster evolution and binary statistical parameters and outcomes for the initial condition models listed in Table~\ref{t1}. Here $f_{\rm esc}$ is the fraction of stars unbound to the cluster. $f_{\rm col}$ is a measurement of the frequency of collision defined as $f_{\rm col}\equiv N_{\rm col}/N_{\rm tot}$ (number) and $f_{\rm col}\equiv M_{\rm col}/(2M_{\rm tot})$ (mass), where $N_{\rm col}$ and $M_{\rm col}$ are the total number of and stellar mass involved in collision events, given the total number and mass of stars initially in the simulations, $N_{\rm tot}$ and $M_{\rm tot}$. $f_{\rm hard}$ is the binary hardening parameter defined in Sec~\ref{s3}. $f_{\rm B}$ ($f_{\rm HDB}$) is the fraction of stars in (hard) binaries. 
    $f_{\rm mul}\equiv 1-f_{\rm B,iso}/f_{\rm B}$ is the fraction of binaries that reside in multiple (up to 4 stars) systems (i.e. with companion single stars or binaries), where $f_{\rm B,iso}$ is the fraction of stars in isolated binaries. $n_{\rm XRB}\equiv N_{\rm XRB}/M_{\rm tot}$ is efficiency of XRBs per unit stellar mass, while $f_{\rm acc}\equiv M_{\rm acc}/M_{\rm tot}$ is the efficiency of accretion onto XRBs, given the total number and accreted mass, $N_{\rm XRB}$ and $M_{\rm acc}$, assuming Eddington accretion with an efficiency $\epsilon=0.125$. $f_{\rm BBH}\equiv N_{\rm BBH}/M_{\rm tot}$ is the efficiency of BBH mergers, where $N_{\rm BBH}$ is the total number of BBHs that merge within a Hubble time. For each model, the first row shows the results with respect to stellar mass, where $f_{\rm BBH}$ is for isolated evolution, while the second row shows the results with respect to star number, where $f_{\rm BBH}$ is for evolution in dense environments whose stellar density profile has a power-law index $\gamma=1.5$. If a quantity is irrelevant to mass or number, its value is given in the first row. For \texttt{tf1e3ta1e6a1m1}, we have shown the results before and after PISNe at $t=2$ and 4~Myr, which correspond to $\sim 6$ (20) and $\sim 11$ (40) relaxation (dynamical) timescales, respectively.}
    \begin{tabular}{cccccccccccc}
    \hline
         Model (mass/isolated) & $f_{\rm esc}$ & $f_{\rm col}$ & $f_{\rm hard}$ & $f_{\rm B}$ & $f_{\rm HDB}$ & $f_{\rm mul}$ & $\langle a\rangle_{\rm med}$ & $\langle a_{\rm HDB}\rangle_{\rm med}$ & $n_{\rm XRB}$ & $f_{\rm acc}$ & $f_{\rm BBH}$ \\
         (number/$\gamma=1.5$) & & & & & & & $\log$ [AU] & $\log$ [AU] & $[\rm M_{\odot}^{-1}]$ & & $[\rm M_{\odot}^{-1}]$\\
         \hline 
         \texttt{tf1e2ta1e5a1m1} & 0.216 & 6.02e-3 & 0.357 & 0.694 & 0.348 & 0.390 & 3.260 & 2.857 & - & 6.79e-6 & 0 \\
         \texttt{(fiducial)} & 0.495 & 4.90e-3 & - & 0.317 & 0.121 & 0.392 & - & - & 1.45e-5 & - & 3.58e-4\\
         \texttt{tf1e2ta1e5a1m10} & 0.294 & 1.05e-2 & 0.266 & 0.566 & 0.259 & 0.440 & 2.847 & 2.322 & - & 8.64e-6 & 0 \\
         & 0.399 & 9.98e-3 & - & 0.417 & 0.167 & 0.430 & - & - & 4.00e-5 & - & 1.28e-3 \\
         \texttt{tf1e2ta1e5a017m1} & 0.309 & 1.24e-2 & 0.292 & 0.551 & 0.230 & 0.451 & 2.913 & 2.262 & - & 5.95e-6 & 0 \\
         & 0.406 & 1.16e-2 & - & 0.422 & 0.160 & 0.443 & - & - & 3.25e-5 & - & 2.06e-3 \\
         \texttt{tf1e2ta1e5a0m1} & 0.297 & 1.16e-2 & 0.311 & 0.565 & 0.243 & 0.430 & 2.943 & 2.319 & - & 1.37e-5 & 0\\
         & 0.399 & 1.09e-2 & - & 0.421 & 0.161 & 0.422 & - & - & 5.50e-5 & - & 1.30e-3 \\
         \texttt{tf1e2ta1e5a05m1} & 0.276 & 8.72e-3 & 0.278 & 0.593 & 0.282 & 0.421 & 2.978 & 2.514 & - & 1.35e-5 & 0\\
         & 0.447 & 7.37e-3 & - & 0.369 & 0.149 & 0.410 & - & - & 5.00e-5 & - & 1.26e-3 \\
         \texttt{tf1e2ta1e5a15m1} & 0.143 & 5.12e-3 & 0.373 & 0.775 & 0.366 & 0.366 & 3.296 & 2.955 & - & 2.30e-6 & 0\\
         & 0.515 & 4.03e-3 & - & 0.226 & 0.076 & 0.382 & - & - & 1.25e-5 & - & 2.58e-4 \\
         \texttt{tf1e2ta1e5a2m10} & 0.254 & 8.58e-3 & 0.370 & 0.608 & 0.322 & 0.404 & 3.001 & 2.487 & - & 0 & 0 \\
         & 0.412 & 8.45e-3 & - & 0.370 & 0.150 & 0.410 & - & - & 0 & - & 5.60e-4 \\
         \hline
         \texttt{tf1e1ta1e5a1m1} & 0.110 & 3.06e-3 & 0.452 & 0.840 & 0.382 & 0.269 & 3.725 & 3.314 & - & 9.45e-7 & 0\\
         & 0.412 & 2.89e-3 & - & 0.386 & 0.153 & 0.310 & - & - & 5.00e-6 & - & 2.28e-4 \\
         \texttt{tf1e3ta1e5a1m1} & 0.326 & 1.76e-2 & 0.258 & 0.571 & 0.259 & 0.476 & 2.777 & 2.171 & - & 1.81e-5 & 0 \\
         & 0.531 & 1.32e-2 & - & 0.292 & 0.098 & 0.441 & - & - & 5.50e-5 & - & 1.18e-3 \\
         \hline
         \texttt{tf1e2ta3e5a1m1} & 0.169 & 3.55e-3 & 0.390 & 0.746 & 0.435 & 0.421 & 3.825 & 3.471 & - & 1.29e-6 & 0 \\
         & 0.476 & 2.64e-3 & - & 0.299 & 0.132 & 0.423 & - & - & 4.48e-6 & - & 2.91e-4 \\
         \texttt{tf1e3ta1e6a1m1\_2Myr} & 0.163 & 3.34e-3 & 0.470 & 0.684 & 0.375 & 0.617 & 4.102 & 3.547 & - & 1.26e-6 & 0\\
         & 0.405 & 2.71e-3 & - & 0.279 & 0.111 & 0.554 & - & - & 4.19e-6 & - & 6.19e-4 \\
         \texttt{tf1e3ta1e6a1m1\_4Myr} & 0.251 & 3.34e-3 & 1.330 & 0.369 & 0.122 & 0.365 & 4.554 & 3.236 & - & 6.23e-7 & 0\\
         & 0.697 & 2.71e-3 & - & 0.139 & 0.020 & 0.241 & - & - & 1.86e-6 & - & 8.15e-5 \\
         \hline
         \texttt{fiducial\_vir} & 0.143 & 2.88e-3 & 0.593 & 0.750 & 0.239 & 0.489 & 3.592 & 3.061 & - & 1.23e-6 & 0\\
         & 0.392 & 2.26e-3 & - & 0.373 & 0.088 & 0.476 & - & - & 5.00e-6 & - & 2.20e-4 \\
         \hline
         \texttt{fiducial\_small} & 0.216 & 1.43e-2 & 0.348 & 0.687 & 0.451 & 0.395 & 2.786 & 2.487 & - & 1.20e-5 & 0 \\
         & 0.497 & 1.01e-2 & - & 0.314 & 0.159 & 0.395 & - & - & 4.75e-5 & - & 6.25e-4 \\
         \texttt{fiducial\_comp} & 0.323 & 3.09e-2 & 0.295 & 0.649 & 0.542 & 0.233 & 2.010 & 1.834 & - & 4.93e-5 & 0\\
         & 0.637 & 1.87e-2 & - & 0.279 & 0.192 & 0.243 & - & - & 1.90e-4 & - & 8.17e-4 \\
         \texttt{fiducial\_greif} & 0.559 & 0.946 & 3.275 & 0.616 & 0.575 & 0.056 & 0.754 & 0.689 & - & 1.55e-4 & 5.50e-5\\
         & 0.462 & 0.414 & - & 0.286 & 0.258 & 0.071 & - & - & 4.60e-4 & - & 6.63e-4 \\
         \texttt{fiducial\_greif\_ncol} & 0.483 & 0.196 & 0.435 & 0.591 & 0.572 & 0.112 & -0.123 & -0.193 & - & 1.01e-3 & 6.38e-4\\
         & 0.715 & 8.34e-2 & - & 0.245 & 0.228 & 0.125 & - & - & 1.03e-3 & - & 1.06e-3 \\
         \texttt{fiducial\_tight} & 0.221 & 5.57e-3 & 0.369 & 0.690 & 0.360 & 0.374 & 3.150 & 2.686 & - & 1.37e-5 & 0\\
         & 0.497 & 3.99e-3 & - & 0.329 & 0.134 & 0.379 & - & - & 3.25e-5 & - & 4.58e-4 \\
         \texttt{fiducial\_close} & 0.286 & 3.33e-2 & 0.492 & 0.649 & 0.446 & 0.355 & 2.723 & 2.379 & - & 1.35e-5 & 0\\
         & 0.579 & 2.46e-2 & - & 0.304 & 0.169 & 0.344 & - & - & 5.00e-5 & - & 5.75e-4 \\
         \hline
    \end{tabular}
    \label{t2}
\end{table*}

\section{Reproducing the results of previous studies}
\label{a2}

\begin{figure*}
\includegraphics[width=1.7\columnwidth]{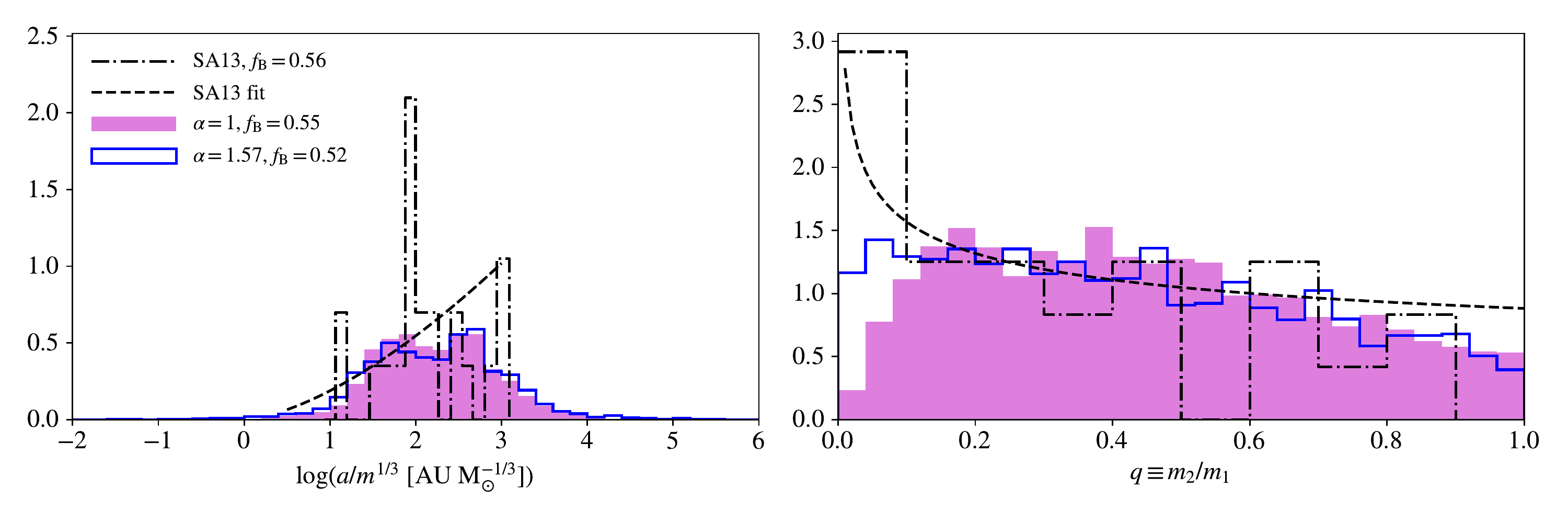}
\caption{Binary statistics from two Phase 1 models with $M_{\rm min}=0.3\ \rm M_{\odot}$, $t_{\rm frag=}=40$~yr, $t_{\rm acc}=5$~kyr, $\alpha=1$ (purple histograms) and 1.57 (blue solid contours), in comparison with the results from \citealt{stacy2013constraining} (SA13, dashed-dotted contours for raw data, dashed curves for pow-law fits of the cumulative distribution functions). Here $m\equiv m_{1}=m_{2}$ is the total mass of binary.}
\label{comp4}
\end{figure*}

\begin{figure*}
\includegraphics[width=1.7\columnwidth]{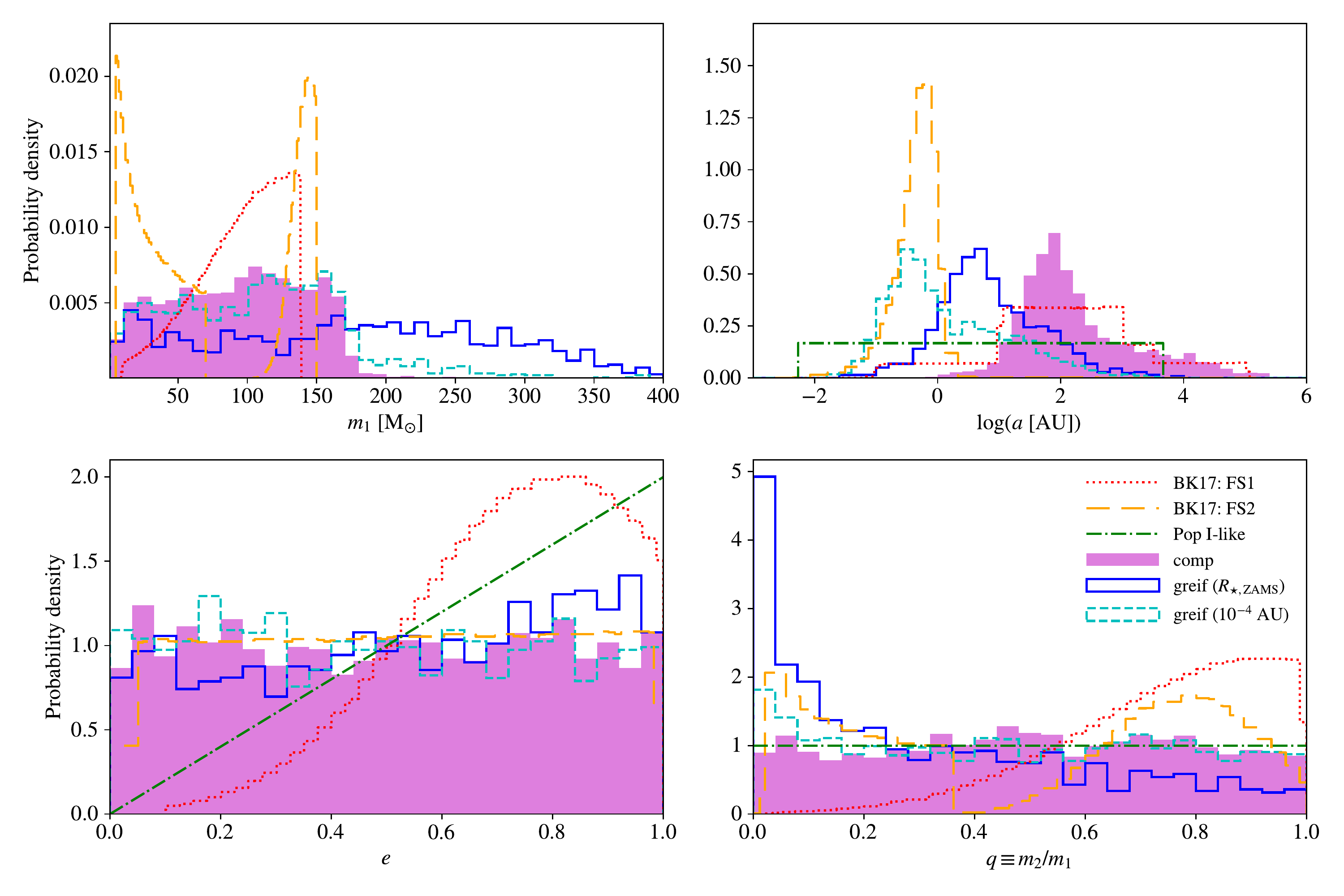}
\caption{Binary statistics from \texttt{tf1e2ta1e5a1m1\_comp} (purple histograms), \texttt{tf1e2ta1e5a1m1\_greif} (blue solid contours) and \texttt{tf1e2ta1e5a1m1\_greif\_ncol} (cyan dashed contours), as defined in Table~\ref{t1}. The first is meant to reproduce the results of the FS1 model (red dotted contours) in \citealt{belczynski2017likelihood} (BK17). While the last two is to reproduce the results of the FS2 model (orange long-dashed contours) in BK17.}
\label{comp5}
\end{figure*}

As mentioned in Sec.~\ref{s4}, the discrepancies between our results with literature results are caused by different
initial conditions, especially for cluster sizes. We here tune our model parameters to mimic the conditions in \citealt{stacy2013constraining} (SA13) and the FS1 and FS2 models in \citealt{belczynski2017likelihood} (BK17) based on \citet{ryu2016formation} (RT16). It turns out that our models can well reproduce their results when the initial conditions are constrained, as already shown in Fig.~\ref{size}.

In SA13, 10 minihaloes are simulated for 5~kyr, and in total 90 Pop~III protostars remain in the disk at the end, with masses $m_{\star}\sim 0.3-30\ \rm M_{\odot}$ and IMF slopes $\alpha\sim 1-1.57$ (see their fig.~6). Protostellar feedback is not included in the simulation. Therefore, SA13 provides snapshots of Pop~III systems in the middle of Phase 1. In light of this, we consider two models with $\alpha=1$ and 1.57, and other parameters fixed as $M_{\rm min}=0.3\ \rm M_{\odot}$, $t_{\rm frag=}=40$~yr and $t_{\rm acc}=5$~kyr, such that the global properties of the cluster match those in SA13. We generate 1000 protostar clusters for each model and calculate the binary statistics, which are compared with the results in SA13 (see their fig.~8 and 9), as shown in Fig.~\ref{comp4}. Our results agree well with those in SA13 in terms of the fraction of stars in binaries ($52-55\%$ vs. 56\%), as well as distributions of separation and mass ratio, especially for $a/m^{1/3}\lesssim 10^{3}\ \rm AU\ M_{\odot}^{-1/3}$. However, our models under-predict binaries with $q\lesssim 0.1$ by up to a factor of $\sim 3$. This implies that the way we assign parent/hosts to newly formed fragments favors low-mass existing fragments too much. In future work, we will explore the situations where the rate of forming fragments is proportional to $m_{\rm frag}^{\beta}$ with $\beta>1$, given the mass of the existing fragment $m_{\rm frag}$.

In BK17 and RT16, the FS1 and FS2 models are initialized according to the properties of protostar clusters in SA13 and \citet{greif2012formation}, with cluster/disk sizes $R_{0}\sim 2\times 10^{3}$ and 7~AU, respectively. Their N-body simulations are run for 5~Myr and $\gtrsim 1$~kyr, respectively, by which the evolution of binary statistics have fully saturated, such that their results are directly comparable to ours during Phase 2. We then consider 3 models with global cluster properties and IMF\footnote{We do not tune the IMF for simplicity. Besides, it has been shown in Sec.~\ref{s4.3} that the IMF has minor effects on binary orbital parameters.} fixed to the fiducial choices and cluster sizes set to the BK17 values (see Table~\ref{t1}). To be specific, we have $R_{0}=2\times 10^{3}$~AU in \texttt{tf1e2ta1e5a1m1\_comp} to match FS1. We have $R_{0}=7$~AU in \texttt{tf1e2ta1e5a1m1\_greif} and \texttt{tf1e2ta1e5a1m1\_greif\_ncol} to match FS2. Here stellar collision is suppressed by setting the stellar radii to $10^{-4}\ll R_{\star,\rm ZAMS}$ in \texttt{tf1e2ta1e5a1m1\_greif\_ncol}, since stellar collision is ignored in RT16. It is shown in Fig.~\ref{comp5} that the distributions of eccentricity and separation in FS1 are well reproduced by \texttt{tf1e2ta1e5a1m1\_comp}. While only \texttt{tf1e2ta1e5a1m1\_greif\_ncol} can roughly reproduce the distribution of separation in FS2, although the fraction of very close binaries with $a\lesssim 1$~AU is underestimated by a factor of $\sim 2$. This highlights the importance of stellar collision, which brings significant differences between \texttt{tf1e2ta1e5a1m1\_greif} and \texttt{tf1e2ta1e5a1m1\_greif\_ncol}/FS2. The remaining discrepancies between FS2 and \texttt{tf1e2ta1e5a1m1\_greif\_ncol} in the distributions of primary mass and mass ratio may be caused by different IMFs, numerical implementations of N-body dynamics and procedures of identifying binaries.

\bsp	
\label{lastpage}

\end{document}